\documentclass[%    
onecolumn,
preprint, % UDOFORMAT
bibnotes,
amsmath,amssymb,
aps,
pra,
floatfix,
]{revtex4-2}
\usepackage[english]{babel}
\usepackage{enumerate}
\usepackage{xspace}
\usepackage{setspace}
\usepackage{yfonts}

\makeatletter\AtBeginDocument{\let\@elt\relax}\makeatother
\usepackage{graphicx}% Include figure files 
\usepackage{dcolumn}% Align table columns on decimal point
\usepackage{bm}% bold math
\usepackage{mathrsfs}
\usepackage{longtable}
\usepackage{stmaryrd}  
\usepackage{hyperref}% add hypertext capabilities
\usepackage{amsthm}\newtheorem{thm}{Theorem}[subsection]
\newtheorem{lem}[thm]{Lemma}
\newtheorem{prop}[thm]{Proposition}
\newtheorem{cor}[thm]{Corollary}
\newtheorem{defn}[thm]{Definition}
\theoremstyle{remark}

\usepackage{dsfont}
\usepackage{bbm}
\usepackage{color} 
\usepackage{mathtools}

\newcommand{\doofy}[1]{\ignorespaces}

\usepackage{changes}
\newcommand{\1}{\mathbf{1}}
\newcommand{\ind}{\mathbbm{1}}
\newcommand{\w}{\mathbbm{w}}
\newcommand{\av}{\mathbbm{a}}
\newcommand{\PW}{\mathbf{P}}

\DeclareMathOperator{\diag}{diag}
\DeclareMathOperator{\cut}{cut}

\renewcommand{\P}{\mathbb{P}}
\renewcommand{\hat}{\widehat}
\renewcommand{\tilde}{\widetilde}
\newcommand{\TN}{\mathbb{T}^{[N]}}
\newcommand{\DeltaU}{u_{0}}

\newcommand{\Z}{\mathbb{Z}}
\newcommand{\e}{\mathbf{e}}
\newcommand{\N}{\mathbb{N}}
\newcommand{\R}{\mathbb{R}}

\newcommand{\E}{\mathbb{E}}

\newcommand{\U}{\mathscr{U}}
\newcommand{\V}{\mathscr{V}}
\newcommand{\s}{\mathcal{S}}%{\operatorname{size}}
\newcommand{\dur}{\mathcal{D}}
\newcommand{\NN}{\mathscr{N}}
\newcommand{\fixF}{\mathscr{F}}
\newcommand{\p}{p}
\renewcommand{\d}{\mathrm{d}}
\newcommand{\MTP}{\mathscr{M}_{\hat{T}_{g}}(\nu)}
\newcommand{\MT}{\mathscr{M}_{\hat{T}_{g}}}
\newcommand{\Vk}{\mathbf{k}}
\newcommand{\omegaVk}{\mathbf{k}}
\newcommand{\Vu}{u}
\newcommand{\Va}{a}
\newcommand{\VU}{U}
\newcommand{\VV}{U'}
\newcommand{\Vp}{p}

\newcommand{\Eq}[1]{Eq.~(\ref{#1})}
\newcommand{\Fig}[1]{Fig.~\ref{#1}}

\usepackage{soul}
\usepackage{todonotes} % UDOFORMAT
\setuptodonotes{backgroundcolor=red!10,linecolor=black!100,size=\tiny}

\makeatletter
\if@todonotes@disabled
     
\else

\fi
\makeatother
\newif\iffiginline
\figinlinetrue

\begin{document}
\title{A rigorous stochastic theory for spike pattern formation in recurrent neural networks with arbitrary connection topologies}%The effect of the connection topology of finite neural networks on their spike patterns and avalanche statistics}% Force line breaks with \\

% UE: Je länger man darüber nachdenkt, umso länger werden die Titel... deswegen höre ihc nach den drei folgenden Vorschlägen besser auf, darüber nachzudenken. b) gefällt mir im Moment noch am besten:
%MHK: Mir gefällt b) auch sehr gut. "exact" liefert eigentlich keine Erkenntnis, aber angesichts der ambitionierten Zeitschriftenwahl sollte man auch etwas auf die Pauke hauen!! 
% a) Activation patterns and avalanche statistics derived from the connection topology of recurrent spiking networks
% b) An exact quantitative theory for spike pattern formation in recurrent networks with arbitrary connection topologies
% c) Understanding avalanche formation and spike pattern generation in recurrent networks with arbitrarily complex positive interaction topologies 
% Maik: Hier ein paar kürzere Vorschläge
% d) Exact avalanche and spike pattern statistics on networks of spiking neurons
% e) Theory of avalanches on networks of excitatory spiking neurons
% f) closed form avalanche assembly distributions on networks of excitatory spiking neurons relate to graph-theoretic measures. 

\author{Maik Schünemann${}^1$, Udo Ernst${}^1$, and Marc Kesseböhmer${}^2$}

\address{
    ${}^1$ Computational Neurophysics Lab, Institute for Theoretical Physics, University of Bremen, Bremen, Germany\\
    ${}^2$ Institute for Dynamical Systems, Dept.\ of Mathematics and Computer Science, University of Bremen, Bremen, Germany}

\date{\today}% It is always \today, today,

% ------------------------------------------------------------------------------

\begin{abstract}
% Was beobachten wir im Gehirn, und wofür könnten diese Vorgänge interessant sein?
Cortical networks exhibit synchronized activity which often occurs in spontaneous events in the form of spike avalanches. Since synchronization has been causally linked to central aspects of brain function such as selective signal processing and integration of stimulus information, participating in an avalanche is a form of a \emph{transient} synchrony which temporarily creates neural assemblies and hence might especially be useful for implementing \emph{flexible} information processing.
% Was wollen wir in diesem Kontext verstehen?
For understanding how assembly formation supports neural computation, it is therefore essential to establish a comprehensive theory of how network structure and dynamics interact to generate specific avalanche patterns and sequences. Here we derive exact avalanche distributions for a finite network of recurrently coupled spiking neurons with arbitrary non-negative interaction weights, which is made possible by formally mapping the model dynamics to a linear, random dynamical system on the $N$-torus and by exploiting self-similarities inherent in the phase space. We introduce the notion of relative unique ergodicity and show that this property is guaranteed if the system is driven by a time-invariant Bernoulli process. This approach allows us not only to provide closed-form analytical expressions for avalanche size, but also to determine the detailed set(s) of units firing in an avalanche (i.e., the avalanche assembly). The underlying dependence between network structure and dynamics is made transparent by expressing the distribution of avalanche assemblies in terms of the induced graph Laplacian. We explore analytical consequences of this dependence and provide illustrating examples.
% Was erreichen wir hiermit?
In summary, our framework provides a major extension of previous analytical work which was restricted to regularly coupled or discrete state networks in the infinite network limit. For systems with a sufficiently homogeneous or translationally invariant coupling topology, we make an explicit link to critical states and the existence of scale-free distributions. 

\end{abstract}

\maketitle

% -------------------------------------------------------------------------------

\section{INTRODUCTION}

% --------- distributed, time-varying neural representations (assemblies) are everywhere -------------
An influential concept in neuroscience introduced in 1949 by Hebb \cite{hebb2005organization} proposes that the brain uses distributed neural representations as code for memories and behavior. Transient activation of neural ensembles, i.\,e.\  the formation and decay of so-called neural `assemblies', would thus represent cognitive entities. Experimental studies in different species and neural systems have provided strong support for this concept by observing assembly formation on a wide range of spatial and temporal scales, and by linking their dynamics to brain function and behavior. Examples include face processing in macaques and humans \cite{freiwald2010functional,tsao2008comparing}, attentional networks in visual cortex~\cite{bastos2015visual}, odor representations in the olfactory system~\cite{mazor2005transient,laurent1996dynamical}, and vocal control in birdsong~\cite{leonardo2005ensemble,lynch2016rhythmic,lipkind2017songbirds}.

% -------- spike-sync. is our focus, allows rapid processing and seems to underly important functions ---
Spike synchronization constitutes a versatile mechanism for assembly formation. It may occur spontaneously on very short time scales and is much more efficient in driving post-synaptic cells than spikes arriving asynchronously~\cite{hahn2019portraits,buzsaki2010neural}. The ability to quickly form or to break up neural ensembles with varying compositions of participating cells supports information processing in different aspects. For instance, synchronization can indicate global dependencies among distributed local information in complex sensory scenes (e.\,g.~\cite{tomen2019role}), while mutual synchronization between different brain areas can rapidly establish or suppress communication channel for selective information processing in dependence on task demands (e.\,g.~\cite{bastos2015visual,grothe2012switching,harnack2015model}).

% ----- cortical networks seem to be optimized for sync-schemes: they call it 'criticality'... -----------
For optimally exploiting these functional opportunities, it has been suggested that cortical networks operate close to a critical state~\cite{beggs2008criticality,bottani1995pulse,bertschinger2004real,gautam2015maximizing,shew2013functional} in which spontaneous synchronization generates neural avalanches engaging large groups of cells (`assemblies') over far distances \cite{plenz2007organizing}. Formation of avalanches is fast since it does not require entrainment over a number of oscillation cycles as needed for synchronizing coupled phase oscillators. Indeed investigations of spontaneous synchronization in the brain revealed typical signatures of a dynamics being close to a critical state
\cite{papanikolaou2011universality,perkovic1995avalanches,sethna2001crackling}, % Theorie -> maximize pattern at critical state
such as power law distributions of avalanche sizes and durations
\cite{beggs2003neuronal,petermann2009spontaneous,yu2014scale} % Experimente -> power laws  
in combination with the observation of a large dynamical range
\cite{shew2009neuronal}. % Experimente -> range

% --- patterns for distributed codes are important, but functional role in crit. unclear ------------
However, it is not always size that matters. In a highly structured network like the brain, the specific composition of an activation pattern is of equal importance. It is the topology and efficacy of synaptic connections originating from the presently active neurons which will determine to what destination a signal will propagate, and if it will be enhanced or attenuated. In consequence, network function is defined by the pattern of neural activity only in combination with the microscopic structure of the network. The large reservoir of possible spike patterns in a system near criticality provides a good opportunity for a versatile processing here, but it is unclear how this property can be functionally exploited. 

In general, interactions and synergies between a (near)-critical dynamics and the microscopic network structure have barely been addressed by theoretical work and are thus not well understood. Studies on neuronal avalanches commonly assume \emph{homogeneous and/or global connectivity}, often in the limit of large networks, and focus mainly on determining the critical power-law exponents \cite{eurich2002finite,kinouchi2006optimal}. Structured networks are usually analyzed by focusing on particular connection schemes with certain fundamental statistical properties, such as small-world networks \cite{de2002self,massobrio2015self}, scale-free networks \cite{cohen2002percolation}, or branching processes \cite{larremore2014critical,di2017simple}, which then allow to compute global characteristics of the avalanche dynamics. For example, assuming a locally tree-like structure \cite{larremore2011predicting,larremore2012statistical,larremore2014critical} made it possible to analytically investigate robustness of the critical exponents against changes of the network topology. In parallel, there has been progress in formally understanding the structure-function relationship for recurrent networks with more general coupling structures. Using the theoretical framework of excitatory and linearly coupled Hawkes processes, analytical closed-form relations between the network adjacency matrix and equilibrium rates as well as spike count covariances were developed\cite{pernice2011structure,jovanovic2016interplay,hu2018feedback}. In contrast to studies relying on global statistical properties of network connectivity, these exact relations hold for \emph{arbitrary} network topologies and allow to relate specific graph motifs and the spectral distribution of the network to the strength and structure of the resulting correlations.  In general, however, the effect of network topology on particular avalanche \emph{patterns} and spike assemblies has not yet been fully elucidated.  

% --------------------------- what we do ------------------------------------------------
Here we bridge this gap by developing a formal framework which allows to rigorously analyze how particular network topologies shape avalanches and spike patterns in randomly driven networks of non-leaky integrate-and-fire units. For this purpose we employ the framework originally introduced by Eurich, Herrmann, and Ernst \cite{eurich2002finite} (in the following termed EHE model), which has successfully been used to formally study neural avalanches in globally coupled homogeneous networks, i.e. with constant or block constant coupling matrices \cite{levina2008mathematical,leleu2015unambiguous,tomen2019role}, and whose basic mathematical properties are well understood \cite{denker2014ergodicity,denker2016avalanche}. We extend the EHE model to arbitrary positive coupling matrices and derive closed-form expressions for the probabilities of arbitrary cell assemblies becoming transiently active in form of an avalanche \cite{plenz2007organizing}. This is possible by means of a suitably defined torus transformations which simplifies the seemingly highly complex spiking dynamics to a random walk on a finite dimensional torus. At the same time, the transform allows us to establish a mathematical link between the avalanche statistics and graph theoretical measures of the EHE network in terms of its adjacency matrix. 

The article is structured as follows: First, the basic model and its extension to arbitrary network topologies will be introduced. Second, we will show that the dynamics of the model is equivalent to shifts on the $N$-torus and derive simple expressions for mean activation and spike covariances in the network. Next we will focus on analyzing spike patterns and derive various closed-form expressions for the probabilities of \emph{particular} avalanche sequences from the corresponding state space volumes. The corresponding mathematical expressions will then be linked to graph-theoretical measures before finally discussing some network examples.

\clearpage
\pagebreak

% ----------------------------------------------------------------------------------------
% MODEL
% ----------------------------------------------------------------------------------------
\section{MODEL STRUCTURE AND DYNAMICS}\label{sec:modelintro}

%---------------------name the model ----------------------------
We employ a generalization of the Eurich-Herrmann-Ernst (EHE) model, which has been widely used to model neural avalanches~\cite{eurich2002finite,leleu2015unambiguous,jung2020avalanche}, and to study avalanche dynamics analytically~\cite{levina2008mathematical,denker2014ergodicity,denker2016avalanche}.

%------------- arbitrary coupling topology ---------------
The model can be described as a randomly driven network of pulse-coupled non-leaky integrate-and-fire neurons. Each unit $i \in [N] \coloneqq \{1,\ldots,N\}$ is characterized by a state $0 \leq u_i < U_i$, hence the \emph{phase space} of the system is given by the $N$-dimensional cube $C \coloneqq \bigtimes_{i\in [N]} [0, U_i)$. States can be interpreted as membrane potentials with 0 representing the resting potential and $U_i$ the individual firing threshold for unit $i$. Units are coupled by the non-negative weight matrix $W = (w_{ij})_{i,j\in [N]}$, with $0 \leq w_{ij}$ specifying the increase of the membrane potential for unit $i$ upon receipt of a spike from unit $j$. The coupling matrix $W$ induces a weighted directed graph $G(W)$  with vertices $[N]$, edges $E(W)\coloneqq\{(j,i)\in {N\times N}|w_{ij} > 0\}$ and weights $(j,i)\mapsto w_{ij}$, see Definition~\ref{def:gw}.
We refer to the induced subgraph with vertices $I\subseteq [N]$ as a subnetwork along $I$ and use $W_I$ for the corresponding weight matrix with rows and columns restricted to $I$.

% ------------------- slow timescale ---------------------
We formulate this model in discrete time, in which external input arrives at each time step to a randomly chosen unit. Avalanches resulting from units crossing their firing threshold occur on a fast timescale and complete before the next unit receives external input (separation of time scales). The external input dynamics is particularly simple: a random unit $k$ is chosen with probability $p_k$ and its state $u_k$ is increased by an amount $\DeltaU \in \R_{\geq 0}$. 

% ------------- fast timescale - internal avalanche dynamics -------------
Should this increase push the state of a unit above the firing threshold, an avalanche starts and evolves on a fast timescale. The avalanche dynamics $\fixF$ consists of repeatedly resetting the currently supra-threshold units by subtracting $\diag(\VU) A(\Vu)$ and distributing internal activation by $W A(\Vu)$. Here $A(\Vu)\coloneqq \delta(\{i\in [N]:u_i \geq U_i\})$ describes the supra-threshold units using $\delta(I) = \sum_{i\in I}\e_i$ with $\e_i$ denoting the $i$-th unit vector in $\R^{[N]}$. The avalanche terminates after $\tau$ steps or generations when all units are below threshold. Note that $U$, $A$ and $u$ are vectors, with $U_i$, $A_i$ and $u_i$ designating their $i$-th components. This will be standard notation from here on until mentioned otherwise.

The dynamics $\fixF$ is formalized as follows, with $F$ describing one generation of an avalanche and $\tau$ defining the termination condition:
\begin{alignat}{2}\label{def:f}
  \fixF&:\mathbb{R}_{\geq 0}^{[N]} \to C, && \Vu\mapsto  F^{\tau(\Vu)}(\Vu) \\
	\text{with } F&: \mathbb{R}_{\geq 0}^{[N]} \to
		\mathbb{R}_{\geq 0}^{[N]},\quad && \Vu\mapsto \Vu - (\diag(\VU)-W)A(\Vu) \\
  \text{and } \tau&:\mathbb{R}_{\geq 0}^N \to
		\mathbb{N}_0, && \Vu\mapsto \min\{n\in \mathbb{N}_{0}\mid F^n(\Vu) \in C\}
\end{alignat}
External input given to unit $k$ and the resulting avalanche dynamics can be formally combined into a single action $T_k$ given by
\begin{align}
  T_k&: C \to C, \; \Vu \mapsto \fixF(\Vu+\DeltaU \e_k) \,\, \text{,} \label{eq:def-Tk}
\end{align}
which also includes the trivial case that no unit crosses threshold and thus there is no avalanche.

Throughout this study, we impose the condition
\begin{align}\label{ass}
  \DeltaU + \sum_{j=1}^N w_{ij} < U_i \text{ for all $i\in [N]$} \text{ ,}
\end{align} 
ensuring that each unit can fire at most once during an avalanche (see Proposition~\ref{prop:at-most-once}). This condition implies that $\diag(\VU) - W$ is strictly diagonally dominant, hence its inverse
\begin{align}\label{eq:defM}
  M \coloneqq (\diag(\VU) - W)^{-1} 
\end{align}
exists. 

Due to the separation of timescales, it is guaranteed that external input does not arrive \emph{during} an avalanche, and thus we can track the detailed pattern of an avalanche by index sets listing which units fired at each generation of the avalanche. For this purpose we introduce the avalanche function $\av$ which provides a vector of index sets 
\begin{align}
  \av(k, \Vu) &\coloneqq (\{j \in [N] \mid 
		(F^{i-1}(\Vu+\e_k \DeltaU))_j \geq U_j\})_{i=1,\ldots, \tau(\Vu+\DeltaU\e_k)} \in \mathcal{A}
\end{align}
as the {\em avalanche} started in state $\Vu$ by giving external input to unit $k$, where $\mathcal{A}$ is the set of all avalanches (see Definition~\ref{def:A}). 
\iffiginline
\begin{figure*}[ht] 
  \includegraphics[width=17.2cm]{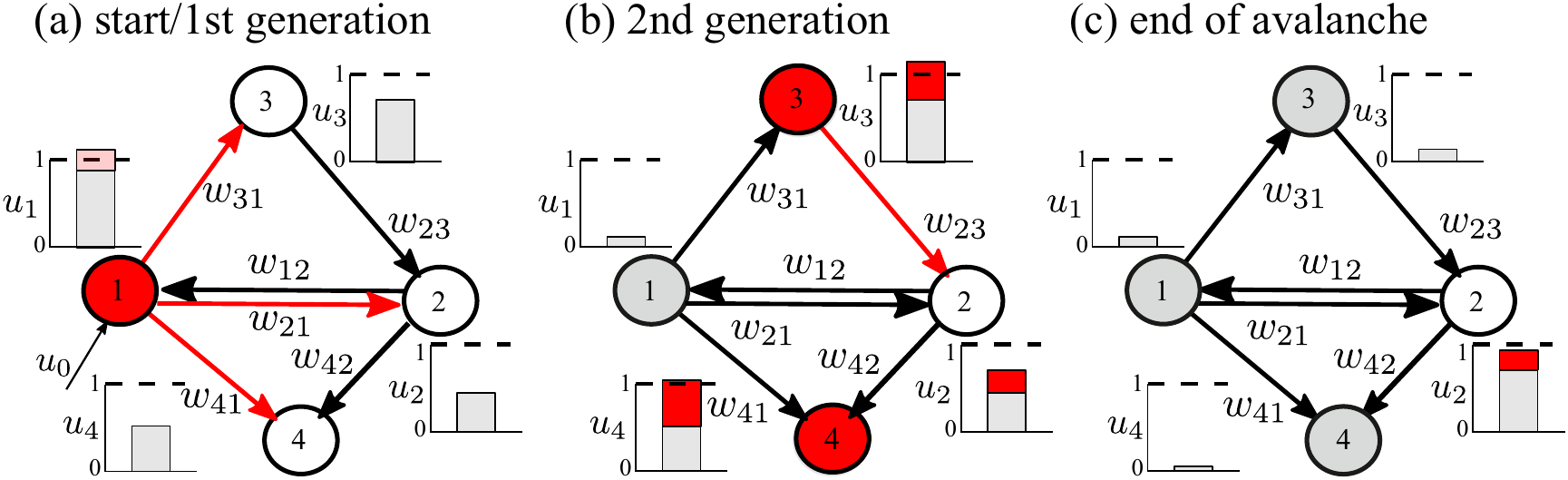}
  \caption{\label{fig:avillus}
	    Spreading of an avalanche in an example network with four units (circles).
        Each unit $i$ has a state $0 \leq u_i < U_i$,
		with 0 representing the resting state and $U_i$ the firing threshold.
		States are visualized as bar graphs with $U_i=1$ for all $i$. 
		Units are coupled by a directed, weighted graph with coupling matrix
        $W=(w_{ij})_{i,j\in [N]}$ where $w_{ij}$ defines connection strength
		(non-zero entries shown as arrows).
		A spike of unit $j$ increases the state of the receiving unit $i$
		by the corresponding interaction strength $w_{ij}$ (red bars).
		White units have not participated in the avalanche yet, and
		red units are currently active and send spikes to all units connected
		(arrows marked in red). Incoming weights are restricted in their magnitude
		such that no unit can be active twice in an avalanche, hence active units
		become quasi-'refractory' (gray) in the next step.
		Panels (a) to (c) depict the spreading of the avalanche $\Va = (\{1\}, \{2,4\})$,
        having a size $\s(\Va)=3$, duration $\dur(\Va) = 2$,
		and an assembly of $\U(\Va) = \{1,2,4\}$. In panel (a), giving external input
		(light red) to unit 1 pushes its state beyond firing threshold and starts
		the avalanche. The avalanche terminates in panel (c) since activation from the
		second generation shown in panel (b) is insufficient to bring any unit above
		firing threshold.
	}
\end{figure*}

\else
\\
-insert Figure 1 here-\\
\fi
For a particular avalanche $\Va=(a_n)_{n=1,...,d} \in \mathcal{A}$, the first generation of $\Va$ always consists of a singleton $a_1 = \{k\},k\in [N]$, if $\Va$ is not the empty avalanche $\Va=()$. The length of the sequence $\Va$ will be denoted by $\dur(\Va)$ and called the {\em duration} of the avalanche. We call the union $\U_{j}$ of the generations 
\begin{align}
  \U_{j}(\Va) &\coloneqq \biguplus_{i=1}^{j} a_i \text{, }
		1\leq j\leq \dur(\Va) \mbox{ and }
		\;\U(\Va)\coloneqq\U_{\dur(\Va)}(\Va)
\end{align}
the \emph{avalanche assembly} (up to generation $j$) and the sum of cardinalities
\begin{align}
	\s(\Va) &\coloneqq \sum_{i=1}^{\dur(\Va)} |a_i| =\lvert\U(\Va)\rvert
\end{align}
its {\em size}.

\Fig{fig:avillus} illustrates in detail the avalanche dynamics on an example network after giving external input to unit $1$, which leads to the avalanche $\Va=(\{1\},\{3,4\})$.

Our goal is to derive the probability distribution of avalanches $a\in \mathcal{A}$ in dependence of the coupling matrix $W$ and external input probabilities $p$. To make this mathematically rigorous, we model the dynamics as a \emph{random dynamical system}, or more precisely as a skew-product dynamical system $T$:
\begin{align}
	T&: \Sigma_N \times C  \to \Sigma_N \times C,\;
		T(\Vk, \Vu) \coloneqq \left( \sigma(\Vk),T_{k_{1}}(\Vu) \right)     
\end{align}
Here, $\Vk=(k_1,k_2,\ldots) \in [N]^{\N} \eqqcolon\Sigma_N$ is a right-infinite sequence over the alphabet $[N]$ modeling the sequence of units receiving external inputs, and $\sigma((k_1,k_2,\ldots)) = (k_2,k_3,\ldots)$ is the left shift operator. In order to turn this model into a random dynamical system, we equip $\Sigma_N \times C$ with the Borel $\sigma$-algebra and a measure $\P$. To model the randomness of the external input, $\P$ will be given as a product measure composed of the time invariant Bernoulli measure $\mathbb{B}_p$ with success probabilities $\Vp$ on $\Sigma_N$ and a measure $\PW$ on $C$. One of our main insights is that if $\DeltaU \notin \mathbb{Q}$, there exists a unique choice for $\PW$ such that $T$ is ergodic with respect to $\P=\mathbb{B}_p\times \PW$ for almost all non-negative coupling matrices, if and only if (Theorem~\ref{thm:ergodicity-almost-everywhere}) every unit is reachable by a path via non-zero coupling weights starting from a unit receiving external input ($p$-reachability, see definition~\ref{def:gw}).

Thus, we will always assume $p$-reachability of the coupling network in the following. With the unique ergodic measure $\P$ and a slight abuse of notation using $\av(\Vk,\Vu)\coloneqq \av(k_1,\Vu)$, the avalanche function $\av$ is a random variable with respect to $\P$, which allows us to study the \emph{avalanche probabilitites} $\P(\av = a)$ for $a\in \mathcal{A}$.
% This result is required for relating avalanche probabilities to volumes in state space. 

\iffiginline
\begin{figure*}[t]
  \includegraphics[width=17.2cm]{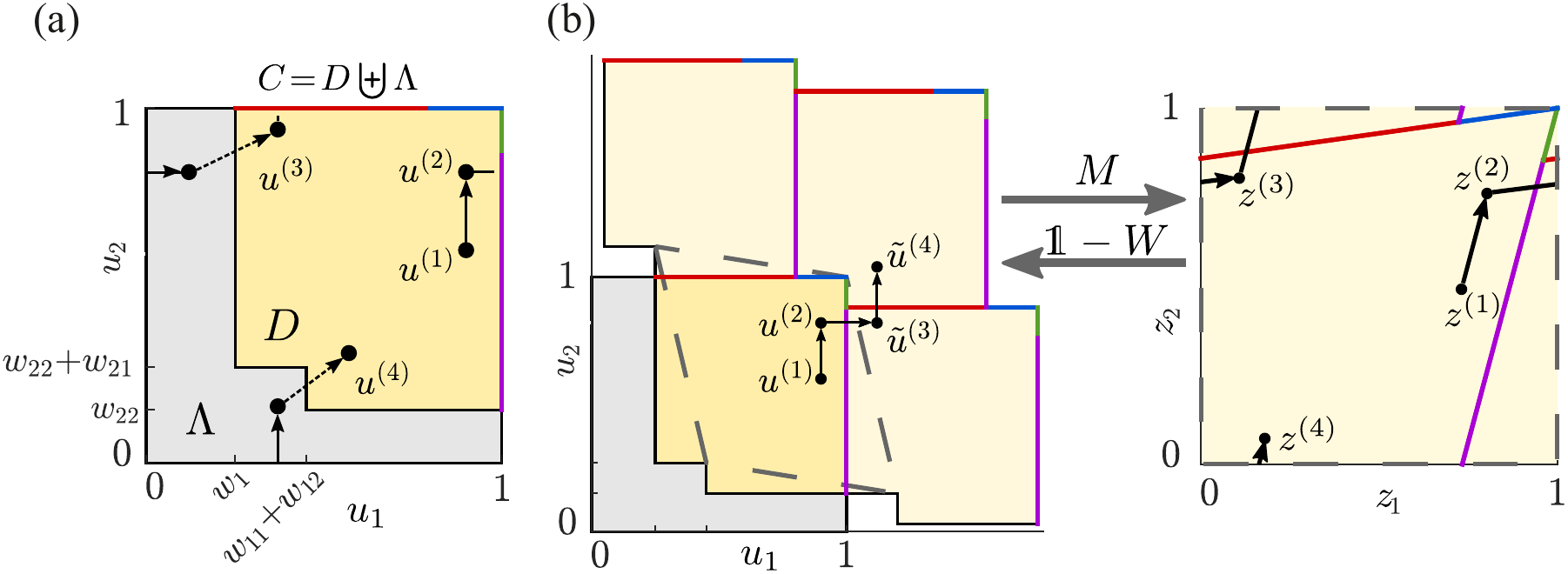} \\
  \caption{\label{fig:trafoillus}\\
	(a) State space for a two-dimensional EHE-model. States $u_1$ and $u_2$
	span the state space $C$ (unit rectangle) which consists of the inhabited
	region $D$ (yellow shading), and the non-inhabited region $\Lambda$ (gray shading).
	Black dots and solid arrows indicate a sample trajectory $u^{(1)},\ldots, u^{(4)}$
	during which external input $u_0$ is provided first to unit \#2, then to unit \#1 and
	finally to unit \#2 again. The length of the solid arrows is $u_0$.
	When the trajectory crosses the right or upper boundary of the unit cube
	(i.e., the firing thresholds $U_1=1$ or $U_2=1$), a unit spikes and its state
	is 'reinjected' at the opposite side of $C$ (spike reset).
	Simultaneously, recurrent activation is distributed to all connected units,
	corresponding to shifts by columns of $W$ (dashed arrows). Distribution of
	recurrent input can continue multiple times until no state is above threshold
	anymore, thus forming multiple generations of avalanches comprising
	different numbers of units.\\
	(b) Torus transformation for a two-dimensional EHE-model. 
	{\em Left:} 
	Copies (bright yellow) of the inhabited region $D$ (dark yellow)
	tesselate the $u_1$-$u_2$ plane.
	Equivalent points to $u^{(3)},u^{(4)}$ in the example trajectory
	introduced in (a) are labeled with $\tilde{u}^{(3)},\tilde{u}^{(4)}$
	in translated copies of the inhabited region.
	They are reached by simple shifts $u_0$, while reset and recurrent activation
	have no effect on the equivalent trajectory (black arrows). 
	The colors of the line segments indicate the avalanche which is 
	triggered when the trajectory crosses the corresponding border.
	In this example, purple, red, green, and blue designate the avalanches
	$(\{1\})$, $(\{2\})$, $(\{1\},\{2\})$, and $(\{2\},\{1\})$, respectively.
	The equivalent points lie on a grid spanned by the column vectors of $\mathds{1}-W$,
	with one unit cell indicated by the dashed gray lines.
	The inhabited region is the image of this unit cell under $\fixF$.
  {\em Right:} 
	Applying the inverse $M=(\mathds{1}-W)^{-1}$ leads to an equivalent dynamical system
	on the torus which consists of translations by column vectors of $M$.
    Points $z^{(i)}$ on the torus are the images of states $u^{(i)}$ in $D$.
}
\end{figure*}
\fi

In the remainder of this section, we illustrate the dynamics of the extended EHE model from a phase space perspective using the system displayed in \Fig{fig:trafoillus}: Iterations of the dynamics $T$ induce a trajectory in phase space $C$. The trajectory $\Vu^{(1)} \to \Vu^{(2)}\to \Vu^{(3)} \to \Vu^{(4)}$ is obtained by three iterations of $T$ starting in $(\Vk, \Vu^{(1)})$ with an external input sequence $\Vk = (2, 1, 2, \ldots)$. 
% (the trajectories with points labeled \(\tilde{u}\) and \(\hat{u}\) will be described in the following sections). 
While external input induces state shifts parallel to the axes, the avalanche dynamics $\fixF$ results in reinjection of points pushed outside of $C$ by subtracting the thresholds $U_i$ of the activated units, and by distributing internal activation which induces shifts along column(s) of $W$. Note that $\av(k_1, \Vu^{(1)})=()$, denoting the empty avalanche, $\av(k_2, \Vu^{(2)})=(\{1\})$ and $\av(k_3, \Vu^{(3)})=(\{2\})$. It becomes apparent that the actions $T_k$ are discontinuous transformations, where simple shifts along the axes are followed by the more complicated avalanche dynamics $\fixF$ if $u_k+\DeltaU \geq U_k$.

Note that during an avalanche, the internal, recurrent activation after reset pushes a state out of a region $\Lambda$, which is indicated in gray shading in \Fig{fig:trafoillus}. As a consequence, the state density becomes zero in $\Lambda$ which we will thus designate as the \emph{non-inhabited region}. The existence of $\Lambda$ is a general feature of the model, as for all $p$-reachable networks there exists an \emph{inhabited region} $D$ which depends on $W$. $D$ acts as a uniform attractor on the phase space in the sense that for $\mathbb{B}_\Vp$-almost all input sequences $\Vk$ the projection of $T^n(\Vk, C)$ onto its second component equals $D$ for all $n\geq n_{0} \in \N$ (see Proposition~\ref{thm:d-inhabited}). In particular, any invariant density of states necessarily vanishes on $\Lambda \coloneqq C\setminus D$ and we can therefore proceed by analyzing the system restricted to $\Sigma_N \times D$ with the associated restricted Borel $\sigma$-algebra.

For the simple two-dimensional system in \Fig{fig:trafoillus}, there are only four possible non-empty avalanches $\av(\Vk, \Vu)$, namely $(\{1\})$, $(\{2\})$, $(\{1\},\{2\})$, or $(\{2\},\{1\})$. These avalanches occur exactly when the state trajectory crosses the lines along the boundary of $C$ colored in purple, red, green, or blue, respectively. Since the unique choice of $\PW$ for this system is always the uniform distribution supported on the inhabited region $D$ (yellow region), also known as the Lebesgue measure on $D$, the probabilities $\P(\av = \Va)$ are proportional to the lengths of the respectively colored boundary segments (Theorem~\ref{thm:ergodicity}).

% ---------------------------------------------------------------------------------------------------------
 
\section{Analysis of avalanche distributions}

For ease of notation, we state in the following the main theorems for the special case $U_i=1$ for all $i\in [N]$ and give references to the general statements and corresponding proofs collected in the appendix.

In this section, we will take a closer look at the intricate dynamics of the skew-product dynamical system $T$. Most importantly, we will derive a linear transformation of the system which allows to represent the complex avalanche dynamics as simple translation dynamics on the $N$-torus. This central idea allows to show relative unique ergodicity of the system and to derive the equilibrium measure on the phase space in dependence of the weight matrix $W$. Using these new mathematical insights we derive closed form expressions for mean firing rates and their variance in the network. In addition, we derive closed form expressions for the probability that a particular unit $i$ participates in an avalanche started with unit $k$, the probability distribution of individual avalanches $\P(\av = \Va)$, and probability distributions of avalanche assemblies $\P(\U(\av) = I)$ in closed-form expressions.

\subsection{Equivalence to an ergodic translation dynamics on the $N$-torus}

%------- current state of the art ----------------------

For globally connected, homogeneous systems it has been shown that the uniform state density supported on the \emph{inhabited region} $D$ is invariant under the dynamics \cite{eurich2002finite,leleu2015unambiguous,denker2014ergodicity,denker2016avalanche}. However, ergodicity of the skew-product system has only been conjectured~\cite{denker2014ergodicity,denker2016avalanche}. Ergodicity is important since it allows to associate (sub)volumes in phase space with the probability to observe particular avalanches. Here, we close this conjecture and extend it to the generalized system in which the units are coupled by an arbitrary non-negative coupling matrix $W$. More specifically, in Theorem~\ref{thm:ergodicity} we derive necessary and sufficient conditions for unique ergodicity of the Lebesgue measure on $D$ relative to a given measure on the shift space $\Sigma_N$ for a general class of translation dynamics on the standard $N$-Torus $\TN\coloneqq \R^{[N]}/\Z^{[N]}$. By constructing a translation dynamics on $\TN$ which is topologically conjugated to the original dynamics $T$, we  use this theory to establish relative unique  ergodicity of the Lebesgue measure supported on the inhabited region $D$ for almost all weight matrices $W$ with $p$-reachable $G(W)$ as long as $\DeltaU$ is irrational (see Theorem~\ref{thm:ergodicity-almost-everywhere}).
  
Here we explain this simplification of the dynamics using the geometric intuition illustrated in \Fig{fig:trafoillus} while referring to the associated proofs for the general case in the appendix. The main idea is the following: Even though the avalanche dynamics is discontinuous (in $\mathbb{R}^{[N]}$) due to the resets of the unit's states after firing spikes, we can summarize the effect of the dynamics $T_k$ on state vector $\Vu$ by using the assembly $\U$ of the resulting avalanche as
  \[T_k(\Vu) = \Vu + \DeltaU \e_k - (\mathds{1}-W)\delta(\U(\av(k, \Vu))) \text{,}\]
where $\mathds{1}$ denotes the identity matrix. There are no explicit thresholds in this equation, because the spike reset is expressed as a simple subtraction of $1$ for every unit becoming active. After $n$ iterations we have
\begin{equation}
	\Vu' \coloneqq \pi_2 T^n(\Vk, \Vu) = \fixF\left(\Vu + u_0\sum_{t=1}^n e_{k_t}\right) 
	= \Vu + u_0\sum_{t=1}^n e_{k_t} - (\mathds{1}-W)\NN^{n}(\Vk, \Vu) \text{, }
	\label{eq:transformex}
\end{equation}
where $\pi_2$ denotes the projection to the 2nd component and 
	\[\NN^{n}(\Vk, \Vu) \coloneqq \sum_{t=0}^{n-1} \delta(\U(\av(T^t(\Vk, \Vu))))\]
% UDO: pi2 weglassen und Tk^n schreiben?
is the \emph{spike count vector} which collects how often each unit fired during $n$ steps of the dynamics $T$ starting from the initial state $(\Vk, \Vu)$.

% ---------M has full rank and induces coordinate system ---------------------
Since $M^{-1}=\mathds{1}-W$ is a diagonally dominant $M$-matrix~\cite{plemmons1977m} in virtue of~\eqref{ass}, it has full rank and the column vectors induce a new coordinate system on $\R^{[N]}$, indicated by the unit cell (dashed parallelogram) in \Fig{fig:trafoillus}. Effectively, the recurrent dynamics displaces a state $\Vu$ along a linear combination of the column vectors in $W$ with integer coefficients. Expressed in the new coordinates, the system's state $\Vu'$ after $s$ avalanches, and the external input dynamics $\Vu+u_0\sum_{t=1}^s \e_{k_t}$ without recurrent feedback and spike reset are always integer coordinates apart. This can easily be seen by transforming $u'$ in \Eq{eq:transformex} into the new coordinate system via multiplication by $M$ (\Eq{eq:defM}),
\[
	M\Vu' = M\left(\Vu + u_0\sum_{t=1}^n e_{k_t}\right) - \NN^{n}(\Vk, \Vu) \text{.}
\]

Geometrically, this property implies that copies of the inhabited region translated by integer coordinates $(\mathds{1}-W)\Z^{[N]}$ tesselate $\R^{[N]}$.

% ---- points with inter coordinate difference can be glued together  --------
% ---- to arrive at a topology homeomorphic to the torus topology     --------       

Considering all points with a difference of $(\mathds{1}-W)\Z^{[N]}$ as being equivalent induces a topology which is homeomorphic to the topology on the standard $N$-torus (which associates all points with a difference of $\Z^{[N]}$). These considerations imply that the dynamics $T$ is equivalent to the dynamics $\hat{T}$ (see Eq.~\ref{eq:def-hatt}) on $\TN$, with the shifts $\DeltaU \e_k$ induced by external input being transformed by the inverse $M =(\mathds{1}-W)^{-1}$. We prove this equivalence in Theorem~\ref{thm:equivalence}.

% Beschreibung der Figur
\Fig{fig:trafoillus} illustrates the described equivalence. By transforming the region in the dashed unit cell to $[0,1)^{[N]}$, we map the state trajectory on the left to its equivalent trajectory on the torus on the right. 
\begin{itemize}
\item
In the old coordinate system, shown on the left, the axes are equivalent to the states of single units. The external input is realized by a shift along the axis of the unit receiving the input, while the recurrent input and spike reset are given by a combination of shifts along the columns of $\mathds{1}-W$. 
\item
In the new coordinate system, the axes are equivalent to the combined effect of recurrent input provided by one unit to all other units. External input is still represented by a shift, but projected onto the new coordinate system via $\DeltaU M \e_k$ it is in general no longer parallel to the coordinate axes. However, recurrent input and spike reset are now just mapping the current state to its equivalent point in a different unit cell. On the $N$-torus with its periodic boundary conditions, recurrent input and spike reset thus map to the identical state, hence making the recurrent dynamics much simpler to handle formally. 
\end{itemize}
The transformation onto the $N$-torus is invertible since the inhabited region $D$ is the image of the region enclosed by the dashed lines under $\fixF$ (see Theorem~\ref{thm:D} and Proposition~\ref{thm:d-inhabited}). With these insights it becomes possible to more easily assess ergodicity by performing the corresponding analysis on the transformed system $\TN$ first, and then to transfer results to the original system $T$. In the following, we briefly state our main insights on ergodicity and refer the reader to the Appendix for the detailed formal treatment.

% UDO: logische Anschlüsse durchgehen...
It turns out that the Lebesgue measure is invariant since every translation $\Vu \to \Vu + \DeltaU M  \e_k$ is bijective on $\TN$. In Theorem~\ref{thm:ergodicity} we show for a more general class of translation dynamics on $\TN$ that the Lebesgue measure is also the \emph{unique} ergodic measure of the system, given a stationary probability distribution of the external input. In addition we find that the system $\hat{T}$ is uniquely ergodic relative to the external input statistics $\mathbb{B}_p$ for almost-all coupling matrices $\mathcal{W}(E)$ with the edge set $E$, if and only if $E$ is $p$-reachable (see Theorem~\ref{thm:ergodicity-almost-everywhere}).

Since ergodicity is invariant under topological conjugacy, this ensures that also $T$ is ergodic with respect to $\P = \mathbb{B}_P\times \PW$ with $\PW = \lambda_D$ denoting the normalised Lebesgue measure supported on the inhabited region $D$.

Let us give an intuition for this remarkable result: Consider the case that only a single unit $k\in [N]$ receives external input. In this case, the equivalent dynamics on $\TN$ is a simple rotation on the $N$-Torus by the vector $\DeltaU M\e_k$. For this classical dynamical system (see e.g. \cite{katok1997introduction}) ergodicity of the Lebesgue measure requires the components of this vector to be irrational and rationally independent, since otherwise the orbits were not dense in $\TN$. If only the unit $k$ receives external input, $p$-reachability requires that there has to be a directed path from $k$ to every other unit in $[N]$. This condition alone already ensures that each component of $M\e_k$ is positive. For this case, Theorem~\ref{thm:ergodicity-almost-everywhere} states that if you fix a network topology, e.g. the sparsity pattern of a $p$-reachable coupling matrix $W$, and construct a coupling matrix $W'$ by choosing random values for the positive entries in $W$, the entries of $\DeltaU M' \e_k$ will indeed be irrational and rationally independent almost surely (i.e., with probability one), hence the resulting system is ergodic.

However, almost sure ergodicity does not exclude exceptions, such as can be seen for the much simpler special case of the previously studied homogeneous coupling matrix $W_{ij}={\alpha}/{N}$ for all $i,j\in[N]$ with $\alpha + \DeltaU < 1$.
It was already noted (see \cite{denker2014ergodicity,denker2016avalanche} that this system is not ergodic if only a single unit receives external input. 
In fact, our theory shows that this system is ergodic if all units receive external input (Corollary~\ref{cor:hom-ergodic}), but not if two or more units do not receive external input (Corollary~\ref{cor:hom-not-ergodic}).

\subsection{Equilibrium rates, spike covariance and mean avalanche sizes\label{sec:equilibrium}}

Topological conjugacy to a simple translation dynamics on the $N$-Torus $\TN$ greatly facilitates analysis of the dynamics of the extended EHE system. In this subsection, we will first derive key properties such as equilibrium rates and spike covariances, and subsequently assess spike propagation probabilities, allowing us to finally compute mean avalanche sizes in closed-form expressions.

%-----------------linear transformations give equilibrium rates and covariances -------------------

Let $Y_0 \coloneqq \lim_{n\rightarrow \infty}{\E (\NN_0^{n})}/{n}\in \mathbb{R}^{[N]}$ and $X_0 \coloneqq \lim_{n\rightarrow \infty}{\operatorname{cov}(\NN_0^{n})}/{n}\in \mathbb{R}^{[N\times N]}$ be the stationary firing rates and their covariances in a network of uncoupled EHE-units. The number of external inputs received by each unit after $n$ steps of the slow external dynamics follows a multinomial distribution with parameters $n$ and probability vector $\Vp$. Setting w.l.o.g.\ the time interval between external inputs to 1, we find $Y_0=\DeltaU \Vp$ and $X_0 = (\DeltaU )^2(\diag(\Vp)-\Vp \, \Vp^{T})$ as the covariance matrix for the spike counts of different units in the limit of long observation times.

The dynamics of $T$ generating the spike count vector $\NN^{n}$ translates to $\hat{T}$ in the sense that its $i$-th component $(\NN^{n})_i$ counts how many times its trajectory has been winding around the side $i$ of the $N$-Torus up to time $n$. We use this fact to obtain the equilibrium firing rates $Y_W$ and covariances $X_W$ as linear transformations of the firing rates and spike count covariances of the uncoupled system (see Theorem~\ref{thm:ratios}, Theorem~\ref{thm:Vw}) via
\begin{equation}
	Y_W = MY_0 = (\mathds{1}-W)^{-1}Y_0, \quad X_W = M^T X_0 M \text{ .}
	\label{eq:equilib}
\end{equation}

%------------------- graph theoretical interpretation of M --------------------- 

This functional form is similar to the analytical rates and covariances for linearly coupled Hawkes processes, which have been used to study in detail the influence of network topology on population activity of neural networks (e.\,g.\ \cite{pernice2011structure,pernice2012recurrent,jovanovic2016interplay,hu2018feedback}). In the following, we restate the corresponding implications of this functional form for $Y_W$, which in this model is closely related to the mean avalanche size.

To do so, we study $W$ as a weight matrix of a directed graph allowing us to give an interpretation of the firing rates in terms of weighted paths. We will further investigate this viewpoint for understanding the probability distributions of avalanches in Section~\ref{sec:relationgraph}.

We denote by $G(W)$ the graph induced by the coupling matrix $W$ with the edge set $E(W)$, see Definition~\ref{def:gw}. Through Taylor expansion $M$ can be written as a Neumann series $M=(\mathds{1}-W)^{-1}=\sum_{l=0}^\infty W^l$. For the directed graph $G(W)$, $(W^l)_{ij}$ equals the product of edge weights summed over all paths from unit $j$ to unit $i$ with exactly $l$ edges. Thus one can interpret \Eq{eq:equilib} for the equilibrium firing rates as summing the influences between units over \emph{all possible paths} in the network.
Moreover, the weighted sum of all paths from node $k$ to node $i$ gives the probability that unit $i$ fires in an avalanche started by unit $k$ (see Theorem~\ref{thm:ratios}):
\begin{align}\label{eq:avprop}
  \P_k(i\in \U(\av)) = \frac{M_{ik}}{M_{kk}}\text{ ,}
\end{align}
where we use the abbreviation
\begin{align}
  \P_k(A) \coloneqq \P(A|\av_1 = \{k\})
\end{align}
for the probability of events $A$ conditioned on the event that external input to unit $k$ started an avalanche.

The sum over the equilibrium firing rates $\1^T Y_W$ can be seen as the average number of units firing in each time step, i.e. in each iteration of $T$, which depends on $\Vp$ and $\DeltaU$ via $Y_0=u_0 \Vp$. However, when we condition on $\av_1 = \{k\}$, these dependencies vanish and we find a closed form which solely depends on $M=(\mathds{1}-W)^{-1}$:
\begin{eqnarray}
  & & \mathbb{E}(\s(\av) \mid \s(\av)>0) = \sum_k \p_k\mathbb{E}(\s(\av)|\av_1 = \{k\}), \nonumber \\
  \mbox{with} & \hspace*{0.2cm} &
		\mathbb{E}(\s(\av)|\av_1 = \{k\}) = \sum_{j\in [N]}\P_k(j\in \U(\av)) = \frac{\sum_{j\in[N]} M_{jk}}{M_{kk}} \text{.}
	\label{eq:mavs}
\end{eqnarray}
The diagonal element $M_{kk}$ in the denominator is equal to the quotient $M_{kk} = |\mathds{1}_{[N]\setminus \{k\}}-W_{[N]\setminus \{k\}}|/|\mathds{1}-W|$ which has a geometrical interpretation (see next section) as the quotient of the volumes of the inhabited regions of the $[N]\setminus \{k\}$ system and the full system and is proportional to the probability that unit $k$ starts an avalanche.

\Eq{eq:mavs} can again be interpreted in terms of the graph $G(W)$: Given that an avalanche is started by the unit $k$, the average size of the avalanche is equal to the weighted sum of all paths in $G(W)$ from $k$ to all units of the graph, normalized by the weighted sum of all paths from $k$ to itself. 

%\mhk{W\"are diese Darstellung interessant? $N_{ik}\coloneqq\frac{M_{ik}}{M_{kk}}$ dann gilt
%  \[\mathbb{E}(\s(\av)|\s(\av)>0)=\1^{t}Np.\]
%MS: $N$ could be confused with the spike count vector $\NN^{s}$. The fractions only occur for \(\E(\s(\av))\) and \(\P_k(\i in \U(\av))\)}

Note that the association of expressions containing $M$ to putative paths in the graph $G(W)$ does not have a one-to-one correspondence to the actual dynamics on the network. For instance, $i \to j \to i \to k$ is a path in $G(W)$ and the product $w_{ki}w_{ij}w_{ji}$ is one summand in $(W^3)_{ki}$. However, since we constrained weights in our model via \Eq{ass}, units only interact via avalanches in which each unit can occur at most once, such that an avalanche $i \to j \to i \to k$ is not possible. We will resolve this apparent conflict between actual dynamics and apparent interpretation of the mathematical expressions in Section~\ref{sec:dependencies}, where we show that the states of recurrently connected units are correlated. It turns out that these correlations increase the probability of eliciting a spike in a connected unit beyond the corresponding entry in $W$, hereby compensating for paths existing, but never taken by an avalanche.

\subsection{Geometrical structure and self-similarity of the inhabited region}

In the previous subsections we showed results for equilibrium first and second order statistics, which we derived from the conjugacy of the original dynamics $T$ to a simple random walk on $\TN$. A deeper look into the statistics of avalanches requires to study the phase space regions on the inhabited region $D$ for which $T_k$ will generate specific avalanches. The self-similar geometrical structure of $D$, which we characterize in this subsection, greatly simplifies the identification of these regions and the computation of their volumes, which will be detailed in the following subsection.

It is convenient to describe regions of interest in phase space as hyperrectangles which are restricted to lower and upper boundaries in the dimensions specified by an index set $I\subset [N]$, and unrestricted in all other dimensions,
\begin{align}
  [a_i,b_i)_{i\in I} \coloneqq \pi_I^{-1}\left(\bigtimes_{i\in I}[a_i,b_i)\right)
  =\{ \Vu\in C \mid  a_i \leq u_i < b_i \text{ for } i\in I \subseteq [N]\}
  \quad ,
\end{align}
where $\pi_I:\R^{[N]}\supset C \to \R^{I}$ denotes the restriction to $C$ of the  natural  projection onto coordinates in $I$.
Note that relative unique ergodicity of the uniform measure on the $N$-torus translates to relative unique  ergodicity of $\PW = \lambda_D$ with $\lambda$ being the Lebesgue measure.The geometry of $D$ is thus closely related to the stochastic properties enforced by $\PW$. The $\PW$-volume of a measurable subset $A \in C$ is given by the quotient of the $N$-dimensional Lebesgue volumes $\lambda(A \cap D)/\lambda(D)$.

Using the conjugacy to the translation dynamics on $\TN$, the volume $\lambda(D)$ for the general case of arbitrary firing thresholds $\VU$ is given by $\lambda(D) = \lvert \diag(\VU) - W \rvert$. The intuition behind this closed-form expression is illustrated in \Fig{fig:trafoillus}: The inhabited region $D$ is the image of the white dashed parallelepiped, which represents a unit cell, under $\fixF$. Since $\fixF$ only induces translation, it is volume-preserving and thus the volume of $D$ is the volume of the unit cell, which is simply $\lvert \diag(\mathbf{U})-W \rvert$, i.e. the determinant of the inverse mapping from $\TN$ to $C$.

The geometrical structure of $D$ is most apparent when studying its complement in $C$, which we termed the non-inhabited region $\Lambda \coloneqq C \setminus D$. In Theorem~\ref{thm:DGamma} we show that this non-inhabited region is given by a union of cylinder sets \(\Gamma_I = [0,(W\delta(I))_i)_{i\in I} \) (see also definition~\ref{def:Lambda}) for all subsets \(\emptyset \neq I \subseteq [N]\):
\begin{align}\label{eq:Dmain}
   \Lambda=\Lambda_{[N]}
	\text{ with }
	\Lambda_I \coloneqq \bigcup_{\emptyset \neq J \subseteq I} \Gamma_I\,.
\end{align}

\iffiginline
\begin{figure*}
  \centering
    \includegraphics[width=17.2cm]{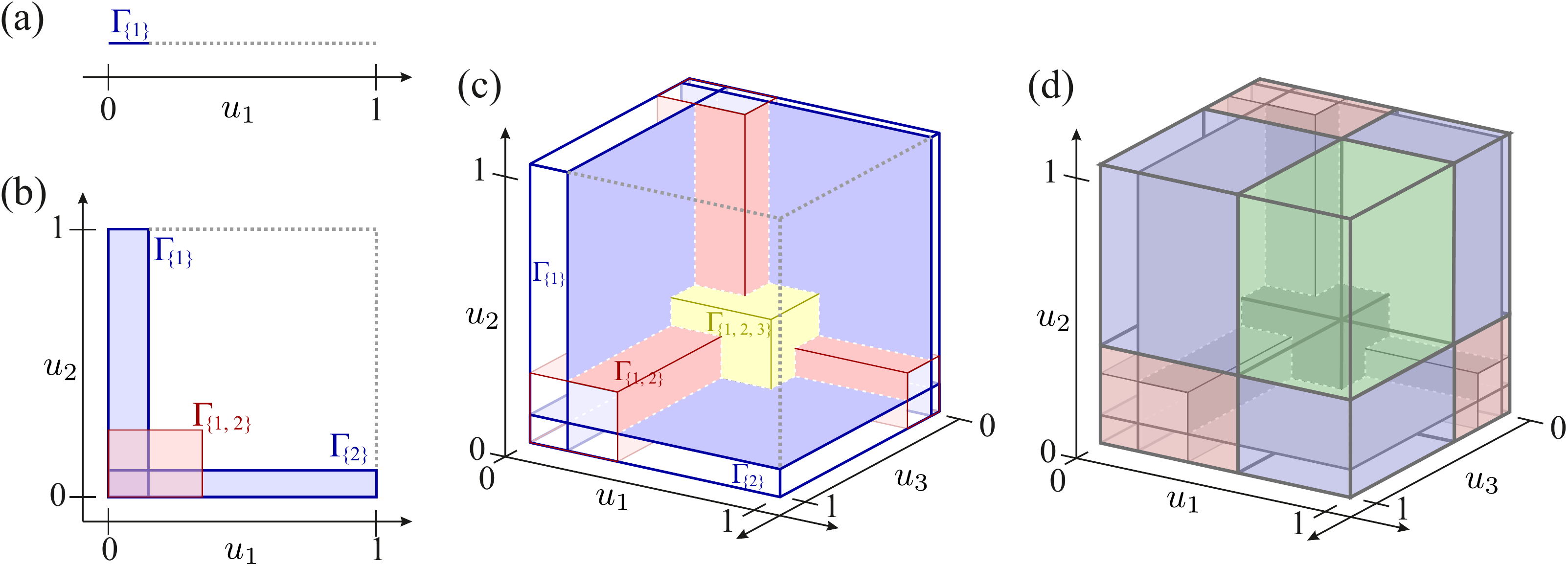}
    \caption{
    Illustration of non-inhabited region for the EHE model in (a) one, (b) two, and (c) three dimensions formed by the union of regions $\Gamma_I$ for index sets \(\emptyset \neq I \subseteq [N]\) defined by \Eq{eq:Dmain}. The colors blue, red, and yellow mark regions with index sets of cardinalities 1, 2, and 3, respectively.
    (d) Decomposition of the three-dimensional state space $C=[0,1)^3$ into the non-overlapping rectangles according to \Eq{eq:decomp}\label{fig:decomp_better}. The intersection of the cube with the region colored in green, blue, or red reduces to a product of a zero-, one-, or two-dimensional non-inhabited region with a cube, rectangle, or interval, respectively.}
\end{figure*}

\fi

\Fig{fig:decomp_better} (a)-(c) illustrates the self-similar geometry of the noninhabited region $\Lambda$ for one- to three-dimensional systems. In the phase space of the one-dimensional system shown in (a), the inhabited region consists of the interval $\Lambda_{[1]} = \Gamma_{\{1\}} = [0,w_{11})$. It is intuitively clear that the density of states has to vanish along this region in the one-dimensional system since after unit $u_1$ crossed the threshold and is reset, it immediately receives internal activation $w_{11}$ pushing it above this value. Since external activation only increases states $u$, the interval $[0,w_{11})$ can not be entered by the dynamics, i.e. $T_{1}(C\setminus [0,w_{11}]) \cap [0,w_{11}) = \emptyset$. Similarly, the noninhabited region for the two-dimensional system shown in (b) is the union of the two-dimensional extension of $\Gamma_{\{1\}}$, the equivalent region $\Gamma_{\{2\}}$ for unit 2, and $\Gamma_{\{1,2\}}$. As in the one dimensional system, the regions $\Gamma_{\{1\}}$ and $\Gamma_{\{2\}}$ mark the regions which can not be entered after unit $1$ and $2$ crossed the threshold, respectively. The additional feature in two dimensions is that both units can also fire together in an avalanche. In this case they receive internal activation from both units and the state after such an avalanche has to be outside of the rectangle $\Gamma_{\{1,2\}} = [0,w_{11}+w_{12})\times [0,w_{22}+w_{33})$. In general, the upper boundaries of any $\Gamma_{I}$ along dimensions specified by index set $I$ consist of the total internal activation which units $I$ receive in an avalanche with $\U(\av) = I$. In three dimensions, illustrated in subpanel (c), the noninhabited regions consists of the eight different subregions $\Gamma_{I},\emptyset \neq I \subseteq [3]$, and due to the recursive construction there is a striking self-similarity in the noninhabited region: The projection of $\Lambda_{[3]}$ to the face $u_3 = 1$ is the two-dimensional $\Lambda_{[2]}$ shown in panel (b) and the projection of $\Lambda_{[2]}$ in to $u_2 = 1$ is the interval $\Lambda_{[1]}$ shown in (a). This self-similarity of $\Lambda$ is used in the following section to identify the phase-space regions where $T_k$ elicits specific avalanches, and to compute the corresponding volumes with respect to the unique relative ergodic measure $\lambda_D$. 

Additionally, the self-similarity inherent in $D=C\setminus \Lambda$ allows to evaluate the \emph{cumulative distribution function} of $\PW$ in closed form for coordinates
$\VV$ with $\VU \geq \VV > W\1$ componentwise: Since condition~\eqref{ass} is fulfilled for the system with modified firing thresholds $\VV$, its inhabited region is given by $D' = [0, U'_i)_{i\in [N]} \setminus \Lambda_{[N]}$. In particular, we have $D' = [0, U'_i)_{i\in [N]} \cap D$ and thus
\[
	\PW(u_1 < U'_1,\ldots,u_N < U'_N) = \frac{\lambda([0, U'_i)_{i\in [N]} \cap D)}{\lambda(D)} 
	= \frac{\lvert \diag(\VV) - W \rvert}{\lvert \diag(\VU)-W \rvert} \text{ .}
\]

Furthermore, this  self-similarity relates the inhabited region $D$ of the full system to the inhabited region $D_I \coloneqq \pi_I(C \setminus \Lambda_I)$ of a lower-dimensional subsystem. The corresponding subsystem is defined by coupling the subset of units in $I\subseteq [N]$ by the submatrix $W_I$ obtained by choosing the rows and columns with indices $I$ from $W$. If the states of the units $[N] \setminus I$ are constrained to be greater or equal to $(W\1)_{[N]\setminus I}$, $\Lambda_{[N]}$  reduces to $\Lambda_I$ (see Lemma~\ref{lem:elimdim}). We denote the $\R^{I}$ Lebesgue volume of the inhabited region $D_I$ for the subsystem on $I \subseteq [N]$ with firing thresholds $\VV_I$ by (see Corollary~\ref{cor:volinhabited}) by
\begin{align}\label{eq:def_vi}
	\V_{I}(\VV) \coloneqq \lambda_I(\pi_I([0,U'_i)_{i\in I} \setminus \Lambda_I)) = \rvert\diag(\VV)_I-W_I\lvert
\end{align}

\subsection{Exact avalanche distributions follow from phase space volumes}

As for the two-dimensional example in \Fig{fig:trafoillus}, because of the ergodicity of $\PW=\lambda_D$, the probability distribution of the avalanches $\av$ is given by the volumes of the corresponding phase space regions. In this subsection we will identify and evaluate the $\PW$-volume of regions corresponding to certain avalanches as a function of an arbitrary coupling matrix $W$.

% ---------------- simple product of intervals along dimensions in the assembly -------------

For a detailed avalanche $a=\av(k, \Vu)$, the corresponding phase space region has a simple structure. Along the dimensions specified by its assembly it factorizes into a hyperrectangle $V_{a}$ according to the three following conditions:
\begin{itemize}
\item
The starting unit $a_1 = \{k\}$ has to be in the interval $[U_k-\DeltaU, U_k)$ for being able to be activated by the external input. 
\item
Similarly, the states of each unit in the $i$-th generation of an avalanche have to be sufficiently low such that internal activation received up to generation $i-2$ did not bring these units over threshold. At the same time, their states have to be sufficiently high such that the additional internal activation from generation $i-1$ succeeds in making the units fire.
\item
%---------------- fixed upper boundaries for the remaining dimensions -----------------------
The condition for the states of the units that do not participate in the avalanche is, due to absence of leaks, just that they are sufficiently low such that the total internal activation they receive in the avalanche does not push them above firing threshold.
\end{itemize}

%-------------- with self-similarity the pav distribution follows -----------------------

Due to the self-similar strudcture of the inhabited region (see previous section and Appendix, section~\ref{sec:self-similarity}), the subregion corresponding to the avalanche $a$ factorizes into a hyperrectangle along dimensions $\U(a)$ and a lower-dimensional inhabited region along dimensions $[N]\setminus \U(a)$ with upper boundaries reduced by $W\delta(\U(a))$ (see Proposition~\ref{prop:rav}).

These considerations lead to the probability distribution of avalanches:
\begin{align}\label{eq:pavinline}
  \P_k(\av=a) &= \frac{V_{a} \V_{[N]\setminus \U(a)}
		(\diag(\VU)-W\delta(\U(a)))}{\V_{[N]\setminus \{k\}}(\diag(\VU))}
		\quad\quad
		\mbox{with}
		\quad\quad
%\end{align}
%\begin{align}
  V_{a} \coloneqq \prod_{j=2}^{ \dur(a) }\prod_{i\in a_j}\sum_{\ell\in a_{j-1}}w_{i\ell} 
\end{align}
%---------- everything is solved now, isn't it -------------------------------
Equation~\eqref{eq:pavinline} completely specifies the probability distribution of detailed avalanches. Distributions over avalanche sizes $\P_k(\s(\av))$, avalanche durations $\P_k(\dur(\av))$, and avalanche assemblies $\P_k(\U(\av))$, as well as the probabilities $\P_k(i\in \U(\av))$ introduced in \Eq{eq:avprop} all follow from this distribution by summation over the corresponding detailed avalanche probabilities.

%---------- no, give me closed form expressions which can be interpreted --------------- 
However, due to the exponentially increasing number of detailed avalanches $\av$ with growing $N$ it is much harder to evaluate and investigate the dependence of these probabilities on the coupling structure: the corresponding sum over detailed avalanches is often difficult to bring into a closed form-expression in terms of the variables of interest, as it is possible for $\P_k(i \in \U(\av))$.  

%--------- there you go :) -----------------------------------------
Nevertheless, we were able to derive a closed-form expression for the avalanche \emph{assembly} distribution which is given by
\begin{align}\label{eq:pavu_inline}
	\P_k(\U(\av) = I) = \frac{\V_{I\setminus \{k\}}(W\delta(I))\, \V_{[N]\setminus I}(\diag(\VU)-W\delta(I))}
	{\V_{[N]\setminus \{k\}}(\diag(\VU))} \text{ .}
\end{align}
Note that $\V_{I\setminus \{k\}}(W\delta(I)) = 0$ if activation from unit $k$ can not spread to all units in $I$.  
%------------- intuition for the closed form, which reference to the appendix -----------
When deriving this expression from \Eq{eq:pavinline}, the sum of the $V_{a}$ over the corresponding avalanches is given in closed form by the single determinant $\V_{I\setminus \{k\}}(W\delta(I))$. We give two proofs of this remarkable identity (see Theorem~\ref{thm:avu}). The first one is a geometric proof that shows that the images of the avalanche regions under the dynamics $T_k$ cluster together and completely fill up the inhabited region along dimensions $\U(a)\setminus \{k\}$ up to the boundaries given by the total internal activation $W\delta(I)$ received by the units during the avalanche. The second, combinatorial proof directly uses the graph theoretical interpretation of the term $\V_{I \setminus \{k\}}(W\delta(I))$, which will be established in the next section.

\section{\label{sec:dependencies}Structure-function relation of avalanche assembly probabilities}

% ------------------ why study this relationship --------------------------

In this section we will interpret the assembly probability distribution in \Eq{eq:pavu_inline} in the context of graph theory, with the aim to distill the features of network connectivity which makes a given assembly likely to become active in form of an avalanche. To make the connections to graph topology easier to recognize, we will set $U = \mathbf{1}$ in this section.

Intuitively, assembly probability grows with increasing density of connections within the assembly, and increasing sparseness of the connections between the assembly and the rest of the network. The following considerations will allow us to assess which existing edges in the assembly network, i.e. the subnetwork along $\U(\av)$, contribute most to its activation probability, and which new edge would be most beneficial for increasing this probability. Such information becomes important when a network needs to be optimized for assembly formation under given biological constraints, such as having to spend energy for formation and strengthening neural connections.

We start by first introducing related graph theoretical concepts, and continue by linking these concepts to the assembly probability distribution in \Eq{eq:pavu_inline}.

%How do our formulas correlate with known graph theoretical measures?
%Especially measures of graph reliability?

\iffiginline
\begin{figure*}[t] 
  \includegraphics[width=12.9cm]{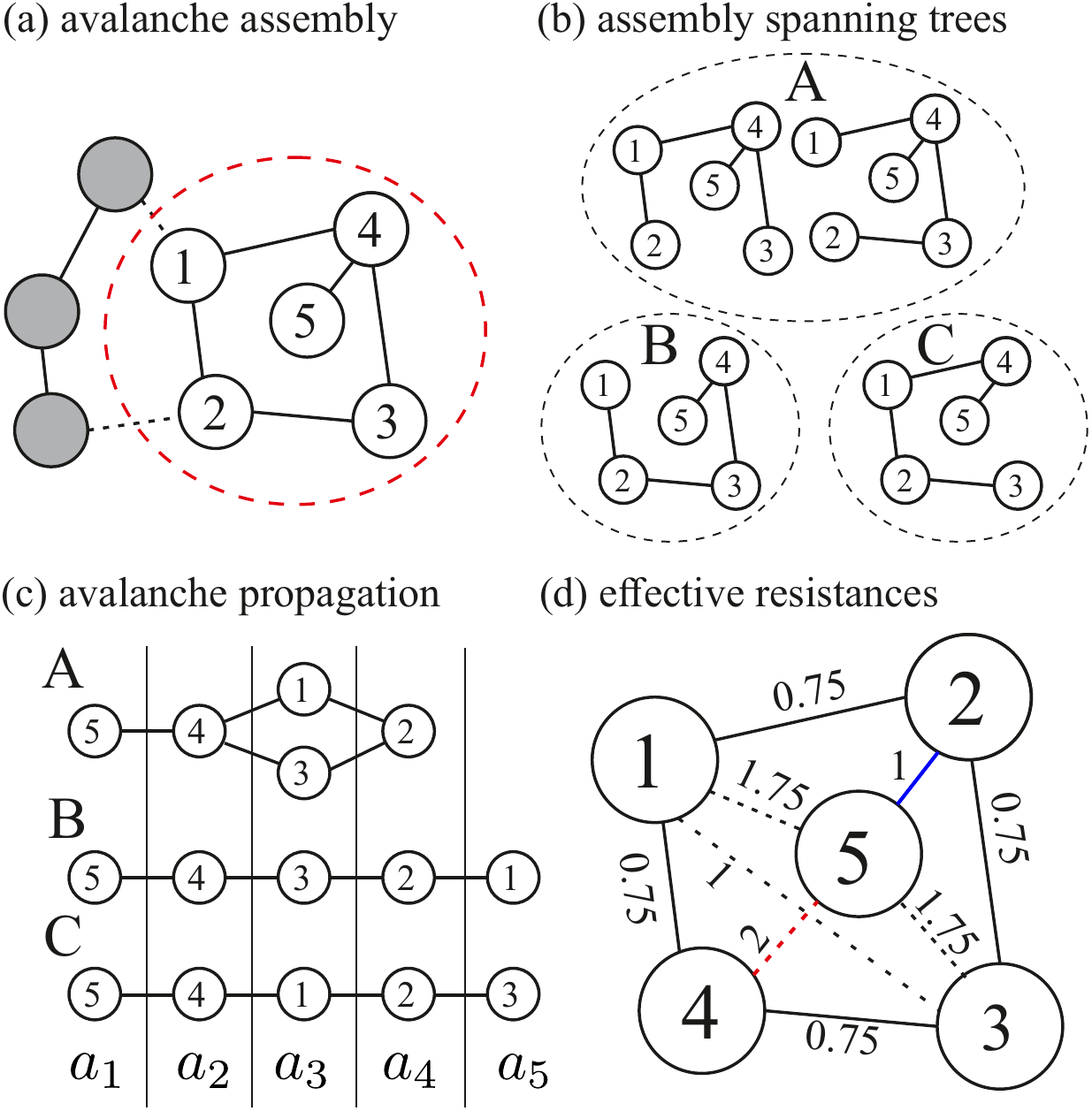}
  \caption{\label{fig:assembly_illus}
		Assembly probabilities relate to spanning trees and resistance distance.
		(a) Illustration of an avalanche assembly.
		Dashed red circle represents the graph cut separating the
        assembly \(I = \{1,2,3,4,5\}\) from the rest of the network (gray units).
        Dashed edges are in the cut set and contribute to the cut weight. For simplicity, the assembly graph is undirected.\\
		% unit weights -> werden erst in d) relevant.
		%
		(b) There are four spanning trees rooted at unit \(5\). They are obtained by
		deleting one of the edges of the \{1,2,3,4\}-cycle.\\
		(c) The three possible ways in which an avalanche starting at unit $5$
		can spread through the assembly. Numbers associate the avalanche 
		with the set(s) of corresponding spanning trees.\\
		(d) Effective resistances between pairs of units for an electrical network
		coupled by resistors with unit conductance along the edges of the graph
		(solid lines). Dashed lines represent edges missing in the assembly network.
		Numbers at existing edges also indicate the fraction of spanning trees
		that would be lost upon edge deletion. For example, assembly activation is 
		impossible without edge $(4, 5)$ (blue line).
		Numbers at non-existing edges indicate the relative number of additional
		spanning trees emerging when the edge is added to the assembly. 
		For example, adding the edge $(2, 5)$ (red line) would double the assembly probability
        by doubling the number of spanning trees.\\
	}
\end{figure*}

\fi

Selecting an assembly subnetwork with units $I\subseteq [N]$ describes a directed graph cut in which all outgoing edges from units in $I$ to units in $[N]\setminus I$ are part of the cut set with (vectorized) cut weight $\cut(I)\in \R^{[N]\setminus I},\cut(I)\coloneqq (W\delta(I))_{[N]\setminus I}$. The weight of the cut set is equivalent to the recurrent input the units in subnetwork $I$ provide to the units outside the subnetwork. \Fig{fig:assembly_illus}(a) illustrates the graph cut between an assembly of five units and the rest of the network.

The cut weight appears directly in \Eq{eq:pavu_inline} in the term $\V_{[N]\setminus I}(1-W\delta(I))$ which computes the phase space volume along dimensions $[N]\setminus I$ of the hyperrectangle $[0,1-(W\delta(I))_j)_{j\in [N]\setminus I}$:
\[\V_{[N]\setminus I}(1-W\delta(I)) = \lambda_{[N]\setminus I}(\{u\in D_{[N]\setminus I} \mid u< U_{[N]\setminus I} - \cut(I)\})\]
This term reveals that at the start of an avalanche, the state of all units that did not participate in it had to have a distance from firing threshold which was at least as big as the cut weight.

% Selecting an assembly subnetwork with units $I\subseteq [N]$ is described by a \emph{graph cut} with the \emph{cut set} $\cut(I) = \{(i,j) \mid  i\in I, j \in [N]\setminus I\}$ which contains all outgoing edges from units in $I$ to units in $[N]\setminus I$. \Fig{fig:assembly_illus}(a) illustrates the graph cut between an assembly of five units and the rest of the network.

% %-- probability that units outside assembly do not fire depends on graph cut weights -----

% The weighted cut set appears directly in \Eq{eq:pavu_inline} in the term $\V_{[N]\setminus I}(1-W\delta(I))$ which computes the phase space volume along dimensions $[N]\setminus I$ of the hyperrectangle $[0,1-(W\delta(I))_j)_{j\in [N]\setminus I}$. This term reveals that at the start of an avalanche, the state of all units that did not participate in it had to have a distance from firing threshold which was at least as big as the sum of weights in the cut set.

\subsection{Assembly probabilities are proportional to weighted number of spanning trees\label{sec:relationgraph}}

%------ assembly probability proportional to weighted number of spanning trees -----------
While the probability that units outside of the assembly \emph{do not} fire is determined by the weight of the graph cut, the probability that the units in the assembly \emph{do} fire is given by the weighted number of \emph{spanning trees} in the assembly subnetwork.

%----- introduce spanning trees --------------------

A directed graph $s_k = (V, E')$ will be called an (outgoing) \emph{spanning tree} of $G=(V, E)$ rooted at unit $k\in V$, if $s_k$ is a subgraph of $G$ which includes all vertices of $G$ and has $|V|-1$ edges $E' \subseteq E$ such that every unit except $k$ has an in-degree of 1, i.\,e.\ there exists a unique path from $k$ to each unit in $V$. For every spanning tree $s$ the product $\prod_{(j,i)\in E(s)} w_{ij}$ is the  {\em weight} of $s$, and for every subset $S$ of spanning trees we write
\[
	\w(S) \coloneqq \sum_{s\in S} \prod_{(j,i)\in E(s)} w_{ij}.
\]
%for the sum of weights of $s$ for all   $s\in S$. 
We further denote the set of all spanning trees rooted at vertex $k$ by $S_k$.

%----- introduce the graph Laplacian ---------------------- 

Spanning trees are well-studied objects in graph theory and closely connected to the \emph{graph Laplacian} $\mathcal{L}(W) =\diag(W\1)-W$. Note that $\diag(W\1)$ contains the weighted in-degrees of each unit on the diagonal. Similar to the adjacency matrix $W$, the graph Laplacian is a matrix representation of a graph and its spectral properties contain information about the graph connectivity of $G(W)$ \cite{cvetkovic1980spectra}. For example, $L$ has a trivial eigenvalue 0 corresponding to the eigenvector $\1$, while the second smallest eigenvalue is nonzero if and only if the graph is connected.
% TODO give a link to more details....

%------ matrix tree theorem connects the term in our formula to the number of spanning trees ---------

The product of the eigenvalues, except for the trivial one, is related to the weighted number of spanning trees by Kirchhoff's \emph{Matrix Tree Theorem}~\cite{bollobas2013modern,chaiken1978matrix}. Specifically, the weighted sum of spanning trees in the subnetwork along a subset $I\subseteq [N]$ starting in $k\in I$ is equal to $\mathcal{L}^{(k)}(W_I)$,  which is the $(k,k)$-cofactor $k\in I$ of the graph Laplacian for the induced subgraph $\mathcal{L}^{(k)}(W_I) \coloneqq |L(W_I)_{I\setminus \{k\}}|$. This is exactly the term $\V_{I\setminus \{k\}}(W\delta(I)) = \lvert\diag(W\delta(I))_{I\setminus \{k\}} - W_{I\setminus \{k\}}\rvert$ in the numerator of \Eq{eq:pavu_inline} and thus we have:
\[\V_{I\setminus \{k\}}(W\delta(I)) = \mathcal{L}^{(k)}(W_I) = \w(S_k)\, ,\]
where $S_k$ is the set of spanning trees in the assembly subnetwork, and taken together, we can rephrase \Eq{eq:pavu_inline} to be proportional to the product of the graph-theoretical terms
\begin{align}
    \P_k(\U(\av) = I) \propto \mathcal{L}^{(k)}(W_I)\lambda_{[N]\setminus I}(\{u\in D_{[N]\setminus I} \mid u< U_{[N]\setminus I} - \cut(I)\})\, .
\end{align}
 The associated correspondence between assembly probabilities and spanning trees is illustrated in \Fig{fig:assembly_illus}(b) for an assembly rooted at unit $k=5$.

%------- spanning trees = ways that an avalanche can spread through the network --------

There is a natural correspondence between spanning trees rooted at unit $k\in I$ of an assembly subgraph with $I$ as set of vertices, and the number of ways in which an avalanche can spread from $k$ through the assembly $I$. This correspondence was formalized for homogeneous networks in~\cite{levina2008mathematical} and is extended here to weighted directed graphs: Each avalanche $a=\av(\Vk, \Vu)$ specifies which units fire at which generation of the avalanche. Each generation specifies one level of the spanning tree, hence the units in $a_{j}$ are separated from the root by exactly $j-1$ edges. The term $V_{a}$ in \Eq{eq:pavinline} has a combinatorial interpretation, since expanding the terms leads to a sum over products of $|I|-1$ edge weights, with each product being the total weight of an entire spanning tree rooted at $k$ which is consistent with the level structure imposed by the detailed avalanche $a$. This correspondence is illustrated in \Fig{fig:assembly_illus}(c). Taken together, the sum over all terms in $V_{a}$ for avalanches $a=\av(k, \Vu)$ with $\U(a) = I$ is the total, weighted number of spanning trees rooted at $k$, leading to a combinatorial proof (Appendix, page~\pageref{proof:comb}) of \Eq{eq:pavu_inline}.

%------ what information does the number of spanning trees provide ----------------

Interestingly, the weighted number of spanning trees has a strong connection to graph reliability measures \cite{khosoussi2016maximizing}, yielding that the uniformly most robust graph maximizes the number of weighted spanning trees. Consequently, the probability for joint firing of an avalanche assembly is optimized when the assembly subnetwork is robust under random edge failure.             
% ------------------------------- synfire example ------------------------------
For example, the connectivity of a synfire chain resembles a bipartite graph which, in the undirected version, maximizes the number of spanning trees under the constraint of fixed number of edges and units \citep{khosoussi2016maximizing}.

\subsection{Effect of links on assembly probability is measured by effective resistance}

%------ now use this correspondence to measure edge importance ----------

We showed that the assembly probability is proportional to its weighted number of spanning trees. From this exact mathematical relation we derive a measure of the importance of individual edges in the assembly network as well as the optimal new connection to form in order to maximize the assembly probability. 
%--------------- explain the intuition behind the measure   --------------------------------

If a single link is removed from the assembly network, the probability $\P_k(\U(\av) = I)$ is reduced by the weighted number of spanning trees that contain this edge. Similarly, for all new connections between assembly units, the number of additional spanning trees made possible by incorporating this edge into the assembly network increases assembly probability. 

For $j,i \in I$ denote the graph obtained by inserting the (additional) edge $(j,i)$ with weight $w_{ij}>0$ in the weighted graph $G(W)$, and $S^{(j,i)}_k$ be the set of all spanning trees rooted at unit $k$ in the modified graph $G^{(j,i)}$. Note that $G^{(j,i)}=G$, if $ (j,i) \in E(G)$. With these definitions we can generalize the concept of resistance distance to weighted directed graphs by introducing the \emph{matrix of  directed $k$-resistances} $\Omega^k$ via \Eq{eq:resistance}:
% TODO:%---- todo modify formula for directed graphs and than simplify to undirected -------------
% \begin{align}\label{eq:resistance}
%   \Omega^k_{ij} \coloneqq \begin{cases}
% 		\w(\{  s\in S_k\mid(i,j) \in E(s)\}) / \w(S_k) &\mbox{ if } (i,j) \in E(G) \\
%     \w\left(S_k^{(i,j)}  \setminus S_k\right) /\w(S_k) &\mbox{ if } (i,j) \notin E(G), 
%   \end{cases} 
% \end{align}
\begin{align}\label{eq:resistance}
  \Omega^k_{ij} \coloneqq  
    \w\left(\left\{  s\in S_k^{(j,i)} \mid(j,i) \in E(s)\right\}\right) / (w_{ij}\w(S_k)).
\end{align}

% What does \Omega respresent in the general case
The entries of $\Omega^k$ specify the effect of adding/removing a single edge from the assembly network on the assembly probability $P_k^W(\U(\av) = I)$ as follows: 
\[
    \P_k^{W'}(\U(\av) = I) = (1 \pm \omega \Omega^k_{ij})\P_k^W(\U(\av) = I)\, ,
\]
where $W'$ is the coupling matrix obtained either by \emph{adding} the directed link with strength $\omega$ from unit $j\in I$ to unit $i\in I$ if $w_{ij}=0$ (plus sign on r.h.s. of equation), or by \emph{removing} the link with strength $\omega=w_{ij}$ (minus sign on r.h.s) from the original coupling matrix $W$.

%-------------- reduction to resistance distance in undirected networks -------------------

If the assembly subnetwork is undirected, i.e. $w_{ij}=w_{ji}$ for all $i,j \in I$, $\Omega^k$ becomes independent of $k$ and reduces to the resistance distance (\cite{klein1993resistance,bapat2003simple}). The matrix entry $\Omega_{ij}$ then corresponds to the effective resistance between units $i$ and $j$ in an equivalent electrical network in which edges represent resistors with conductances given by the edge weights. $\Omega$ can be computed efficiently and has many applications extending far beyond electrical networks, for example for studying commute times in random walks~\cite{chandra1996electrical,lyons2017probability}.

%------------- illustration with example network --------------------------------------------

\Fig{fig:assembly_illus}(d) annotates the effective resistances in the illustrated simple assembly network. While a failure of one of the edges in the circle connecting units $1,2,3,4$ would destroy three out of the four assembly spanning trees, all spanning trees rely on existence of the edge between units 4 and 5. The optimal new edge to add to the assembly network in order to maximally increase the probability of the assembly avalanches would be the edge between 2 and 5 with effective resistance $\Omega_{2,5} = 2$, thus tripling the assembly probability.

%------ motivation for looking at membrane potential correlations and branching processes -------

In the next two subsections, we switch the focus from how network topology influences assembly probabilities to how it induces correlations between membrane potentials of recurrently connected units, and how these correlations affect the \emph{dynamics} (branching) of an ongoing avalanche. These investigations allow to state conditions on the networks on which the EHE-model reduces to a simple percolation process, and to identify its universality class.

\subsection{States of recurrently connected units are stochastically dependent}

% ---- start by describing that the geometry of the inhabited region determines correlations ----
We showed that the unique ergodic measure $\PW$ of the system is given by the Lebesgue measure $\lambda_D$ supported on the inhabited region $D$. This not only allowed us to determine phase space volumes that represent certain avalanche probabilities, but can also be used to determine stochastic dependencies between states of different units. For a uniform measure, these dependencies can be deduced entirely from the geometrical structure of the support $D$.

If the uniform measure is supported on a rectangle, i.\,e.\ if it factorizes into simple intervals, then the units' states are statistically independent. However, this is typically not the case for the generalized EHE-model, as can be seen from the structure of the inhabited region displayed in \Fig{fig:trafoillus}: The interval of possible values for $u_1$ in the inhabited region is smaller if the state $u_2$ is close to its allowed minimum, while the interval gets bigger when the state $u_2$ is near firing threshold. Thus, the geometry of the inhabited region $D$ (or equivalently, the non-inhabited region $\Lambda_{[N]}$) reflects a negative correlation
between the states $u_1,u_2$, which decreases the probability to find both units in low states, and in turn facilitates that the units fire together in an avalanche.

But exactly which features of the graph topology influence the volume and geometric structure of the inhabited region $D$? It turns out that its volume is completely determined by the eigenvalues of $W$ or -- equivalently -- by circle motifs occurring in $G(W)$, and by the fact that the inhabited region factorizes along the \emph{strongly connected components} of $G$. Strongly connected components are subnetworks in which each unit is reachable from each other unit.
%--------- product of displaced eigenvalues ------------------ 
In more detail, the spectrum of the adjacency matrix $W$ determines the phase space volume of the inhabited region by
\begin{align}\label{eq:psvev}
  \lvert D \rvert = \lvert \mathds{1}-W\rvert= \prod_{i=1}^N(1-\lambda_i) \, ,
\end{align}
where $\lambda_i$ are the eigenvalues of $W$.
%-------- product of cycle motifs --------------------------
There is a combinatorial interpretation of $|\mathds{1}-W|$ which allows to identify cycles in $W$ as the relevant feature determining the volume of the inhabited region (see Corollary~\ref{cor:loop}):
\begin{align*}%\label{eq:ivw}
  % \lvert \mathds{1}-W\rvert = \sum_{i=1}^N\sum_{L\in\mathcal{L}(W)}(-1)^{\#(L)}\w (L), 
	|\mathds{1}-W| = \sum_{n=1}^N\sum_{\mathcal{L}\in\mathscr{L}_n(W)}(-1)^{\#(\mathcal{L})}\prod (\mathcal{L}) 
\end{align*}
where $\mathscr{L}_n(W)$ is the set of all linear directed subgraphs $\mathcal{L}$ of $W$ with $n$ nodes, $\#(\mathcal{L})$ denotes the number of connected components of $\mathcal{L}$, and $\prod(\mathcal{L})$ is the product of all edge weights in $\mathcal{L}$. Note that each component of a linear directed subgraph is a directed cycle.

%----- self-loops vs other cycles 

There is an important distinction between self-loops and directed cycles connecting at least two units. The effect of self-loops on the inhabited phase space is equivalent to lowering the corresponding firing thresholds (Proposition~\ref{prop:self-weights}), whereas a recurrent coupling between more than one unit induces a stochastic dependency between the recurrently connected units (i.\,e.\ units in the same strongly connected component) as described previously.

%--------- factorizes over strongly connected components ----------

However, only the states of units which are recurrently connected are stochastically dependent in this model. The set of strongly connected components partitions the units in a graph in such a way that the connections between these components form a directed acyclic graph (DAG). In fact, the inhabited region factorizes into a direct product of the inhabited regions along the strongly connected components of $W$ (Theorem~\ref{thm:unit-correlations}). This correspondence between network topology and phase space structure is illustrated in \Fig{fig:avillus_phase_space} for different three-unit network motifs.

\iffiginline
\begin{figure*}[t] 
  \includegraphics[width=17.2cm]{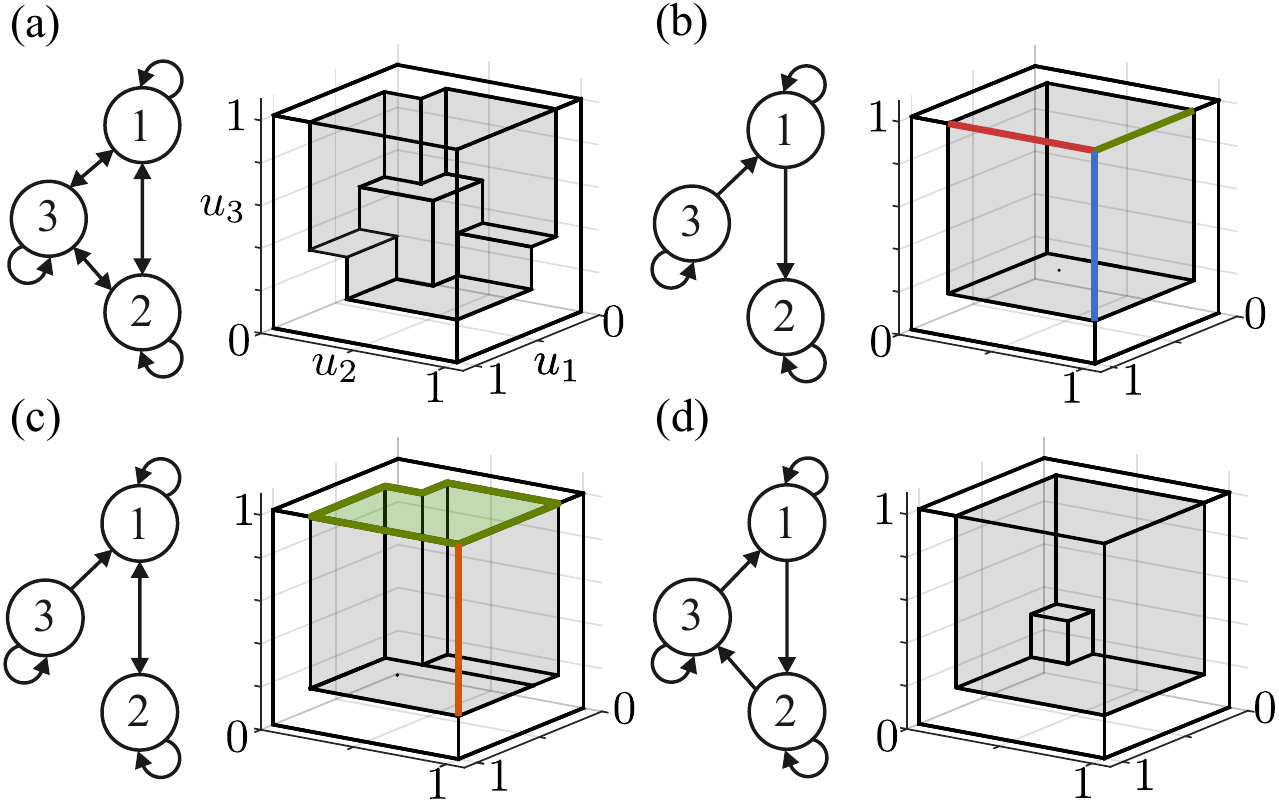} 
  \caption{\label{fig:avillus_phase_space} 
		Relation between phase space structure, graph motifs, and state correlations.
		The top row shows the phase space structure for weight matrices satisfying
		the graph motifs depicted in the bottom row (all edge weights are set to 0.2).
		The boundary between inhabited and non-inhabited region is shaded in gray,
		with the inhabited region being the complement of the non-inhabited region
		$\Lambda$ in the unit cube.
    (a) complete digraph. The self-similarity of the phase space structure is apparent
		at the faces of the cube, on which two-dimensional inhabited regions emerge
		(cf. to \Fig{fig:trafoillus}).
    (b) Graph motif without recurrent connections between different units.
        The inhabited region factors
		into a product of three intervals (red, blue, and green lines).
    (c) Strongly connected components are \(\{1,2\}\) and \(\{3\}\).
    Correspondingly, the inhabited region decomposes into a direct
    product of the two-dimensional \(\{1,2\}\) inhabited region (green area) and
    an interval along \(u_3\) (red line).
    (d) The circle motif connects all units recurrently. Like in (a),
    the inhabited region does not factorize.
	}
\end{figure*}
\fi

% ---- these correlations lead to larger branching probabilities -----------------
These considerations show that the inhabited region $D$ is the full cube $C$ if the coupling network is a DAG and that in this case all states are stochastically independent. The conditional branching probabilitity \Eq{eq:avprop} is particularly easy to interpret in this case: $M_{ik}$ is the finite sum of paths between nodes $k$ and $i$ weighted by the product of their edge weights in the DAG, and $M_{kk}=1$. Increasing an edge weight in a DAG only increases the numerator in \Eq{eq:avprop}. 

However, if there are recurrent connections in the coupling matrix, the number of paths between two nodes is not necessarily finite anymore. Since we have a limit on the recurrent feedback via \Eq{ass}, avalanches can also not spread along paths with units occurring more than once. In consequence, there is a discrepancy between the interpretation of $M_{ik}$ as a sum over all putative paths in the network, and the much smaller number of paths that can actually be realized by propagating avalanches. This discrepancy is resolved by the denominator $M_{kk}$ which is bigger than one in recurrent networks: Geometrically, $M_{kk}$ is the quotient of the ($\NN$-1)-dimensional volume of the (hyper-)face $u_k=1$ of the inhabited region and the $\NN$-dimensional volume of $D$. 
As we have shown above, the inhabited region shrinks with increasing recurrent weights and increases correlations in the unit's states which in turn increases $M_{kk}$. Thus, these correlations lead to an additional increase in the branching probabilities during avalanches which compensates for the lower number of possible paths along which an avalanche can spread in this model.

\subsection{Avalanche branching process and relation to directed percolation}

% ------------ now avalanche dynamics --------------------------------

In this subsection we investigate how state correlations influence the \emph{dynamics} during an avalanche i.\,e.\ the branching of the avalanche through the network. These considerations allow us to show the EHE-model reduces to a compact directed percolation process on DAGs.

% Neural avalanches are often studied using branching processes on
% networks instead of spiking neuron models. For the homogeneous
% EHE-model it was shown that the avalanche size statistics converges in
% distribution the statistics obtained by a Watson-Galton branching
% process. In this section we describe how different network topologies
% shape the avalanche branching process associated with the linear
% spiking neuron model.

%---------describe correspondence to discrete units -----------------------

In order to describe the branching process associated to the spreading of an avalanche in the model, we tag the units during the ongoing avalanche as either off, active or refractory. Let $\Va =\av(k, \Vu)$ with $\P(\av=\Va) > 0$. At generation $2 \leq j \leq \dur(\Va)$, the \emph{active} units $a_{j-1}$ are the units that crossed the threshold in the previous generation. Since recurrent feedback has an upper threshold defined by \Eq{ass}, all previously active units can not fire again in the ongoing avalanche $\U_{j-2}(\Va)$ and thus become 'quasi'-\emph{refractory} from the next generation on. The remaining units $[N] \setminus \U_{j-1}(\Va)$ are in the \emph{off}-state until they eventually become active.

%-------state the resulting update equation --------------------------

The probability that the set of units $a_j$ becomes active in generation $j$ of the avalanche, given the units which fired in previous generations $a_1, \ldots, a_{j-1}$ is given by the following equation (Theorem~\ref{thm:branching}):
\begin{align}\label{eq:branching}
  &  \!\!\!\!\!\!\!\!\!\!\!\!\!\!\!\!  \P(\av_j=a_{j}|\av_1=a_{1},\ldots,\av_{j-1}=a_{j-1}) \nonumber\\
	&= \prod_{k \in a_j}\left (\sum_{\ell\in a_{j-1}}w_{k\ell} \right )
  \times \frac{\V_{[N]\setminus \U_{j}(a)}(\1 - W\delta(\U_{j-1}(\Va)))}{
  \V_{[N] \setminus \U_{j-1}(a)}(\1  - W\delta(\U_{j-2}(\Va)))}
\end{align}

%------more complicated than simple watson-galton model obtained for hom. system -----

For the homogeneous EHE-model it was shown~\cite{levina2008mathematical} that the avalanche size statistics converges in its distribution to the statistics obtained from a Galton-Watson branching process. However, the general branching process described by \Eq{eq:branching} is much more involved as  it requires the  memory of the units triggered in previous avalanche steps, and since updates of individual units $\P(\av_j=a_j|\av_1=a_1,\ldots,\av_{j-1}=a_{j-1})$ for $a_j \in [N]\setminus \U_{j-1}(\Va)$ are not statistically independent due to correlations between their states.

If the coupling matrix $W$ represents a DAG, the inhabited region is the complete cube $C$ and the branching equation simplifies to 
\begin{align*}
  &  \!\!\!\!\!\!\!\!\!\!\!\!\!\!\!\!  \P(\av_j=a_{j}|\av_1=a_{1},\ldots,\av_{j-1}=a_{j-1}) \nonumber\\
	&= \prod_{k \in a_j}\left (\sum_{\ell\in a_{j-1}}w_{k\ell} \right )
   \prod_{k\in [N]\setminus \U_{j}(a)}(1-(W(\delta(\U_{j-1})))_k) \bigg /  \prod_{m\in [N] \setminus \U_{j-1}(a)}(1-(W\delta(\U_{j-2}(a)))_m) \text{ .}
\end{align*}
In this case, the region of states consistent with the avalanche propagation up to step $j-1$ of the avalanche is given by a simple hyperrectangle. and the probability that units fire in step $j$ of the avalanche is for each unit $k$ equal to $\sum_{\ell\in a_{j-1}}w_{k\ell}/(1-(W\delta(\U_{j-2}(a)))_k)$.
The dependence of the branching probability for step $j$ on the previously active units $\U_{j-2}(a)$ vanishes on networks for which all paths from the starting unit $a_1$ to an arbitrary unit $k$ have the same number of steps. This is the case, for example, if the coupling network is a directed tree or a regular percolation network. In this case, the branching probability becomes particularly simple: 
\begin{align*}
\P(\av_j=a_{j}|\av_{j-1}=a_{j-1}) = \prod_{k \in a_j}\left (\sum_{\ell\in a_{j-1}}w_{k\ell} \right )
	\prod_{k \in [N]\setminus a_j}\left (1-\sum_{\ell\in a_{j-1}}w_{k\ell} \right ).
\end{align*}
This indicates a simple branching process, in which each unit $k$ fires (transitions to the active state) independently on other units with probability $\sum_{\ell \in a_{j-1}}w_{k\ell}$. As an example, consider the (infinite) (1+1)D lattice, in which each unit $v_{t,x}, t\in \N,x\in \Z$  receives input only from its two 'parents' $v_{t-1,x-1},v_{t-1,x+1}$ with connection strength $w$. In this case, the probability that the node $v_{t,x}$ fires at the current step of the avalanche depends on how many of its parent nodes fired in the previous step of the avalanche. If none/exactly one/both of its parents was active in the previous step of the avalanche, the branching probability is $0/w/2w$, respectively. This shows that the EHE-model on the (1+1)D lattice is equivalent to the Domany-Kinzel model~\cite{PhysRevLett.53.311,kinzel1985phase} with $p_2=2p_1$, and thus has its critical point in the limit $w=1/2$ in which it displays \emph{compact directed percolation}, which belongs to the exactly solvable universality class of branching-annihilating random walks~\cite[Section 3.2]{hinrichsen2000non}.

\section{Application to structurally simple networks \label{sec:shift-invariant-networks}}

%-------- motivation ---------------------------
Our mathematical framework provides a novel gateway for better understanding collective behavior and synchronization statistics in recurrent excitatory networks. To demonstrate its advantage, we apply our framework in this section to structurally simple examples, including a planar network with periodic boundary conditions and distance-limited connectivity. Here we study deviations from mean field behavior in dependence of changes in coupling topology analytically, and show that scaling exponents of the mean avalanche size depend on the maximal coupling distance. The limiting case of all-to-all couplings leads to a homogeneous system without self-weights, for which we derive not only the mean avalanche size but also the avalanche size distribution analytically. In addition we illustrate topology-induced effects on the avalanche size distribution in small networks by rewiring a ring network into a small world network, and by transforming a ring network into a line network by deletion of a single edge. Furthermore, we quantify the effect of an inhomogeneity between intra-network and inter-network coupling weights on the avalanche size distribution of two all-to-all coupled subnetworks. 

%------ homogeneous case is particulary simple --------------------

\subsection{Homogeneous network without self-weights} 
 In this section we derive the analytical avalanche size distribution and the mean firing rate for a homogeneous network without self-weights. We denote the coupling matrix of this network with zeros on the diagonal and otherwise constant entries $\alpha/(N-1)$ by $W^{\text{h}}(\alpha)$. Using our mathematical framework, it is easy to extend the known results for homogeneous networks with self-weights \cite{eurich2002finite,levina2008mathematical} to calculate the avalanche size distribution and the mean avalanche size in dependence of $N$ and $\alpha$. For completeness, we also show how our framework reproduces the known expressions for the avalanche size distribution and mean avalanche size of the homogeneous network with self-weights in the appendix, section~\ref{sec:app-homogeneous}.

%----------- derivation of the S(av) distribution ---------------------

Due to the symmetry in the homogeneous network, every assembly of size $n$ has equal probability and thus the avalanche size distribution is obtained from the assembly distribution by counting the number of assemblies of a given size. Let $W^{\text{h}} = ((1-\delta_{ij})w)_{i,j\in [N]}$ for some $w\geq0$. Fixing an arbitrary starting unit, there are $\binom{N-1}{n-1}$ possible assemblies of size $n$. Every assembly graph is itself a complete graph.
Due to this symmetry, the probability of $\s(\av)=n$ follows from the probabilities of assemblies with $|I|=n$ units for the special case of homogeneous matrices. We will give closed form expressions for the determinants $\V_{I\setminus \{k\}}(W^{\text{h}}\delta(I))$ and $\V_J(\VV)$ for general $\VV \in \R_{>0}^{[N]},\emptyset \neq J\subseteq [N]$ occuring in~\eqref{eq:pavu_inline}.
The first term is the $(k,k)$ minor of the assembly subgraph Laplacian which is equal to the number of spanning trees in the assembly subgraph rooted at $k$. By Cayley's Theorem, which is the special case of the Matrix Tree Theorem for complete graphs, the number of spanning trees in a complete graph with $n$ units is $n^{n-2}$. Each spanning tree consists of $n-1$ edges and is thus weighted by $w^{n-1}$. For the more general $\V_{J}(\VV)$ we obtain $\V_{J}(\VV) = (\VV+w)^{|I|}-|I|w(\VV+w)^{|I|-1}$ by using Proposition~\ref{prop:self-weights} and the corresponding expression~\eqref{eq:V-hom} for homogeneous matrices with self-loops. Using the parametrisation $w = \frac{\alpha}{N-1}, 0 \leq \alpha < 1$, the avalanche size distribution of nonempty avalanches in the homogeneous network without self-weights is given by:
\begin{align}\label{eq:pavs-hom-ws}
  p_h(n) & = \P^{W^h}(\s(\av) = n|\s(\av) > 0) \nonumber \\    
		& = \binom{N-1}{n-1}n^{n-2}\left (\frac{\alpha}{N-1} \right)^{n-1} 
    \frac{\left (1-\frac{(n-1)\alpha}{N-1}\right  )^{N-n-1}(1-\alpha)}{\left (1+\frac{\alpha}{N-1} \right )^{N-2}
		\left (1-\frac{(N-2)\alpha}{N-1}\right )}
\end{align}
%--------------------- mean field avalanche size exponent -----------------
Using the Stirling approximations for the factorial $n!\approx \sqrt{2 \pi n}(\frac{n}{e})^n$ and for the binomial coefficients $\binom{L}{l} \approx \frac{L^l}{l!}$ for $n\gg 1,L \gg l$, we obtain a power law scaling of the avalanche size distribution in the limit $N\rightarrow \infty,\alpha \uparrow 1$ with exponent $3/2$, which is the expected mean-field limit. 

%------------------- mean avalanche size -----------------------------------  

With $M^{\text{h}}$  given by the expression
\[
	M^{\text{h}}_{ij}=\left ((\mathds{1}-W^{\text{h}})^{-1}\right)_{ij} = \frac{w+ \delta_{ij}(1-(N-1)w)}{1-(N-2)w-(N-1)w^2}
\]
and \Eq{eq:mavs} we obtain the mean avalanche size as
\[
	\mathds{E}^{W^\text{h}}(\s(\av)|\s(\av)> 0) = \frac{N-1+\alpha}{(N-1)-(N-2)\alpha} \, .
\]
For large $N$, the expected avalanche size is approximated by $\E^{W^\text{h}}(S(\av)) \sim {1}/{(1-\alpha)}$ and scales like a power law with exponent $-1$ in dependence of $1-\alpha$.

%--------- deviations from mean field behavior for limited coupling weights
\subsection{Planar network with periodic boundary conditions and translation-invariant distance-limited connectivity}

%-------- explain the network structure ----------------------------
Consider a two-dimensional grid of units with periodic boundary conditions and varying coupling distance $l$. On the $L \times L$ periodic grid, an edge exists between each pair of distinct units at positions $(i_1,i_2)$ and $(j_1,j_2)$ if the distance $\max\{|i_1-j_1 \mod L|,| i_2- j_2 \mod L|\} \leq l$. For $l=1$, each unit is connected to its eight neighbors with distance one, while $l \geq \lceil (L-1)/2 \rceil$ leads to the fully connected graph, which we have treated in the previous section. For simplicity, we impose a uniform edge weight of $w = \alpha/((2l+1)^2-1), \alpha\in [0,1)$. We denote this connectivity by
the coupling matrix $W^l(\alpha)$.

%-- what we do - closed form analytical mean avalanche scaling exponents ---
We will examine the avalanche size distribution and the scaling of the mean avalanche size, for which we obtain a closed-form analytic solution for every $l$ - in dependence of the coupling distance $l$.

For $l < \lceil (L-1)/2 \rceil$, coupling is not all-to-all anymore. In this case, there is no apparent symmetry between avalanche assemblies of the same size that would simplify computing the avalanche size distribution. Even though we can evaluate each specific assembly probability analytically using \Eq{eq:pavu_inline}, we did not find a closed-form expression for the sum over assemblies of a given size. However, we found a closed-form expression for the mean avalanche size in dependence of $L$ and the coupling strength $w = \alpha/((2l+1)^2-1)$.

%------------ using shift invariance for eigenvalues of W ------------

From shift invariance and periodic boundary conditions, matrix vector multiplication of the $L^2 \times L^2$ matrix $W$ represents a two-dimensional convolution with a rectangular $(2L+1) \times (2L+1)$ point spread function. The eigenvalues of $W$ are thus given by the two-dimensional Fourier transformation of its point spread function and determine the phase space volume through \Eq{eq:psvev}. We denote the eigenvalues of $W^\ell$ by $\lambda_i^{(\ell)}, i=1,\ldots,L^2$.

%----------- they give us the mean avalanche size in closed form -------------

We will now determine the mean avalanche size given by \Eq{eq:mavs}. To do this, we need to know the diagonal elements of $M^\ell \coloneqq (\mathds{1}-W^\ell)$ and the sum of entries in its rows. Note that all entries of $M^\ell$ are positive, since the graph given by $W^\ell$ is connected, and that $M^\ell$ inherits the shift invariance from $W^\ell$. Thus, all diagonal elements $M^\ell_{kk}$ are the same and equal to 
\[
    M^\ell_{kk} = \operatorname{trace}(M^\ell)/N = \sum_{i=1}^N \frac{1}{N(1-\lambda_i^{(\ell)})}\, .
\]
The second equation follows since each eigenvalue $\lambda_i^{(\ell)}$ of $W^\ell$ corresponds to an eigenvalue $(1-\lambda^{(\ell)}_i)^{-1}$ of $M^\ell$. The special structure of $M^\ell$ also leads to a closed form for its column-sum norm $\|M^{\ell}\|_{1} \coloneqq \max_{j\in [N]}\sum_{i=1}^N |M^{\ell}_{ij}|$. It is the inverse of a diagonally dominant M-matrix, which allows to use \cite[corollary 4]{moravca2008bounds} to find $\|M^{\ell}\|_1 = (1-\alpha)^{-1}$.

Taken together, we arrive at a closed form expression for the mean (non-empty) avalanche size 
\begin{align}\label{eq:mavs-shift-invariant-torus}
  \mathds{E}^{W^\ell}(\s(\av)|\av_1 = \{k\}) = \frac{\|M^\ell\|_1}{M^\ell_{kk}} = \frac{N(1-\alpha)^{-1}}{\sum_{i=1}^N(1-\lambda^{(\ell)}_i)^{-1}} = \mathds{E}^{W^\ell}(\s(\av)|\s(\av)>0)\, .
\end{align}
Note that eigenvalues $\lambda_i^{(\ell)}$ depend on $\alpha$. Note that \Eq{eq:mavs-shift-invariant-torus} holds for arbitrary non-negative shift-invariant coupling matrixes with $\alpha<1$ being the sum of incoming edge weights to each unit and $\lambda_i$ the eigenvalues of the corresponding point spread function.

\iffiginline
\begin{figure*}
  \includegraphics[width=12.9cm]{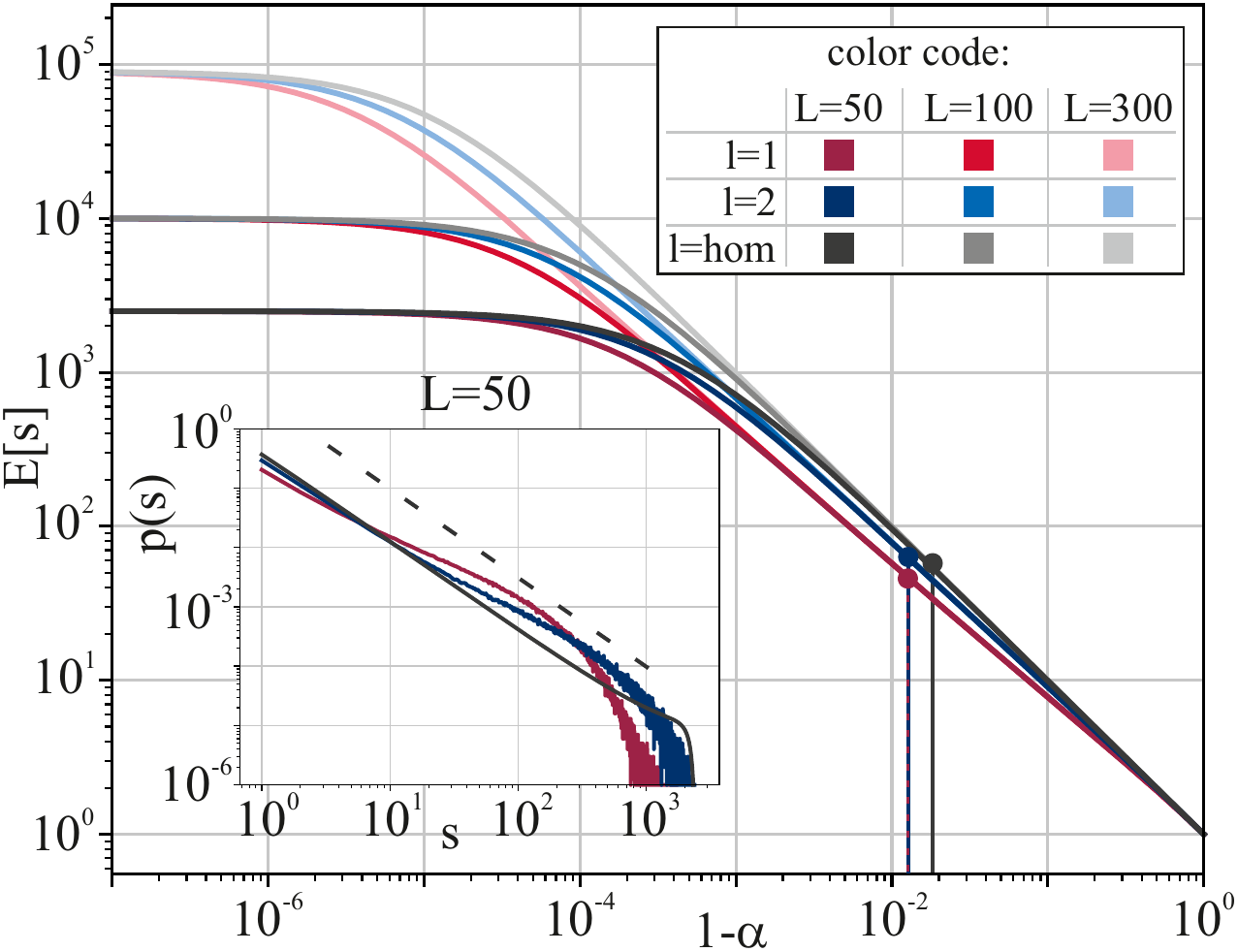}
  \caption{\label{fig:mavs-scaling}
		Scaling of mean avalanche size $E[s] \coloneqq \mathbb{E}(\s(\av)\mid \s(av) > 0)$ with coupling strength $\alpha$ for uniform translation-invariant coupling on a periodic two-dimensional grid, evaluated for different grid lengths $L$
        and coupling distances $l$ as well as the homogeneous network. The scaling is depicted as a log-log plot of $E[s]$ in dependence of $1-\alpha$, where $\alpha$ denotes the sum of incoming weights to each unit. Colors red, blue and gray indicate coupling distances $l$ of
        1, 2 and a global coupling ('hom'), respectively. Brightness of the colors denotes different
        grid lengths $L$ (dark, medium and bright for $L=50, 100, 300$, respectively).
        At the 'critical' values of $\alpha$ for $L=50$ indicated by the dashed lines and filled dots,
        the corresponding avalanche size distributions $p(s) \coloneqq \P(\s(\av) = s \mid \s(\av) > 0)$ (inset) resemble power laws.
		For $l=1,2$, avalanche size distributions were obtained
		from a simulation of $10^6$ avalanches, while the analytical
		distribution from~\Eq{eq:pavs-hom-ws} is shown for the homogeneous network
		for the critical coupling strength given by~\Eq{eq:alpha-crit-hom-ws}.
		The dashed straight line in the inset has slope $-3/2$.
	}
\end{figure*}
\fi

\Fig{fig:mavs-scaling} displays mean avalanche sizes $\mathds{E}^{W^{\text{h}}}(n)$ for the globally homogeneous network without self-interactions, and mean avalanche sizes $\mathds{E}^{W^\ell}(n)$ for networks with limited coupling distance, as functions of $1-\alpha$ for grid sizes $L=50,100,300$. For $l=1$, each unit is connected to its 8 nearest neighbors with uniform weight $\alpha/8 $, for $l=2$ it is connected to the 24 neighboring units up to distance 2 with weight $\alpha/24$. While for fixed $l$ the graph remains sparse with strong edge weights, the homogeneous coupling is dense and the edge weights scale like $1/N$. Increasing the coupling distance leads for fixed $L$ to on average larger avalanches for all $\alpha\in(0,1)$. This is expected due to the greater number of units that are reachable during each step of an avalanche, which lowers the chance for the avalanche to stop. In the limit $\alpha \rightarrow 1$, the mean avalanche size reaches the system size $L^2$ regardless of the coupling scheme (homogeneous or distance-limited). While the network topology does not change the mean avalanche size in the limit $\alpha \rightarrow 1$, it affects the scaling exponent. As shown above, the mean avalanche size scales according to $(1-\alpha)^{-\epsilon}$ with $\epsilon = \epsilon\left (W^{\text{h}}\right )=1$. This scaling exponent grows with shrinking coupling distance, such that $0 <\epsilon\left (W^{l=1}\right ) < \epsilon\left (W^{l=2}\right ) < \epsilon\left (W^{\text{h}} \right )= 1$. This result is intuitively plausible, since a limited coupling distance imposes topological constraints on the spread of an avalanche. This constraint makes larger avalanches less likely to occur, while smaller avalanches are observed more frequently due to the larger interaction strength between two units with decreasing $l$.

%--------- changes in scaling also obtained in numerical avs distributions -------
Differences in scaling exponents are also apparent in the \emph{critical} avalanche size distributions (\Fig{fig:mavs-scaling}, inset) which exhibit a power-law characteristics. While the slope for the homogeneous network approximates the mean field exponent $-3/2$, it becomes less negative with limited coupling distance.

\subsection{Analytical avalanche size distributions in small or structurally simple networks}

% --- assembly distribution and mean avalanche size given in closed form
% --- for arbitrary networks, but size distribution has to be computed by summing
% --- over assemblies in the general case

While our framework provides a closed form expression for the mean avalanche size and for the probability of any avalanche assembly for networks with arbitrary non-negative couplings, the avalanche size distribution has to be obtained by summing over all $2^N$ assemblies in the network (and over at most $N$ choices for the unit starting the avalanche). This is only feasible for small networks with fewer than $20$ nodes. However, networks with specific structure may allow a more efficient computation of the avalanche size distributions, for example by exploiting symmetries -- like in the homogeneous network, where each assembly of a given size is equally likely to occur -- or by taking advantage of sparsity in the network which limits the number of connected assemblies.

% --- effect of network connectivity in sample small networks
In this section, we showcase how the network connectivity changes the avalanche size distributions using example networks which are sufficiently small or structurally simple, using a uniform driving probability $p=(1/N)\cdot\1$. Our considerations are accompanied by the formal treatment detailed in section~\ref{sec:app-simple} of the Appendix, which is complemented by some analytical insights for the ring network and Erdös-Renyi networks.

\iffiginline
\begin{figure*}
  \centering
    \includegraphics[width=17.2cm]{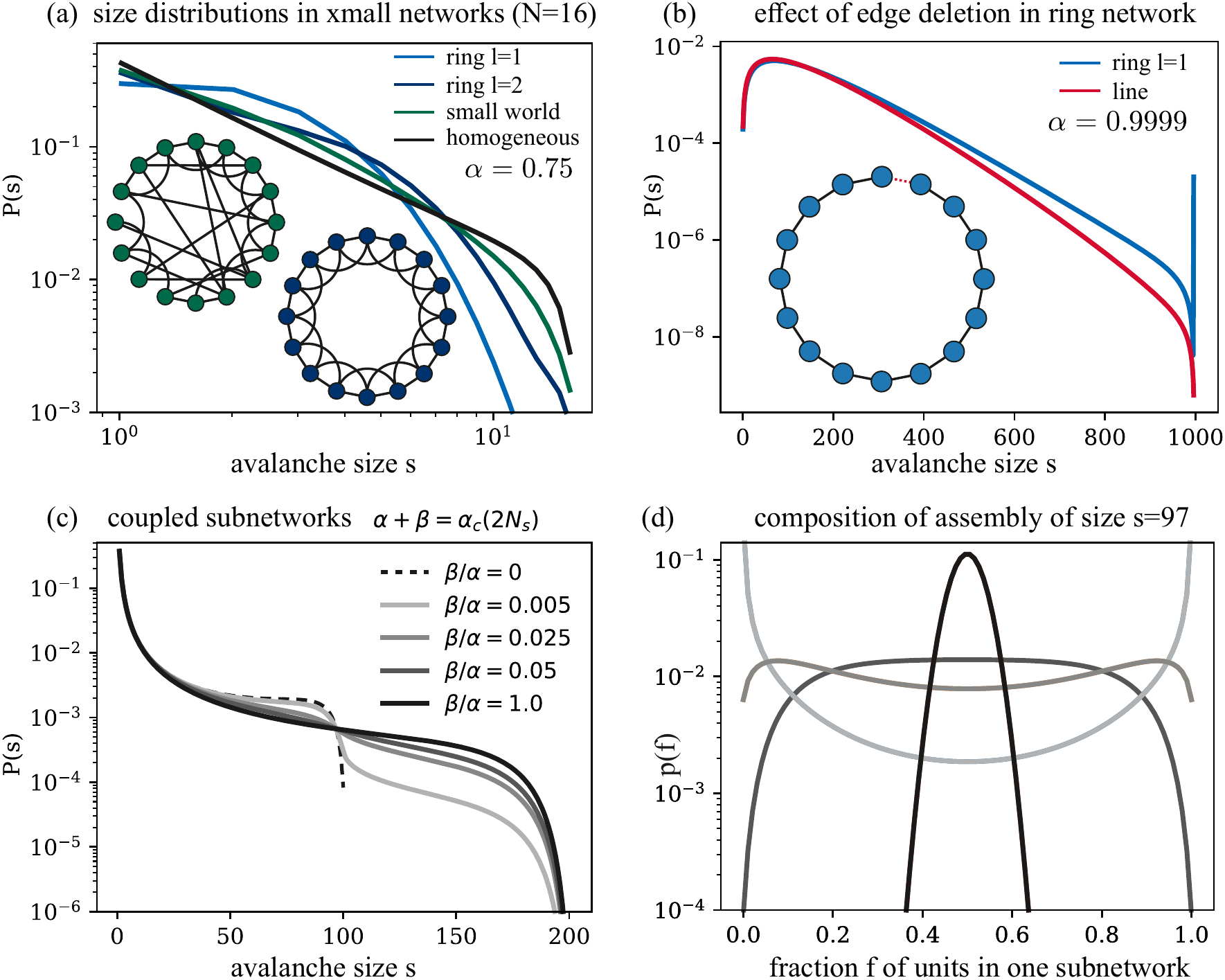}
    \caption{
    Avalanche size statistics $P(s) \coloneqq \P(\s(\av)=s|\s(\av) > 0)$ in small or structurally simply networks.
    (a) Effect of rewiring edges from ring to small world network.
    Analytical avalanche size statistics for the $l$-nearest neighbor coupled ring networks, a homogeneous network and a small world network. Coupling strengths in all networks are normalized so that each row sum of each matrix is $\alpha=0.75$.
    The $l=2$ ring network and the chosen realization of the small world network are shown in the inset. Panel (b) shows the effect of removing a single edge from the $l=1$ ring network (illustrated in the inset, removing the red edge turns ring into line network) with $\alpha=0.9999$ and $N=1000$. Whereas the avalanche size distribution for the ring network is bi-modal with a peak for global avalanches, the distribution of the line model is unimodal and decays exponentially for large avalanches. Panel (c) shows the avalanche size distribution for a network consisting of two homogeneous subnetworks of size $N_s = 100$ for different levels of inhomogeneity $\beta/\alpha$ at a constant row sum $\alpha+\beta = \alpha_c(2N_s)$.
    %For small $\alpha_c$ the avalanche size distributions shows inflexion points at $n=100$.
    Avalanche assemblies of a fixed size differ only in the number of units from each subnetwork. The probability $p(f) \coloneqq P\left(\left|\U(\av)\cap [N_s]\right|/N_s = f\mid\s(\av) = 97\right)$ that an assembly of size $s=97$ consists of a certain fraction $f$ of units in one subnetwork is shown in panel (d).
    \label{fig:applications_avs}}
\end{figure*}
\fi

We start with examples of small networks of size $N=16$, comparing ring networks with $l$-nearest neighbor couplings to a small world network and a homogeneous network. For the $l$-nearest neighbor ring model, each unit is connected to its $l$ neighbors on each side. The homogeneous network is connected all-to-all with equal weights.  \Fig{fig:applications_avs}, panel (a) shows the corresponding avalanche size distributions with the insets illustrating the 2-nearest neighbor ring network and the particular realization of the small world network. To illustrate changes induced by the topology, all coupling matrices were normalized such that the sum of incoming edge weights to each unit is equal to $\alpha=0.75$. The probabilities for large avalanche sizes increase from the more sparsely coupled ring networks, over the small-world network to the homogeneous network. Since the small world network was generated by randomly rewiring edges with a probability of $0.3$ from the $l=2$ ring network to form a Watts-Strogatz graph~\cite{watts1998collective}, this process decreased the mean path length which in turn also facilitated larger avalanches to occur. Note that the avalanche distribution we show is not an average over an ensemble of small-world networks, since our framework allowed to compute it for a \emph{particular} realization of a small-world topology, which is displayed in the inset.

% --- global effect of turing ring into line network by deletion of a single edge

In general, altering just a single edge can have large effects on the global dynamics and associated avalanche distributions. This point is illustrated for the extreme case of deleting a single edge from a strongly coupled one-nearest neighbor ring network with $N=1000$ units in panel (b). Due to the sparsity of this network, assemblies always form connected line segments. This makes if efficient to compute the avalanche size statistics from the assembly probabilities in \Eq{eq:pavu_inline} for line segments leading to the avalanche size distribution in \Eq{eq:avs_ring}. The avalanche size distribution of the ring network with $\alpha=0.9999$ (where $\alpha$ is again the sum of incoming edge weights to each unit) is bi-modal with peaks at around $s=70$ and at the global avalanche size $s=1000$. We obtained the line network by deleting a single (undirected) edge from the ring network and without changing any of the remaining coupling weights. The avalanche size distribution of this network is given by \Eq{eq:avs_line}. In contrast to the ring network, the avalanche size distribution of the line network is uni-modal and decays exponentially for large avalanche sizes. This effect is due to the restriction on the spreading of an avalanche imposed by the missing edge, and can be understood intuitively: If the avalanche starts at one end of the line, it can only spread in one direction and has to complete $N-1$ iterations to become a global avalanche. In contrast, an avalanche can always spread into two directions simultaneously in the ring network, and does have to complete about only half of the number of iterations to activate all units. 

A formal understanding arises from the observation that there are $N$ spanning trees in the ring network of size $N$, and only one in the corresponding line network. Since removing recurrent connections can only increase the volume of the inhabited region, we immediately deduce from \Eq{eq:pavu_inline} and Kirchhoff's matrix tree theorem that the probability of a global avalanche in the ring network of size $N$ is at least $N$ times as high as the probability for the corresponding line network.

% --- Effect of Inhomogeneities in the coupling matrix for two coupled subnetworks 

In addition to network connectivity, weight inhomogeneities affect the avalanche size distribution. To illustrate this, consider two homogeneous subnetworks of size $N$, which are coupled in an all-to-all fashion with intra-network coupling weight $\alpha/N$ and inter-network coupling weight $\beta/N$. Such a network topology is an ubiquitous structure in the brain, where two strongly coupled local populations or areas interact globally via (potentially) weaker connections. The coupling matrix for this network is a $2 \times 2$ block matrix with blocks of size $N\times N$ and values $\alpha/N$ on the diagonal and $\beta/N$ on the off-diagonal. Note that the row sums of this matrix have a value of $\alpha+\beta$. 
Due to its regular structure, each avalanche assembly can be characterized just by the number of participating units from each subnetwork. In addition, the determinants in \Eq{eq:pavu} for this block matrix can be reduced to determinants of $2 \times 2$ matrices. This reduction is detailed in the appendix, section~\ref{sec:hom-coupled}. By calculating these determinants, we find the avalanche size distribution of this network as the expression given in \Eq{eq:avs-two-coupled-hom}.

% Ab hier neues Bild in panel (d)
Let us discuss some implications of varying the inter-network connection strength $\beta$ on the avalanche dynamics and assembly formation. With $\alpha_c(N) \coloneqq 1-1/\sqrt{N}$ we denote the total, critical coupling strength for a homogeneous network of size $N$ for which it exhibits a power-law avalanche size distribution. For $\alpha = \beta = \alpha_c(2N)/2$ we obtain a critical homogeneous network of size $2N$. We now introduce an inhomogeneity into the weight matrix by varying $\beta/\alpha$, while keeping the row sums constant at $\alpha+\beta = \alpha_c(2N)$. \Fig{fig:applications_avs}, panel (c) shows the avalanche size distribution for different values of $\beta/\alpha$. If $\beta=0$, avalanches cannot spread from one subnetwork to the other and the avalanche distribution for the full network with $2N$ units is just the same as for a homogeneous network of size $N$ with a supercritical coupling of $\alpha_c(2N) > \alpha_c(N)$. 
For non-zero, but weak inter-network coupling weights $0<\beta \ll \alpha$, avalanches up to $s=2N$ are possible and the avalanche size distributions show an inflexion point at around $s=N$. However, at still very strong inhomogeneities with $\beta/\alpha = 0.05$, the avalanche size distribution (dark gray line) quickly becomes very similar to the one of the homogeneous network (black line, $\alpha=\beta=\alpha_c(2N))$.

% Jetzt die Assemblies
In contrast to the small differences in the size distributions observed for a wide range of $\beta$ values, weight inhomogeneities have a larger effect on the avalanche assemblies, i.e. how likely an assembly of a given size is composed of a certain fraction of units from a single network. For the avalanche size $s=97$, which is where the distributions shown in panel (c) intersect, panel (d) shows this assembly distribution in dependence of the fraction of units from the first subnetwork.
For $\beta/\alpha = 0.005$ (blue line), the most likely composition of an avalanche of size $s=97$ is that all participating units stem from either the first or from the second subnetwork. Increasing $\beta$ shifts the two peaks of the assembly distribution closer together until the distribution becomes unimodal with a single peak at $0.5$ (green and red lines). The assembly distribution for the homogeneous network $\alpha=\beta$ is simply a hypergeometric distribution (arising from $97$ draws out of a population of $200$ neurons out of which $100$ are from the first subnetwork) since each assembly has the same probability. In contrast, the shape of the assembly distribution for $\beta/\alpha = 0.05$ is much wider and is approximately constant from $0.3$ to $0.7$, indicating a much higher variability of assembly compositions due to the inhomogeneity in the network.

\section{Discussion}

% --- was haben wir nur getan?
In this study we generalized a well-established model class for neural avalanches~\cite{eurich2002finite,leleu2015unambiguous,jung2020avalanche,levina2008mathematical,denker2014ergodicity,denker2016avalanche} to arbitrary network topologies and non-negative connection weights, and performed a thorough analysis of its dynamics. 

% --- zentrale Idee: Torus-Transformation und Ergodizität
Mathematical analysis of this neural model has always been a challenge due to the discontinuities of the avalanche dynamics at spiking threshold. Even though remarkable progress has been made~\cite{denker2014ergodicity,denker2016avalanche}, ergodicity of the homogeneous skew-product system remained a conjecture and formal treatment of non-homogeneous coupling topologies and reduced number of units which receive external input was out of reach. We were able to drastically simplify analysis by exploiting an invariance of the fast-scale avalanche dynamics. Formally, our model reduces to a simple translation dynamics with respect to a topology that turns out to the equivalent to the topology of the $N$-torus for general positive coupling matrices $W$, as long as their eigenvalues stay below one. This torus transformation removes the discontinuities of the avalanche dynamics and is the central idea behind our study. This allowed us to show that for almost all coupling matrices, the Lebesgue measure supported on a subset of the phase space $[0,1)^N$ is the unique ergodic measure relative to the given time-invariant Bernoulli drive if and only if all units can be reached by a path starting from a unit receiving external input in the induced graph. In addition, we studied the geometry of the support for the Lebesgue distribution and uncovered its self-similar structure.

% --- Lawinenverteilungen
% Due to the separation of timescales between external drive and resulting avalanches,
Our framework keeps track of avalanches as sequence of index sets specifying which units fired at which avalanche generation. The ergodic measure, along with the self-similar structure of its support, allowed us to derive avalanche distributions analytically by identifying the regions in phase space which lead to specific avalanches. In addition, we found a closed form for the distribution of units involved in avalanches, the assembly distribution. To our knowledge, this approach provides the first detailed investigation of assembly distributions for a recurrent network model with spiking neurons.

% --- Anwendung des Frameworks
For demonstrating the benefits of our approach, we analyzed a structurally simple example for a non-homogeneous coupling topology. Specifically, we considered a shift-invariant uniform connectivity with limited coupling distance on a two-dimensional lattice with periodic boundary conditions (two-dimensional torus). Our framework captures deviations of the scaling exponent of the mean avalanche size from the corresponding mean field value analytically. Furthermore, we assessed the corresponding scaling exponent of the critical avalanche size distributions numerically. These results illustrate changes in scaling exponents in dependence of the coupling topology from a predominantly local coupling exhibiting coalescence\cite{zierenberg2020description} to an all-to-all homogeneous network, where the mean field exponents are attained. Future studies could use our analytical framework to study changes in scaling exponents by other features of the coupling topology like synaptic density and feedback loops which have been shown to change scaling exponents in neuronal cultures~\cite{yaghoubi2018neuronal}.

\emph{Implications for ensemble codes.}
One main result of our analysis is a closed-form expression of the probability that an ensemble of units fires in short temporal succession in form of the assembly of an avalanche. This form of transient synchronization helps to transmit signals in a fast and reliable manner, since it is more efficient in driving postsynaptic cells than spikes arriving asynchronously~\cite{hahn2019portraits,buzsaki2010neural}. Functionally, assemblies can be used for establishing whole coding schemes, as recently formalized in a computational system called Assembly Calculus~\cite{papadimitriou2020brain}. The generalized EHE model could in this context serve as a physiologically more realistic realization of such a coding scheme.

But most importantly, reoccurring sequences of spike patterns with a particular composition of participating units were indeed observed in experiments \cite{hemberger2019reliable,torre2016synchronous,bellay2021selective,miller2021long,hahnloser2002ultra,long2010support,harvey2012choice,pastalkova2008internally}, indicating a robust formation of assemblies during signal processing. 

Having a formal framework is thus essential for interpreting such data, and for understanding how coding with synchronous neural ensembles is enabled by external input and constrained by network connectivity. 

\emph{Relation to graph theory.} 
One major insight from our analysis is the existence of close links between assembly formation and graph theoretical properties of the synaptic connections, which we will discuss in the following.

We found that the adjacency matrix of an assembly subnetwork determines assembly probabilities as a function of the eigenvalues of the corresponding \emph{graph Laplacian}. In particular, the probability that external input to unit $k$ starts an avalanche encompassing a given assembly is proportional to the $(k,k)$-cofactor of the graph Laplacian. Via the well-known \emph{Matrix Tree Theorem} (also known as Kirchhoff's Theorem, which generalizes \emph{Cayley's Theorem} to weighted digraphs) this property is related to the graph theoretical concept of \emph{spanning trees}, making the assembly probability proportional to the (weighted) number of spanning trees for the assembly network. The spanning trees themselves are directly related to the different pathways individual avalanches can spread through the assembly network.

With respect to network function, the weighted number of spanning trees in a graph can be seen as a measure of robustness. Let us consider the elementary setting of a simple graph with a fixed number of units and edges. If every edge can fail independently with a given probability, a uniformly most reliable graph has to maximize the number of spanning trees, i.\,e.\  be $\mathcal{\tau}$-optimal~\cite{rela2019uniformly}. In this sense we could extend the Hebbian principle from pairs of neurons to assemblies: What \emph{robustly} wires together, fires together. Thus our framework provides an explicit objective function for reliable and robust \emph{assembly formation}. In addition, we analytically determined the impact of single edge failure and the gain of formation of a new edge on assembly probability. For unidirected graphs, the resulting measure turns out to be equivalent to the well-known \emph{resistance distance} \cite{klein1993resistance,bapat2003simple}.

The consequence of these mathematical results for brain function is a general prediction that the Laplacian spectrum would relate more directly to the occurrence of collective synchronous events than the adjacency matrix of the underlying anatomical network. In other words, the strength of a direct connection between neural populations is less indicative for the magnitude of their effective interaction than the sum of all direct and indirect (weighted) pathways between those two units. Interestingly, it was demonstrated just recently that functional brain connectivity is indeed best predicted from the Laplacian of the structural connectivity which was extracted from diffusion tensor imaging data~\cite{abdelnour2018functional}.

Furthermore, equations for equilibrium rates and their covariances~\Eq{eq:equilib} are consistent to the corresponding results for Hawkes processes~\cite{pernice2011structure}, i.\,e.\  (linearly) coupled Poisson processes. For these processes, structure-function relationships have been studied in detail \cite{pernice2011structure,jovanovic2016interplay,hu2018feedback}. Corresponding results, such as which graph motifs most strongly influence equilibrium rates, translate directly to our model.

\emph{Relation to branching and percolation processes.}
Neural avalanches are often studied in simplified models with discrete states and a dynamics defined as a branching process on a graph ~\cite{larremore2011effects,larremore2014critical,kinouchi2006optimal}, thus offering opportunities for a rigorous analytical treatment. In comparison, the dynamics of the EHE model is far more complex. Although we demonstrated that an equivalent branching process for the EHE model can in principle be defined, stochastic dependencies between the membrane potentials of units belonging to the same strongly connected component of the network makes its formal description complicated. On directed acyclic graphs these dependencies disappear, and (only) there do the edge weights represent branching probabilities. In particular, the probability that a unit becomes active will be proportional to the sum of incoming edge weights from currently active units.

In addition, we provided a direct relationship to percolation processes by showing that avalanches of the EHE-model for the particular choice of the (1+1)D lattice propagate equivalent to the \emph{directed compact percolation} process which belongs to the universality class of branching-annihilating random walks~\cite{hinrichsen2000non}.

\emph{Model generalization.}
Models are constructed for capturing the generic behaviour of a real system, while being ideally as simple as possible to allow for a comprehensive understanding and analysis of the underlying mechanisms. In this sense we believe that our formal framework provides a major advance over previous work. It is still sufficiently simple for a rigorous analysis, but allows studying assembly formation and avalanche dynamics in arbitrary, inhomogeneous networks. This is the \emph{generic} case for neural systems in the brain, and assuming homogeneity in these situations will lead to misleading results or apply only to small subsystems for which this condition is approximately fulfilled.

For making analytical treatment possible, the extended EHE model retains some simplifying assumptions from the original framework \cite{eurich2002finite}: it does not have leak conductances, there is no "hard" reset after a spike, and it assumes a separation of time scales. In the following, we will consider implications of lifting these assumptions on the mathematical treatment, and discuss how our results can be expected to generalize to physiologically more realistic neural units and networks.

\textbf{a) Separation of time scales.} In order to unambiguously identify the detailed progression of avalanches we assume a separation of time scales in this model, which means that external input only occurs after an ongoing avalanche has terminated. This is a common assumption in avalanche models~\cite{eurich2002finite,kinouchi2006optimal,larremore2014critical}. A weakening of this assumption would allow several avalanches to coexist and to merge. It is known from field-theoretical treatment that allowing external drive during avalanches leads to changes in the scaling relation like for example the avalanche size~\cite{di2017simple}. This phenomenon was recently studied in detail~\cite{das2019critical} in a model very similar to our framework with the result that size distribution exponents in the critical state decreased with increasing relaxation of the time scale separation.

\textbf{b) Spike reset and refractory period.} In the EHE-model resetting a unit's state $u$ after emission of a spike is done by simply subtracting the firing threshold. If instead the units were reset to zero as in other integrate-and-fire models, the spike's impact on the progression of $u$ would no longer be linear. In consequence, the colored boundaries in state space (\Fig{fig:trafoillus}) would still act as portals, but with an additional absorbing condition. This condition would ensure that the state remains `glued' to the resting state $u=0$ after transitioning through the boundary, thus effectively dissipating the excess synaptic input delivered in the current generation of an avalanche. It would still be possible to study the system on the torus, however, with the penalty of having a discontinuous dynamics at the boundaries. A `hard' reset would also induce additional state correlations which would need to be countermanded by additional randomness, as e.\,g.\ a stochastically varying drive $\DeltaU$, for obtaining a smooth invariant measure which potentially can be treated analytically.

Interestingly, by simultaneously lifting time scale separation \emph{and} introducing a 'hard' spike reset, the state dynamics will again become closer to the EHE system. Since an avalanche will now be spread over several milliseconds, it is likely that a smaller part of its total synaptic input will arrive when the neuron is just spiking and insensitive to those inputs. In consequence, a smaller fraction of recurrent feedback would be lost.

\textbf{c) Leak conductances.}
In real neurons leak conductances make the membrane potential decay towards its resting value. Introducing leaks in the EHE model would thus lead to a non-homogeneous invariant measure which increases towards the resting potential. In addition, state trajectories would be able to enter the formerly \emph{non-inhabited} region and hence violate validity of the torus transformation. However, the resulting effects will be sufficiently small if assuming a strong external drive in comparison to a weak intrinsic leak~\cite{levina2008mathematical} , such that we can expect our main result in~\Eq{eq:pavu_inline} to still hold approximately.
% Further problems would occur if time scale separation is lifted. Leaks will then also affect the states of units in between generations in an avalanche. In that case, the order in which the avalanche progresses through the network would influence its impact on units not participating in it. Formally, a network of leaky integrate and fire units could be analysed as a Markov process. However, finding closed-form expressions for ergodic invariant measures would potentially be very challenging.

\textbf{d) Inhibitory units.}
Similar to the influence of leaks, the inclusion of inhibitory units removes the strictly non-inhabited region in state space and the independence assumption underlying~\Eq{eq:pavu_inline}. Inhibition can easily lead to violations of ergodicity. One example is a network of two populations with strong intrinsic excitatory connectivity which are mutually coupled by inhibition. This ubiquitous connection motif could establish a winner-take-all network, in which one of the populations engages in strongly reverberating activity which completely inhibits activation of any unit in the other population. Clearly, extending our framework to networks with inhibition poses the biggest challenge for future studies. However, we believe that in situations with not too strong inhibition on a global scale, reasonable approximations can be made. This could be the case e.g. in normalization schemes where the excitatory drive is on average a little higher than the inhibitory suppression.

\emph{Perspectives.}
The main contribution of our study is the extension of an analytical framework for assembly formation in recurrent networks from homogeneous couplings to networks with arbitrary (positive) connectivity, now allowing rigorous treatment of avalanche dynamics in a much larger class of systems than in previous studies. For future research, a logical next step would be to investigate temporal aspects such as the statistics of avalanche duration, correlations between subsequent avalanches, and inter-event statistics and its relation to known brain rhythms such as gamma oscillations \cite{miller2019scale}. We think that for these aspects, analytical treatment is within reach and could nicely complement our results on assembly formation.

On a more general level, we believe the novel framework introduced here might support a paradigm shift in research on neural criticality. In highly inhomogeneous systems subject to a substantial and structured external drive, we can not expect to observe the 'usual' signatures of criticality even if the system is at the brink of some phase transition or at an optimal point for information processing. However, such a situation is actually the rule, and not the exception when investigating active processing in the brain. Being able to handle these more general situations is the advantage of our theory. In consequence, power laws and criticality played only a secondary role in our study, while instead we focused on detailed assembly formation. Combined with structured inputs from 'meaningful' external stimuli~\cite{tomen2019role} we expect our tools in future studies to provide new insights into how avalanche formation and -- potentially -- criticality serve information processing and brain function.

\begin{acknowledgments}
This work was supported by the DFG priority programs SPP 1665 (ER 324/3-2) and SPP 2205 (ER 324/5-1).
We thank Federica Capparelli and Nergis Tömen for insightful discussions at the initial stage of this project.
\end{acknowledgments}

%
% LEFTOVERS
%
% - cite extensions of the model - Levina and Herrmann \protect\cite{levina2007criticality}?\\
\raggedbottom
    
\pagebreak
\section{Appendix}

This appendix is organized as follows: The order of the sections in this part is the same as the order in which the topics are treated in the main text. While the main text focuses on the most important results and related intuitions, the appendix provides the corresponding rigorous mathematical treatment and technical details. Although the appendix itself is structured to be self-contained, we advise the dedicated reader to go through the corresponding sections in the main text and the appendix in conjunction.

For convenience of the readers, we first repeat the basic definitions of the generalized EHE model. In section A we then state some general properties of the model and its avalanche dynamics $\mathcal{F}$ which are important for all subsequent sections.

In section B we show that dynamics of the model is homeomorphic to a simple translation on the $N$-dimensional torus $\TN$ which greatly facilitates any formal treatment, allowing to determine under which exact conditions the system is ergodic (section C), and permitting to compute expected firing rates and spike count covariances explicitly (section D).

Section E develops a description of the self-similar structure of the 'inhabited region' $D$ in the model's state space. This description and the notion of ergodicity (section C) is a prerequisite for calculating avalanche probabilities in section F, which is followed by section G detailing the relation of the obtained equations to graph theoretical terms.

The final section H exemplifies how this mathematical framework can be used to derive avalanche size statistics for various networks with different topology and regular structure.

% , we state some general properties of the model such as that each unit can fire at most once in an avalanche and properties of the avalanche dynamics $\mathcal{F}$ in section~\ref{sec:appA}. In section~\ref{sec:ergodicity} we show that the model is homeomorphic to a simple translation dynamics on $\TN$. Section~\ref{sec:reluniqeergo} develops ergodic theorems for generalized translation dynamics on $\TN$ including exact conditions for which the Lebesgue measure is the unique ergodic measure relative to an ergodic measure for the external input dynamics.  
% Expected firing rates and spike count covariances are obtained directly from the equivalent system on $\TN$ in section~\ref{sec:app-firing-covariances}, while the self-similar structure of the inhabited region $D$, which is developed in section~\ref{sec:app-geomerty}, is needed to calculate the avalanche probabilities in section~\ref{sec:app-avalanche}. Section\ref{sec:app-topology} details the relation of the obtained equations to graph theoretical terms and section~\ref{sec:app-simple} illustrates how this mathematical framework can be used to derive avalanche size statistics in some simple networks with regular structure. 

\emph{Notation and definition summary:} We start by briefly summarizing the model and notation of its dynamics \Eq{def:f}--\Eq{eq:def-Tk} in the remainder of this section:

\begin{alignat}{2}
	T_k&: C \to C, && \Vu \mapsto \fixF(\Vu+\DeltaU \e_k) \text{,} \\
	\text{with } \fixF&:\mathbb{R}_{\geq 0}^{[N]} \to C, && \Vu\mapsto  F^{\tau(\Vu)}(\Vu) \\
	\text{and } F&: \mathbb{R}_{\geq 0}^{[N]} \to
		\mathbb{R}_{\geq 0}^{[N]},\quad && \Vu\mapsto \Vu - (\diag(\VU)-W)A(\Vu) \\
  \text{and } \tau&:\mathbb{R}_{\geq 0}^{[N]} \to
		\mathbb{N}_0, && \Vu\mapsto \min\left\{n\in \mathbb{N}_{0}\mid F^n(\Vu) \in C \right \} \, .
\end{alignat}
While $F$ describes one generation of an avalanche, $\fixF$ subsumes an entire avalanche with $\tau$ being its duration. Using these definitions, $T_k$ describes one iteration of the model upon receiving external input to unit $k$ (which might or might not trigger an avalanche). 

The connection weights $w_{ij}$ from matrix $W$ are subject to the constraints 
\begin{eqnarray}
	\DeltaU + \sum_{j=1}^N w_{ij} & < & U_i \quad \mbox{for all} \quad i\in [N]
	\quad ,
	\label{eq:constrw}
\end{eqnarray}
ensuring that each unit can fire at most once during an avalanche (see Proposition~\ref{prop:at-most-once}).
It also ensures the existence of 
\begin{align}
  M \coloneqq (\diag(\VU) - W)^{-1} \, .
\end{align}

The avalanche function $\av$ is defined as 
\begin{align}
  \av(k, \Vu) &\coloneqq \left(\left\{j \in [N] \mid 
		\left(F^{i-1}(\Vu+\e_k \DeltaU)\right)_j \geq U_j\right\}\right)_{i=1,\ldots, \tau(\Vu+\DeltaU\e_k)} \in \mathcal{A}\, .
\end{align}
where $\mathcal{A}$ is the set of all avalanches (see Definition~\ref{def:A}). $\Va=()$ denotes the empty avalanche. The length of the sequence $\Va$ will be denoted by $\dur(\Va)$ and called the {\em duration} of the avalanche. We call the union $\U_{j}$ of the generations 
\begin{align}
  \U_{j}(\Va) &\coloneqq \biguplus_{i=1}^{j} a_i \text{, }
		1\leq j\leq \dur(\Va) \mbox{ and }
		\;\U(\Va)\coloneqq\U_{\dur(\Va)}(\Va)
\end{align}
the \emph{avalanche assembly} (up to generation $j$) and the sum of cardinalities
\begin{align}
	\s(\Va) &\coloneqq \sum_{i=1}^{\dur(\Va)} |a_i| =\rvert\U(\Va)\lvert
\end{align}
its {\em size}.

% ========================================================================================================================

\subsection{General properties of the model}\label{sec:appA}

In this section we introduce some common notation and general properties of the model which are  used throughout the appendix. We start by showing that the model is well defined, i.e. that the avalanche duration $\tau(u+\DeltaU \e_k) <\infty $ for all \(u\in C,k\in [N] \). In fact, as long as~\Eq{eq:constrw} holds for the coupling matrix \(W\), each unit can fire at most once during an avalanche. Thus, unions of different generations $a_i$ of an avalanche $a=\av(k,u)$ are disjoint i.e. \(\U(a) = \biguplus_{i=1}^{\dur(a)} a_i\).

\begin{lem}
  \label{prop:at-most-once}
  Assuming \Eq{eq:constrw}, for \(u\in C\), \(k\in [N]\),  we have that 
  each unit can fire at most once during an avalanche and in particular,
	its duration \(\tau(u+\e_k \DeltaU) \leq N\). 
\end{lem}

\begin{proof}
  We will give a proof by contradiction. Let \(u\in C,k\in [N]\) be
  arbitrary and set \(a=\av(k,u)\).
	Let unit $j$ be part of generations $r$ and $s$,
	\(j\in a_{r},j\in a_{s} \) with \(1\leq r < s \leq \dur(a)\)
	such that the components of \(a_{1,\ldots,s-1} \) are pairwise disjoint,
	i.e., no unit has fired twice in the generations up to $s-1$, and unit $j$ would
	fire a second time in generation $s$. It follows that 
	\begin{align*}
		\left(F^{s-1}(u +  \DeltaU\e_k)\right)_j &= \bigg(u+ \DeltaU\e_k - (\diag(\VU) - W)
		\sum_{l=1}^{s}A\left(F^{l-1}(u+\DeltaU\e_k )\right)\bigg)_j  \\
		&
		\leq  \DeltaU  + \sum_{l=1}^N w_{jl}  < \VU_j \text{,}
	\end{align*} 
	which contradicts \(A_j(F^{s-1}(u+\DeltaU \e_k)) = 1\).
	It follows that the index sets are pairwise disjoint and	
	$\U(a) = \biguplus_{i=1}^{\dur(a)} a_i$.
	% \(\U(\av(k,u) ) = \biguplus_{i=1}^{\red{\dur}(\av(k,u))} \av(k,u)_i\). 
	
\end{proof}

Lemma~\ref{prop:at-most-once} allows us to write \(T_k\) in a more compact form using $M^{-1}=\diag(\VU)-W$:
\begin{align}
	\label{cor:total-action}
	T_k(u) = u + \DeltaU \e_k -
		M^{-1}\delta (\U(\av(k,u)))\;\;\;  (u\in C,\,k\in[N]).
\end{align}

Thus, \(\fixF\) projects from \(\R^{[N]}\) back to \(C\) by subtracting integer combinations of columns of \(M^{-1}\). In the next lemma, we introduce some properties of \(\fixF\):

\begin{lem}\label{lem:fixF-minimality}
	For \(u,v\in \R^{[N]}_{\geq 0}\) we have the following properties for $\fixF$:
  \begin{enumerate}[(1)]
  \item
		\(\fixF(u) = u - M^{-1}n\) for some \(n\in \N^{[N]}\)
		if and only if \(u - M^{-1}n \in C\)
		and for every \(n'\lneq n\) (component wise) we have \(u - M^{-1}n' \notin C\),
  \item
		\(\fixF(\fixF(u) + v) = \fixF(u+v)\)
  \item
		\(\fixF\left(W\1  + M^{-1}\R^{[N]}_{\geq 0}\right) =
		\fixF\left(W\1  + M^{-1}[0,1)^{[N]}\right) \), where  $\1  = \sum_{k=1}^N \e_k$. 
	\end{enumerate}
\end{lem}

\begin{proof}
	\begin{enumerate}[(1)]
	\item
		We show the `if' direction by contradiction:\\
		Let \(n'\lneq n\) be such that \(\fixF(u) = u - M^{-1}n\) and
		\(u - M^{-1}n' \in C\). Hence, \(\fixF(u)+M^{-1}(n-n') = u - M^{-1}n' \in C\).
		After \(\ell \leq \tau(u)\) iterations we have
		\(F^{\ell}(u) = u-M^{-1}n^{(\ell)}\) with
		\(n^{(\ell)} \coloneqq \sum_{k=0}^{\ell-1}A\left(F^k(u)\right)\). 
		Since \(n'\lneq n\) there exists an iteration \(t\in \N \) and index
		\(j\in [N]\) such that \(n^{(t)} \leq n'\) but \(n^{(t+1)}_j > n'_j\).
		Since \(n'_j = n^{(t)}_j\) and, for $I\subseteq [N]$, we have
		\(\left(M^{-1}\delta(I)\right)_i > 0\) if and only if \(i\in I\), it follows
    \[\left(u-M^{-1}n'\right)_j \geq \left(u-M^{-1}n^{(t)}\right)_j = F^t(u) \geq \VU_j \text{. }\]
		This contradicts \(\fixF(u)-M^{-1}n' \in C\).\\
		To show the `only if' direction in (1), let us assume that
		\(x= u-M^{-1}n\in C\) and \(u - M^{-1}n' \notin C\) for all \(n'\lneq n\).
		%. Consider \(\fixF(u)\). Since
		Then the stopping condition for the fixed point iteration defining
		\(\fixF(u)\) is fulfilled for the first time at \(x\), thus \(x = \fixF(u)\).
	\item
		Note that \(u = \fixF(\fixF(v) + w) = v + w - M^{-1}(n_1 + n_2) \) for some
		\(n_1,n_2 \in \N^{[N]}\) and \(u+M^{-1}n' \notin C\) for all
		\(0\neq n' \leq n_1 + n_2\). Thus with (1), we have \(u = \fixF(v + w)\).
	\item
		We fix \(z\in [0,1)^{[N]}\). Then we deduce from (1) that
		\(\fixF\left(W\1 +M^{-1}z\right) = W\1 +M^{-1}(z-n)\) and \(W\1 +M^{-1}(z-n+n')\notin C\)
		for all \(0\neq n'\leq n\). Now, every \(x\in \R^{[N]}\) can be decomposed into
		\(x = \lfloor x \rfloor + z\) with \(z = x - \lfloor x \rfloor\in [0,1)^{[N]}\). 
		Since \(\left(W\1  + M^{-1}\left(\R^{[N]}_{\geq 0} \setminus [0,1)^{[N]}\right)\right) \cap C =
		\emptyset \) we find \(\fixF\left(W\1  + M^{-1}z\right) + M^{-1}\tilde{n} \notin C\)
		for all \(0\neq \tilde{n} \leq n+\lfloor x \rfloor\), which implies
		\(\fixF\left(W\1 +M^{-1}z\right) = \fixF\left(W\1 +M^{-1}x\right) \).
	\end{enumerate}
\end{proof}

We will generalize \Eq{cor:total-action} to multiple steps $s$ of applying \(T\) in the following Corollary. There we introduce the \emph{spike count vector} $\NN^{s}(\omegaVk,u)$ which collects how often each unit fired (i.e., participated in an avalanche) during \(T^s(\omegaVk,u)\). This quantity will later be used to determine the equilibrium firing rates and spike count covariances of the model in dependence of the interaction matrix \(W \).             

\begin{cor}\label{cor:equivalence}
  Let \(\NN^{s}(\omegaVk,u)\) be the \emph{spike count vector} after $s$
	applications of \(T\) starting at \((\omegaVk,u) \) defined as
  \begin{align}\label{eq:ns}
    \NN^{s}(\omegaVk,u) \coloneqq \sum_{t=0}^{s-1} \delta\left(\U\left(\av\left(T^t(\omegaVk,u)\right)\right)\right)\text{ .}
  \end{align}
	For $k\in[N]$ we have for $s\in\N_{0},\omegaVk \in \Sigma_N,u\in C$,
  \begin{align}\label{eq:col21eq}
    \pi_2 T^s(\omegaVk ,u) = u+\DeltaU \sum_{t=1}^s\e_{\omegaVk_t} - M^{-1} \NN^{s}(\omegaVk,u) = \fixF\left(u+\DeltaU \sum_{t=1}^s\e_{\omegaVk_t}\right) \text{,}
  \end{align}
  where \(\pi_2 \) denotes the projection onto the second component (here \(C \)).
\end{cor}

\begin{proof}
  The claim follows by induction. The starting case \(s=1\) is provided by
	\Eq{cor:total-action}. Now suppose \Eq{eq:col21eq} holds for \(s-1\).
	Then we have
  \begin{align*}
    \pi_2 T^s(\omegaVk ,u) &= \fixF\left(u+\DeltaU \e_{\omegaVk_s}+
		\DeltaU \sum_{t=1}^{s-1}\e_{\omegaVk_t} - M^{-1} \NN^{s-1}(\omegaVk,u)\right) \\
			&= u + \DeltaU \sum_{t=1}^{s}\e_{\omegaVk_t} - M^{-1}\left(\NN^{s-1}(\omegaVk,u) +
				\delta\left(\U\left(\av\left(T^{s-1}(\omegaVk,u)\right)\right)\right)\right) \\
			&= u + \DeltaU \sum_{t=1}^{s}\e_{\omegaVk_t} - M^{-1}\NN^{s}(\omegaVk,u) =
				\fixF\left(u+\DeltaU \sum_{t=1}^se_{\omegaVk_t}\right) 
  \end{align*}
\end{proof}

The following definitions will allow us to link properties of the avalanche dynamics
to the coupling structure contained in the weight matrix, and aid us in assessing
ergodicity of the system.

\begin{defn}\label{def:gw}
  \begin{enumerate}[(1)]
  \item
		For a given coupling matrix \(W\in \R^{[N \times N]}\) we define the directed graph
		\(G(W)\) with vertices given by the units \([N]\) and edge set \(E=E(W)\), %induced by the coupling matrix \(W\) 
		\begin{align}
			G(W) \coloneqq ([N],E(W))\text{ where } E(W) \coloneqq
			\{(j,i) \in [N]\times [N] \mid  w_{ij} > 0\}.
		\end{align}
		This graph is naturally weighted through $W$ by assigning $(j,i)\mapsto w_{ij}$.
	\item
		Further, we define the set of all coupling matrices \(W'\)
		with the same sparsity pattern as \(W\) to be
		\begin{align}
			\mathcal{W}(W)\coloneqq \left\{W'=(w'_{ij})\in \R^{[N\times N]} \mid W' \mbox{ respects \Eq{ass} and }  w_{ij} = 0 \iff w'_{ij}=0 \right\}.
		\end{align}
	\item
		For a probability vector \(p\in [0,1]^{[N]}\) we call the coupling matrix
		\(W\) (or equivalently the associated graph $G(W)$ or the set
		$\mathcal{W}(W)$) \emph{$p$-reachable}, if and only if for every unit in
		\(k\in [N]\) there exists a driven unit  
		\(\ell\in L\coloneqq \{\ell \in [N]  \mid  p(\ell) > 0\}\) and a
		path (which can also be the empty path) starting in \(\ell\) along edges in \(E(W)\) terminating in \(k\).
	\end{enumerate}
\end{defn}

\begin{lem}
	The coupling matrix \(W\) is \emph{$p$-reachable}
	if and only if \(M\delta(L) > 0\) (component wise).
\end{lem}

\begin{proof}
 % To show that $p$-reachability is equivalent to \(M\delta(L) > 0\) 
 First note that
  \(M^{-1} = \diag(U)-W\) and by \Eq{ass} we have \(\lVert W \rVert < \max(U)\).
	Thus we obtain with a Neumann series expansion
  \[
		(\diag(U)^{-1}M^{-1})^{-1} = (\mathds{1}-\diag(U)^{-1}W)^{-1} =
		\sum_{k=0}^{\infty}\diag(U)^{-k}W^k \, ,
	\]
	implying that all entries of \(M\) are non-negative. Since the entry
	\((W^k)_{i,j}\) is the sum of products of edge weights along all paths of length \(k\)
	in $G(W)$ from unit $j$ to unit $i$, we conclude that \((M\delta(L))_i > 0 \)
	if and only if there exists a path (which can also be the empty path)
	in \(G(W)\) from some unit in \(L\) to the unit \(i\).

  % In the same way, \((M\delta(L))_{i} >  \delta(L)=(\diag(U)^0W^0)\delta(L)  \) 
	% if and only if \((\sum_{k=1}^{\infty}\diag(U)^{-k}W^{k}\delta(L))_{i} > 0\),
	% i.\,e.\ there exists a non-empty path from a unit in \(L\) to \(i\).\udo{Okay!}
\end{proof}

% ========================================================================================================================

\subsection{Equivalence to a simple translation dynamics on the $N$-torus\label{sec:ergodicity}}

The non-smooth dynamics of spike propagation and membrane potential reset represented by \(\fixF\) complicates mathematical analysis of the model. However, \Eq{cor:total-action} shows that the whole effect of the internal dynamics is summarized by a shift along integer coordinates of the columns of \(M^{-1}\). We use this central observation to significantly simplify our dynamical system by restricting its phase space to the \emph{inhabited region} \(D\) which we set to the image of the $N$-torus \(\TN\) under the quotient map \(\theta\) (see \Fig{fig:trafoillus}, $D$ is the image of the unit cell marked with a dashed gray outline under \(\fixF\)). On \(D\), each iteration step \(T_k\) is a bijection and is conjugated via the mapping \(\theta\) to the shift \(z\mapsto z+\DeltaU M\e_k\) on \(\TN\). This is formalized in Theorem~\ref{thm:D} which establishes topological equivalence between the complex dynamics \(T\) and a much simpler translation \(\hat{T}\) on the $N$-Torus.

\begin{defn}
	We define the skew-product dynamical system \(\hat{T}\) on the $N$-Torus 
	\(\TN %mathbb{T}^N = \R^{[N]} /  \mathbb{Z}^N
	\)  by
  \begin{align}\label{eq:def-hatt}
		\hat{T}: \Sigma_N \times \TN  \to
		\Sigma_N \times \TN,\; \hat{T}(\omegaVk, z) \coloneqq
		\left ( \sigma(\omegaVk), z + \DeltaU M\e_{\omegaVk_1} \right ) \text{ ,}
  \end{align}
	with $\sigma$ being the left-shift operator and $\omegaVk_1$ designating
	the first entry of the external input sequence $\omegaVk$.
\end{defn}

% \begin{defn}
%   For \(a,b \in \mathbb{R}^N\) we define the equivalence relation
%   \begin{align}\label{eq:def-simw}
%     a \sim_M b :\iff a-b \in M^{-1}( \mathbb{Z}^{[N]}), % \text{ for some } z\in  \text{ .} 
%   \end{align}
%   which identifies points which are shifted by integer multiplies of columns of \(M^{-1} = (\diag(\VU)-W)\).
% \end{defn}
% %\end{defn}
% Note that \(\fixF\) preserves the equivalence class with respect to \(\sim_M\), i.\,e.\ \(\fixF(u)\sim_M u \), leading to the following theorem.

\begin{thm}\label{thm:D}
	We define the \emph{inhabited region} \(D\) by 
	\begin{align}\label{def:d}
		D\coloneqq \theta\left(\TN\right) \text{, where $  \theta: z\mapsto \fixF \left(W\1 +M^{-1}z\right)$},
	\end{align}
	where %$\1  = \sum_{k=1}^N \e_k$ 
	we equip \(D\) with the quotient topology induced by the quotient map
	\(\theta\) in which way $\theta$ becomes a homeomorphism. For every
	\(k\in [N]\), the map \(T_k\) is a bijection from \(D\) to \(D\) and
	\begin{align}\label{eq:conjugacy}
		T_k \circ \theta(z) = \theta(z + \DeltaU M \e_k)
	\end{align}
\end{thm}  

\begin{proof}
  Since \(\theta\) is a surjective map from \(\TN\) to D, we can
  inherit the \(\TN\) topology to the set \(D\) as the quotient
  topology induced by \(\theta\), i.e. the open sets on \(D\) are the
  images of open sets on \(\TN \) under \(\theta\). In addition,
  \(\theta\) is injective since \(\fixF(u)\) translates only by
  integer coordinates of \(M^{-1}\) thus no two \(z_1,z_2 \in \TN\)
  can be mapped to the same point by \(\theta\). This makes \(\theta\)
  a homeomorphism from \(\TN\) to \(D\).

  Since \(\theta\) and \(z\mapsto z + \DeltaU M \e_k \) are bijections,
	we only have to verify \Eq{eq:conjugacy} which follows with
	Lemma~\ref{lem:fixF-minimality} as the following calculation shows:
  \begin{align*}
		T_k \circ \theta(z) &= \fixF\left( \DeltaU \e_k + \fixF\left(W\1  + M^{-1}z\right)\right)\\ 
		&= \fixF\left(W\1  + M^{-1}(z + \DeltaU M\e_k)\right) = \theta(z+\DeltaU M\e_k) \text{ .}
	\end{align*}
\end{proof}

In the following, we will thus restrict \(T\) to \(\Sigma_N \times D\). The name inhabited region for $D$ stems from the fact that for all starting points \(u\in C\) and almost all input sequences \(\omegaVk \in \Sigma_N\), the iterated dynamics will eventually map to $D$, i.~e.\ \(T^n(\omegaVk,u) \in D\), for all \(n\in \N\) large enough. We show this in the following proposition:

\begin{prop}\label{thm:d-inhabited}
  Let the graph \(G(W)\) be $p$-reachable. Then for \(\mathbb{B}_p\)-almost all
	input sequences \(\omegaVk \in \Sigma_N\) there exists an iteration number
	\(n_{0}\in \N\) such that \(\pi_2 T^n(\omegaVk,C) = D\) for all $n\geq n_{0}$.
  % \ms{Alternative: Let \(G(W)\) be $L$-reachable.
  % Then for all $p'\in \R_{\geq 0}^{[N]},\sum_{i=1}^Np_i = 1$ with
	% $\{\ell\in [N] |p'_{\ell}>0\} = L$ and \(\mathbb{B}_{p'}\)-almost all
	% \(\omegaVk \in \Sigma_N\) there exists
  % an \(n_{0}\in \N\) such that \(\pi_2 T^n(\omegaVk,C) = D\) for all $n\geq n_{0}$.}
\end{prop}

\begin{proof}
  From Corollary~\ref{cor:equivalence} we have with \(c(u) \coloneqq M(u-W\1 )\)
  \[
		\pi_2 T^n(\omegaVk,u) = \fixF\left(u+\DeltaU \sum_{t=1}^n \e_{\omegaVk_t}\right) =
			\fixF\left(W\1  + M^{-1}\left(c(u) +
			\DeltaU M \sum_{t=1}^n \e_{\omegaVk_t}\right)\right) \text{ .}
	\]
  Let \(c_{\text{min}} = \inf_{u\in C}c(u) \in \R^{[N]}_{\geq 0}\).
	Since \(\lim_{n\to \infty}\left(\sum_{t=1}^n\e_{\omegaVk_t}\right)_i = \infty\)
	for all \(i\) with \(p_{i} > 0\) for almost all \(\omegaVk \in \sigma_N\)
	we have \(\lim_{n\to \infty}\left(M\sum_{t=1}^n\e_{\omegaVk_t}\right)_i = \infty\).
	Thus there exists an \(n\in N\) such that
	\(c_{\text{min}}+ \DeltaU\left(M \sum_{t=1}^n \e_{\omegaVk_t}\right)_k \geq 0\) for all
	\(k\in [N] \) which implies \(\pi_2 T^n(\omegaVk,C) \subseteq D\) with
	Lemma~\ref{lem:fixF-minimality}, part (3).
  Since each \(T_k\) is a bijection on \(D \subseteq C\) we also have
	\(D \subseteq \pi_2 T^n(\omegaVk,C)\).
	
  % Since
  % \((M \sum_{t=1}^n \e_{\omegaVk_t})_i \to \infty\) for \(n\to \infty \)
  % there exists an \(n_0\in \N \) such that
  % \((c + \DeltaU M \sum_{t=1}^{n_0} \e_{\omegaVk_t}) \geq 1\)
  % componentwise and thus with Lemma~\ref{lem:fixF-minimality} (3) we
  % have
  % \[\pi_2 T^{n_0}(\omegaVk,u) = \fixF(W\1  + M^{-1}(c + \DeltaU M \sum_{t=1}^{n_0} \e_{\omegaVk_t})) \in D\]
\end{proof}

Now we show that on this inhabited region, the system \(T\) is \emph{topologically conjugated} to the system \(\hat{T}\) which has the whole \(N\)-Torus as its phase space. 

\begin{thm} \label{thm:equivalence}
	The dynamical systems \(T\) on \(\Sigma_N \times D\) and
  \(\hat{T} \) on $\Sigma_N \times \TN$ are topologically conjugated
	via the homeomorphism $\phi\coloneqq \operatorname{Id}\times \theta$, i.e.\
  $ T \circ \phi =\phi \circ \hat{T}$.
\end{thm}

\begin{proof}
  Using Theorem~\ref{thm:D}, \(\theta\) is a homeomorphism from
  \(\TN\) to \(D \) thus \(\phi\) is a homeomorphism from
  \(\Sigma_N\times \TN\) to \(\Sigma_N \times D\). Conjugacy follows from
  \begin{align*}
    T \circ \phi(\omegaVk,z) = (\sigma(\omegaVk),T_{\omegaVk_1}(\theta(z))) =
		(\sigma(\omegaVk),\theta(z+\DeltaU M \e_{\omegaVk_1})) =
		\phi(\omegaVk,z) \circ \hat{T} \, .
  \end{align*}
\end{proof}

% ========================================================================================================================

\subsection{Relative unique ergodicity for skew product dynamical systems\label{sec:reluniqeergo}}

\subsubsection{General case}

Ergodicity is useful for directly relating volumes in phase space to probabilities for particular avalanches. In this section we establish {\em unique ergodicity relative to a given shift-invariant probability measure} on \(\TN\) for a general class of translation dynamics which includes \(\hat{T}\). Since we have shown in the previous section that the simple translation dynamics \(\hat{T} \) on \(\Sigma_N \times \TN\) is topologically conjugated to the system \(T\), ergodicity of \(\hat{T}\) implies ergodicity of \(T\). The unique relative ergodic measure will turn out to be a product measure  given by the shift-invariant probability measure times the normalised  Lebesgue measure on the $N$-torus, which is transported to the normalised Lebesgue measure with support on \(D\) for the system \(T\).

Specificially, for a continuous function $g:\Sigma_N \to\R^{[N]}$ we consider the skew product dynamical system
\begin{align*}
	\hat{T}_{g}:\Sigma_{N} \times\TN \longrightarrow \Sigma_N \times
	\TN,(\omegaVk, z) \mapsto (\sigma(\omegaVk),{g}(\omegaVk)+z) \quad . 
\end{align*}
Note that function $g$ in $\hat{T}_{g}$ describes a more general dynamics than in the EHE system $\hat{T}$ and can depend on more than just the the first component of the input sequence $\omegaVk$. Furthermore, note that every translation on the $N$-torus defines a bijection which leaves the normalised Lebesgue measure \(\lambda\) on the $N$-torus invariant.

%\udo{Wir wollten das Folgende umschreiben und erst zum Schluss auf den "`neuen"' Begriff fuer Ergodizitaet kommen. Habe im Folgenden versucht, das umzusetzen, aber sollte gecheckt werden:}
 
Let us denote the set of $\hat{T}_{g}$-invariant Borel probability measures by $\MT$. For a fixed shift-invariant probability measure $\nu$ on $(\Sigma_{N},\mathcal{B})$ we denote the subset $\MT$ of elements with marginal $\nu$ by $\MTP\coloneqq \{\mu\in \MT: \mu\circ\pi_{1}^{-1}=\nu \}$. We always have   $\nu\otimes \lambda\in \MTP$ and we find, that if  $\nu\otimes\lambda$  is ergodic for  $\hat{T}_{g}$, if and only if $\nu$ is ergodic for $\sigma$ and   $\MTP=\{\nu\otimes\lambda\}$. The ergodicity of $\nu$ follows from the fact that $ \sigma$ with the invariant measure $\nu$ is a measure theoretical factor of $\TN$ with respect to invariant measure $\nu\otimes \lambda$. To infer that $\MTP$ is a singelton,  fix some $\mu\in \MTP$.  For every  $t\in \TN$  the  translation $\tau_{t}:\Sigma_{N}\times \TN\to \Sigma_{N}\times \TN$, $(\omegaVk,x)\mapsto (\omegaVk,x+t)$    commutes with $\hat{T}_{g}$ and hence $\mu_{t}\coloneqq \mu\circ \tau_{t}^{-1}$,  as well as  its averaged version $\overline{\mu}\coloneqq\int_{\TN}\mu_{t}\;\d\lambda(t) $, define again  elements of $\MTP$.  For every integrable function $f$ we have
\begin{align*}
\int f \;\d\overline{\mu} &= \int_{\TN} \int f  \;\d{\mu_{t}}d\lambda (t)=\int_{\TN} \int f(\omegaVk,x+t) \;\d{\mu}\d\lambda(t) \\
&= \int \int_{\TN} f(\omegaVk,x+t) \;\d\lambda(t)\d{\mu(\omegaVk,x)}
= \int \int_{\TN} f(\omegaVk,t) \;\d\lambda(t)\d{\mu(\omegaVk,x)}\\
&= \int   f \;\d(\mu\circ\pi_{1}^{-1}\otimes\lambda) 
= \int   f  \;\d(\nu\otimes\lambda). \end{align*}
Consequently, $ \overline{\mu} =\nu\otimes\lambda$.  Since  $\nu\otimes\lambda$ is assumed to be ergodic and therefore extremal in $\MT$, we have that $\mu_{t}=\nu\otimes\lambda$ for almost all $t\in \TN$. That is, for  almost all $t\in \TN$,  $\mu=\mu_{t}\circ (\tau_{-t})^{-1}=\nu\otimes\lambda$,  uniqueness follows. 

For the converse implication, suppose that $\nu\otimes\lambda$ is not ergodic. Then we find two distinct measures $\mu_1$ and $\mu_2$ in $\MT$ and $c\in (0,1)$ such that $\nu\otimes\lambda=c\mu_1+(1-c)\mu_2$. Then for the first marginal we have $\nu=(c\mu_1+(1-c)\mu_2)\circ\pi_1^{-1}=c\mu_1\circ\pi_1^{-1}+(1-c)\mu_2\circ\pi_1^{-1}$. Then either $\nu$ is not ergodic, or if $\nu$ is ergodic, we conclude that $\nu=\mu_1\circ\pi_1^{-1}=\mu_2\circ\pi_1^{-1}$ and thus $\mu_1$ and $\mu_2$ actually belong to $\MTP$. In this case, $\MTP$ is not a singleton.

This observation gives rise to the following definition: If $\nu\otimes\lambda$ is ergodic for $\hat{T}_{g}$, or equivalently $\MTP= \{\nu \otimes \lambda \}$ and $\nu$ is ergodic, we call the random dynamical system {\em uniquely ergodic relative to} $\nu$.

To find necessary and sufficient conditions for unique ergodicity relative to an ergodic measure $\nu$, we write functions on $\TN$ as Fourier decompositions. This  allows  us to express shifts induced by external input as simple multiplications (i.~e.\ phase shifts) in Fourier space. In the following all equations involving measurable functions are meant to hold almost everywhere with respect to the relevant measure.
%In this way we establish conditions on the modes of a $\hat{T}_{g}$-invariant function and allows to derive conditions for which nonzero modes vanish.

\begin{thm}\label{thm:ergodicity} 
% Let $\Sigma$ be a subshift of finite type over the alphabet $[N]$ with a shift-invariant ergodic probability measure $\P$.  For a   %family of isometries $(g_{\omegaVk}:\TN\to \TN)_{\omegaVk\in \Sigma}$  such that 
% continuous function $g:\Sigma_N   \to\R^{[N]} $  we consider the random dynamical system
%  \begin{align*}
%    \hat{T}_{g}:\Sigma \times\TN \longrightarrow \Sigma_N \times \TN,(\omegaVk ,t) \mapsto (\sigma(\omegaVk),{g}({\omegaVk})+t) 
%  \end{align*}
  With the above notation and such that $\nu$ is ergodic for the base transformation $\sigma$, we have that
  $\hat{T}_{g}$ is uniquely ergodic relative to $\nu$ if and only if, for all \(k\in \Z^{[N]}\setminus\{0\} \) there
  is no complex-valued measurable function $R\neq0$ on $(\Sigma_{N},\mathcal{B})$ such that 
  %for$\nu$-almost every $\omegaVk \in \Sigma_N$,% $\vert R(\omegaVk)\vert=1$  and
	\begin{align}\label{eq:solution}
	R=	\exp\left(2\pi ik^{T} g \right)\cdot R\circ\sigma.
	\end{align}
\end{thm}
 We note that the ergodicity of $\nu$ with respect to $\sigma$ implies that any solution $R$ of \Eq{eq:solution}  has constant modulus, and we may therefore assume without loss of generality that $\vert R\vert=1$.
% \udo{Was ist $\lambda$?} --> Lebesgue, intuitiv Volumen so wie wir es kennen und daher 1
%\udo{Man koennte auch hier statt $x$ lieber $z$ verwenden, als Indikator, dass man im $N$-Torus rechnet...}\red{$x$ nach
%  $z$ ueberall...}

\begin{proof}
 For the proof we adopt ideas of Furstenberg from \cite{MR133429}, where, unlike here, the base transformation 
is assumed to be uniquely ergodic and the fibers are given by the circle:  
To prove that our assumption in  \Eq{eq:solution} implies that  $\nu \otimes \lambda$ is ergodic with respect to $\hat{T}_{g}$,  fix a square-integrable function $f\in L^{2}_{\nu \otimes \lambda}$ with $f\circ\hat{T}_{g}=f$. Setting $\zeta_{k}(x)\coloneqq \exp(2\pi ik^{T} x)$, we can write $f$ as a Fourier series via $f(\omegaVk,x)=\sum _{k\in \Z^{[N]}} c_{k}(\omegaVk)\zeta_{k}(x)$ for appropriate square-summable coefficients $(c_{k}(\omegaVk))_{k\in \Z^{[N]}}$. Since $\nu \otimes \lambda$ is a product measure we have $c_{k}\in L_{\nu}^{2}$ for each $k\in \Z^{[N]}$. The invariance of $f$ gives
\[
	f(\omegaVk,x) = \sum_{k\in \Z^{[N]}} c_{k}(\omegaVk)\zeta_{k}(x)
	=\sum_{k\in \Z^{[N]}} c_{k}(\sigma(\omegaVk))
	\exp\left(2\pi i k^{T}g(\omegaVk) \right)\zeta_{k}(x)=f\circ \hat{T}_{g}(\omegaVk,x)
\]
and we deduce $c_{k}(\omegaVk)= c_{k}(\sigma(\omegaVk)) \exp\left(2\pi i k^{T} g({\omegaVk})\right)$ for all $k\in \Z^{[N]}$. If $c_{k}=0$ for all $k\neq0$ then $f(\omegaVk, z)=c_{0}(\omegaVk)$ and by the ergodicity of $(\Sigma_{N},\mathcal{B},\sigma,\nu)$ we have that $c_{0}$ is a constant function and so is $f$. If $f$ is not constant, then $c_{k}$ does not vanish for at least one $k\in \Z ^{[N]}\setminus\{0\}$, by ergodicity of $\sigma$ and since $\vert c_{k}\vert= \vert c_{k}\vert\circ\sigma$ we have that $\vert c_{k}\vert$ equals a positive constant  function. Consequently, our assumption is violated for $R\coloneqq c_{k}$. Our condition therefore implies that $f$ is constant and hence $\hat{T}_{g}$ is ergodic with respect to $\nu \otimes \lambda$.

% \udo{Sieht gut aus, sollten trotzdem nochmal die Schritte zusammen nachvollziehen...}

% \udo{Die Abschnitte bis zum naechsten Corollar muessen wir definitiv noch einmal zusammen nachvollziehen. Was zum Geier ist $\lambda$? ...ahhh, Lebesgue-Mass}

Conversely, we assume that our condition is not fulfilled. Then for a solution $R\neq0$ of \Eq{eq:solution} for some $k\in \Z^{[N]}\setminus\{0\}$, we have $R\zeta_{k}$ is non-constant and $\hat{T}_{g}$-invariant. Hence, $\nu \otimes \lambda$ is not ergodic with respect to $\hat{T}_{g}$. 
\end{proof}

\subsubsection{Ergodicity in the EHE model}
If the random dynamical systems is given by the EHE model (i.~e.\ $\hat{T}_g=\hat{T}$) and the underlying measure in the base is Bernoulli, then our condition in \Eq{eq:solution} simplifies as follows.
\begin{prop}\label{prop:condition-ergodicity}
	If the shift space $\Sigma_{N}$ is equipped with the Bernoulli measure
	$\nu\coloneqq \mathbb{B}_{p}$ with \(L=\{\ell\in [N]  \mid  p_{\ell} > 0\} \)
	and $g(\omegaVk)\coloneqq u_{0}M\e_{\omegaVk_{1}}$, we have that the condition
	in \Eq{eq:solution} is equivalent to the condition that for all
	$k\in \Z^{[N]}\setminus \{0\}$ there exists an $\ell\in L$ such that
  \begin{align}\label{eq:ConditionA}
		\exp\left(2\pi i \, \DeltaU k^{T} M \e_{\ell}\right)\neq 1.
  \end{align}
 \end{prop}

\begin{proof} We first show that the condition in \Eq{eq:solution} implies the condition of the corollary by contraposition: If for some $k\in \Z^{[N]}\setminus \{0\}$ we have $ \exp\left(2\pi i \, \DeltaU k^{T} M \e_{\ell}\right)=1$ for all $\ell\in L$, then $R=1$ solves condition \eqref{eq:solution}.

Conversely, suppose that for some \(k\in \mathbb{Z}^N\setminus\{0\} \), there exists a  measurable function $R$ on $(\Sigma_{N},\mathcal{B})$ with $\vert R\vert =1$ such that \Eq{eq:solution} holds. Let us set $f:\ell \mapsto \exp\left(2\pi i u_{0} k ^{T}  M\e_{\ell} \right)$, then we have $R(\sigma(\omegaVk))=f(\omegaVk_{1})R(\omegaVk)$. Integrating both sides with respect to the Bernoulli measure $\nu$ gives
\[
	\int R\,\d\nu =\sum_{\ell\in [N]}f(\ell)p_{\ell}\int R\,\d\nu.
\]
Since $\int R\,d\nu\neq 0$ we get $\sum_{\ell\in [N]}f(\ell)p_{\ell}=1$.
% UDO: Komplexe Vektorsumme, geht nur, wenn alle Komp. in dieselbe Richtung und
% UDO: nach 1 zeigen...
By convexity of the unit circle and $1$ being an extremal point, this is only possible if $f(\ell)=1$ for all $\ell\in L$. This shows that \Eq{eq:ConditionA} is fulfilled for $k$.
\end{proof}

The following Corollary states a sufficient condition for unique relative ergodicity of the system \(\hat{T}\):

\begin{cor}\label{cor:sufficient}
  The system \(\hat{T}\) is uniquely ergodic relative to \(\mathbb{B}_{p}\)
	if the components of \(\DeltaU M \delta(L)\) and 1 are rationally independent,
	i.e. \(\DeltaU k^T M \delta(L) \in \mathbb{Z}^{[N]} \) for \(k \in \Z^{[N]}\) implies \(k=0\).
\end{cor}  

\begin{proof}
  Suppose that ergodicity does not hold for \(\hat{T}\).
	With Proposition~\ref{prop:condition-ergodicity} it follows that there exits
   \( k\in \Z^{[N]}\setminus\{0\}\) such that
  \(\exp(2\pi i \DeltaU k^T M \e_\ell) = 1\) for all \(\ell\in L\). In
  particular, this implies that
    \[\prod_{\ell\in L}\exp(2\pi i \DeltaU k^T M \e_\ell) = \exp(2\pi i \DeltaU k^T M \delta(L)) = 1 \text{ .}\]
    However, this contradicts the assumption.
\end{proof}

\begin{thm}\label{thm:ergodicity-almost-everywhere}
  Assume that \(\DeltaU\) is irrational. The system \(\hat{T}\) is
  uniquely ergodic relative to \(\mathbb{B}_p\) for almost all 
  \(W'\in \mathcal{W}(W)\) if and only if \(W\) is $p$-reachable.
	% If \(W\)
  % is not $p$-reachable \(\hat{T}\) is not uniquely ergodic relative to
  % \(\mathbb{B}_p\) for all \(W'\in \mathcal{W}(W)\).
  % \mhk{diese Aussage gilt ja eigentlich auch gleichm\"a{\ss}ig f\"ur alle $p$ mit derselben Mengen $L_{0}$, oder? Vielleicht f\"uhren wir dann $L$-reachable ein und die Aussage gilt f\"ur alle $p$ with $L_{0}=L(p)$, ??!??}
  % \ms{Hier kommt es nur auf den Support an, ja. Allerdings gibt es bei  Proposition~\ref{thm:d-inhabited} kein $n_0$ dass f\"ur alle $p$ mit $L_0$ garantiert, dass das System sich auf dem Attraktor $D$ befindet. --- Das lie{\ss}e sich aber durch die korrekte Anordnung der Quantoren ausdr\"ucken, oder?
  % ----- hab das mal als alternative mit hin geschrieben.}
\end{thm}

\begin{proof}
	If \(W\) is not $p$-reachable, then \((M\delta(L))_j = 0\) for some unit
	\(j\in [N]\) which implies that \(\exp(2\pi i \DeltaU k^T M \e_\ell) = 1\)
	for the mode \(k=\e_j\) and all \(\ell\in L\). Thus, \(\hat{T}\) is not
	uniquely ergodic relative to \(\mathbb{B}_p\) by
  Proposition~\ref{prop:condition-ergodicity} for all \(W'\in \mathcal{W}(W) \).

  Now assume that \(W \) is $p$-reachable and let $E \coloneqq E(W)$
	denote its edge set. 
	First, let \(P = \{\ell \in L  \mid  (M\delta(L))_\ell -M_{\ell \ell}= 0\}\)
	be the set of units which directly receive external input,
	but have no incoming paths starting from units receiving external input.
	
%	\udo{Die folgende Seite ist ein ziemlicher Wust - kann man das verstaendlicher machen und besser gliedern?}
	
	For each unit \(m\in P\) and \(\ell\in L\) we have
    \(\e_{m}^{T}M\e_{\ell} = \e_{m}^{T}\e_{\ell}\) so that the \(P\)
    coordinates of every \(k\in \Z^{[N]}\) not fulfilling the condition in
    Proposition~\ref{prop:condition-ergodicity} have to be zero, since \(\DeltaU \notin \mathbb{Q}\). 

    Unique relative ergodicity of \(\hat{T}\) with respect to
    \(\mathbb{B}_p\) and \(W'\in\mathcal{W}(W) \) would follow from
    Corollary~\ref{cor:sufficient}, if for all \(k\in \Z^{[N]}\setminus\{0\}, k|_P = 0\)
    and  \(c(W')\coloneqq\DeltaU M'\delta(L) \) with \(M'\coloneqq (\mathds{1}-W')^{-1}\)
    %\(M'\coloneqq (\diag(U)-W')^{-1}\) 
    the scalar product \(k^Tc(W')\notin \Z\).

    We want to show that this  property holds almost everywhere with respect to the $\vert E\vert$-dimensional Lebesgue measure  \(\lambda\) on \(\mathcal{W}(W)\) considered as a subset of
    \(\R^{E}\).
For this we will cover the  complement with respect to this property  by sets
    \[A_{k,z} = \{W'\in \mathcal{W}(W) \mid  k^Tc(W') = z\}\]
    with \(k\in \Z^{[N]}\setminus\{0\},k|_P = 0\), \(z\in \Z\) and show that  each $A_{k,z}$ is a \((|E|-1)\)-dimensional submanifolds of \(\mathcal{W}(W)\) and thus a null sets with respect to  \(\lambda\).
     Indeed, let  
   \[F_{k}:\R^{E}\supset\mathcal{W}(W) \to \R ,(w'_{1},\ldots ,w'_{|E|})\mapsto k^{T}c(W'),\] where $W'\in \mathcal{W}(W)$ is uniquely determined by its non-zero entries $(w'_{1},\ldots ,w'_{|E|})$. 
   For each  $k\in \Z^{[N]} \setminus\{0\},k|_P = 0$, the function $F_{k}$ is continuously differentiable and (by the implicit function theorem) for each $z\in\Z$ the set \(A_{k,z}\) defines an \((|E|-1)\)-dimensional submanifold if $\nabla F_{k}(W')\neq0$  for every $W'\in \mathcal{W}(W)$ with   $F_{k}(W')=z$.

   In order to show $\nabla F_{k}(W')\neq0$, consider the directional derivatives
   \[\partial_{ \mathbf{W}} c(W')  = (d/ds)\,c(W'+s\mathbf{W} ) = \DeltaU M' \mathbf{W}  M' \delta(L)\]
   in direction of the matrix \(\mathbf{W} \) with sparsity pattern dominated by \(W\), i.\,e.\ for all \(i,j\in[N]\) we have \(\mathbf{W} _{i,j} = 0\)  if \(W_{i,j} =0\) and we can consider \(\mathbf{W}\) as an element of  \(\R^{E}\). %\((j,i) \notin E\). 
   If we could show
   $\{\partial_{ \mathbf{W}} c(W') :\mathbf{W} \in \R^{E}\}=\R^{[N]\setminus P}$, then, for all
   $k\in \Z^{[N]}\setminus\{0\},k|_P = 0$, we would clearly have   $\nabla F_{k}(W')\neq0$. To verify the latter equality, we fix an arbitrary \(v\in \R^{[N]}\) with \(v|_P = 0\) and 
   construct  a matrix \(\mathbf{W} \) with sparsity pattern dominated by \(W\) such that
   \(\DeltaU M' \mathbf{W} M' \delta(L) = v\), or equivalently \(\mathbf{W} M' \delta(L)= \DeltaU^{-1}(\mathds{1}-W')v =: y\),
   as follows: For every unit \(i\in [N]\setminus P \) pick exactly one  \(j\in [N]\) with \({W} _{i,j} >0\)
%   incoming edge \((j,i)\in E\) 
and set
   \(\mathbf{W} _{i,j} = y_i/(M'\delta(L))_j\), all other entries 
   \(\mathbf{W}_{i,[N]\setminus\{i\}}\) of the $i$-th row are chosen to be zero. 
   In particular, for  all \(i\in [N]\setminus P \) and  \(j\in [N]\), we have  \(\mathbf{W} _{i,j} = 0\)  if \(W_{i,j} =0\).
   For the remaining rows indexed by \(i\in P\) we  set \(\mathbf{W}_{i,j}=(\mathds{1}-W')_{i,j}v_j/(\DeltaU M'\delta(L))_j \) for all  \(j\in [N]\).  Since $v_j= 0 $ for $j\in P$ we have $\mathbf{W}_{i,j} = 0$ if $W_{i,j} = 0$ for all \(j\in [N]\). This choice guarantees that \(\mathbf{W}\) has  a sparsity pattern dominated by \(W\) and establishes the desired equality since for 
  \(i\in [N]\setminus P\) we have \(   (\mathbf{W}M'\delta(L))_i=\sum_{j\in [N]}  \mathbf{W}_{i,j}(M'\delta(L))_j=y_{i} \) and for
    $i\in P$  
   \begin{align*}
   (\mathbf{W}M'\delta(L))_i &= \sum_{j\in [N]}\frac{(\mathds{1}-W')_{i,j}v_j}{\DeltaU (M'\delta(L))_j }(M'\delta(L))_j = \sum_{j\in [N]}\DeltaU^{-1}(\mathds{1}-W')_{i,j}v_j 
  % \\&=\DeltaU^{-1}((\mathds{1}-W')v)_p
  =y_{p}.
   \end{align*}
   To conclude, almost sure ergodicity follows by considering a countable cover  
   \[\lambda(\{W'\in \mathcal{W}(W) \mid  \exists  k\in \Z^{[N]}\setminus\{0\},\exists z\in \Z \colon k^Tc(W') = z\}) \leq \!\!\!\!\sum_{k\in \Z^{[N]}\setminus\{0\} \atop z\in \Z}\!\!\!\lambda(A_{k,z}) = 0.\]
  \end{proof}

\subsubsection{Special case: the homogeneous EHE model}

In the following corollaries, we establish two simple conditions to check unique ergodicity for the homogeneous EHE model with constant coupling matrix.

\begin{cor}
  \label{cor:hom-ergodic}
  The homogeneous EHE-Model
  (i.\,e.\ \(w_{ij} = w \in \R_{\geq 0}\), \(\VU_i=1\) for all
  \(i,j=1,\ldots,N\)) is uniquely ergodic relative to the Bernoulli measure
	$\mathbb{B}(p)$, if \(p\) is a strictly positive probability vector and both
  \(\DeltaU,\DeltaU/(1-Nw)\notin \mathbb{Q}\).
\end{cor}

\begin{proof}
	We will use the characterization given in \Eq{eq:ConditionA}.
	%, we will give a proof by contradiction.
  Suppose that for $k\in \Z^{[N]}$ we have that \(\DeltaU  k^{T} M \e_j\in \Z\)
	for all     
  \(j \in [N]\). In particular, we then have \(\DeltaU  k^{T}M(\e_j-\e_i) \in \mathbb{Z}\).
  For the homogeneous EHE-Model, the entries of \(M\) are given by 
  \[M =(\mathds{1}-W)^{-1} =\mathds{1}  + \frac{1}{1-Nw}W.
    %\begin{cases}w_d \;\coloneqq \frac{1-(N-1)w}{1-Nw} &\mbox{ if } i = j \\ w_{od} \coloneqq \frac{w}{1-Nw} &\mbox{ otherwise.} \end{cases}  
  \]
  For all $i,j\in [N]$ we have
  \begin{align*}
    \DeltaU k^{T}M(\e_j-\e_i) %&%=   \DeltaU \left (k_j(w_d-w_{od}) + k_i(w_{od}-w_d) \right) \\
    &=   \DeltaU (k_j - k_i) \in \mathbb{Z}. 
    \end{align*}
    %with \(z \in \mathbb{Z}\). 
    Since \(\DeltaU  \notin \mathbb{Q}\), it  follows that \((k_j - k_i) = 0\) for all $i,j\in [N]$ and hence \(k=(\ell,\ldots,\ell)^{T}\) for some \(\ell \in \Z\).  Consequently, 
    \[\DeltaU k^{T} M\e_1  = % \ell (w_d - (N-1) w_{od}) =
    \frac{\DeltaU \ell}{1-Nw} \in \mathbb{Z} \text{,}\]
    implying $\ell =0$. This shows that   condition~\eqref{eq:ConditionA} is fulfilled. 
\end{proof}

\begin{cor}
  \label{cor:hom-not-ergodic}
The homogeneous EHE-Model is not uniquely ergodic relatively to  $\mathbb{B}(p) $, if  $p_{i}=p_{j}=0$ for two distinct indices $i,j\in [N]$.
\end{cor}

\begin{proof}
  Fix $i,j \in [N]$ with $i\neq j$,  $p_{i}=p_{j}=0$ and set
  \(k \coloneqq \e_i - \e_j \). Then we have \(\DeltaU  k^{T} M\e_\ell = 0\) for every
  \(\ell \in [N],p_\ell > 0\). 
\end{proof}

\subsection{Expected firing rates and spike count covariances}\label{sec:app-firing-covariances}

In this section we show that the equilibrium firing rates and spike count covariances are linear transformations of the rates and covariances of the external input process. Since the firing rate for unit \(i\) is directly antiproportional to \(U_i\), we suppress this dependency by $w.l.o.g.$ analyzing the special case \(\VU=\mathbf{1}\). 

In order to clearly indicate its dependency on the weight matrix, we denote the spike count vector after $s$ iterations of the EHE dynamics from \Eq{eq:ns} as $\NN^{s}_W$. The results in this section hold for all $W$ and $p$ for which the system \(T\) is ergodic and probabilities are evaluated with respect to the uniquely relative ergodic measure \(\P = \mathbb{B}_p \times \PW\), with \(\PW = \lambda_D \) implicitly depending on \(W\) through the structure of the inhabited region $D$. The same applies to the matrix $M=\mathds{1}-W$.

We denote the asymptotic mean firing rates and the $N\times N$ spike count covariance matrix of the ergodic system with coupling matrix \(W\) by
\(Y_W \coloneqq \lim_{s\rightarrow \infty}{\E\left(\NN_W^{s}\right)}/{s}\) and  \(X_W \coloneqq \lim_{s\rightarrow \infty}{\E \left(\NN_{W}^{s}\right)}/{s}\).

%\udo{Namensclash: $\operatorname{cov}$ auch fuer Volumen verwendet, $y$ in Haupttext und hier in Grossschreibung. Vorschlag: $Y$ fuer Raten, $X$ fuer Kovarianz.}

With the choice \(W=0\), \(T\) is equivalent to the random walk induced by the external input. In the next two Theorems we calculate \(Y_0,X_0\) and show that \(Y_W,X_0\) for a general \(W\neq 0 \) are given by linear transformations of \(Y_0,X_0\).

\begin{thm}\label{thm:ratios}
  The equilibrium firing rate is a linear transformation of the firing rate of the uncoupled (\(W=0\)) system. 
  \begin{align}
    Y_0 &=\lim_{s\rightarrow \infty}\frac{\E\left(\NN_0^{s}\right)}{s} = \DeltaU p \text{, }\\
    Y_W &=\lim_{s\rightarrow \infty}\frac{\E\left(\NN_W^{s}\right)}{s} = MY_0 \text{. }
    \end{align}
    The probability that unit $i$ fires given that an avalanche is started by unit $k$ is given by
    \begin{align}
        \P_k(i\in \U(\av)) = \frac{M_{ik}}{M_{kk}} \text{ .}
    \end{align}
  \end{thm}

  \begin{proof}
    From Corollary~\ref{cor:equivalence} we get
    \begin{align*}
  \NN^{s}_W(\omegaVk ,u) = M\bigg(\DeltaU  \sum_{t=1}^s \e_{\omegaVk_t}-\pi_2T^s(\omegaVk ,u)\bigg). % \text{, where } M = (\diag{U}-W)^{-1}
  \end{align*}
  by solving \Eq{eq:col21eq} for $\NN_W^{s}$.
  Let \(H(s)(\omegaVk,u) \coloneqq \lfloor M(u+\DeltaU\sum_{t=1}^s\e_{\omegaVk_t})\rfloor - \lfloor Mu \rfloor\).
    By compactness of \(D\), we have
    $\left|\NN_W^{s}(\omegaVk,u)-H(s)(\omegaVk,u)\right| < c$ uniformely for all \(\omegaVk \in \Sigma_{N},u\in D,s\in \mathbb{N}\). Thus
    \begin{align*}Y & = \lim_{s\rightarrow \infty}\frac{\E\left(\NN^{s}_W\right)}{s}
                     = \lim_{s\rightarrow \infty}\frac{\E (H(s))}{s} \\
                     &= \lim_{s\rightarrow \infty}\frac{\E\left(\left\lfloor \DeltaU M \sum_{t=1}^s\e_{\omegaVk_t} \right\rfloor \right)}{s} = \DeltaU Mp \text{.}
    \end{align*}
    Since $M$ is the identity matrix for \(W=0\) (and \(\VU=\1\)), this asserts \(Y_0 = \DeltaU p,Y_W = MY_0\).
  In addition, \(N_W(s)\) and \(H(s)\) are Birkhoff sums of \(f(\omegaVk,u) = \delta(\U(\av(\omegaVk,u)))\)
  and \(g(\omegaVk,u) = \lfloor M(u+\e_{\omegaVk_1}) \rfloor - \lfloor Mu \rfloor \), respectively.
  By Hopf's ratio ergodic theorem~\citep[Thm. 2.4.24]{KMS} we have \(\E (f) =\E (g) \). The identity 
  \(g(u,\omegaVk)_{i} = 1\) is equivalent to completing a revolution around direction $i$ of the $N$-Torus   
  \((M(u) \mod 1)_i + (\DeltaU Me_{\omegaVk_1})_i \geq 1\).  From \(M(u) \sim \lambda_{[0,1)^N} \) the probability of \(g(u,\omegaVk)_{i} = 1\) given  \(g(u,\omegaVk)_{\omegaVk_1} = 1\) is \(\frac{M_{i\omegaVk_1}}{M_{\omegaVk_1 \omegaVk_1}}\).
  Since \(i\in \U(\av(\omegaVk,u)) \iff \delta(\U(\av(\omegaVk,u))))_i = 1\) and \(\av(\omegaVk,u)_1 = \{k\} \iff (\omegaVk_1=k \land \delta(\U(\av(\omegaVk,u)))_k = 1\), we have
  \begin{align*}
   \E\left ( \ind_{\{i\in \U(\av)  \}} \mid \av_1 = \{k\} \right )  = \frac{M_{ik}}{M_{kk}} = \P_k(i\in \U(\av))
    \end{align*}
  \end{proof}

\begin{thm}\label{thm:Vw}
	The asymptotic $N \times N$ spike count covariance matrices are given by
	\begin{align}
		X_0
			&= \lim_{s\rightarrow \infty}\frac{\operatorname{cov}\left(\NN_0^{s}\right)}{s}
			= \DeltaU^2( \diag(p) - p p^{T})\\
                X_W
			&= \lim_{s\rightarrow \infty}\frac{\operatorname{cov}\left(\NN_W^{s}\right)}{s}
			= M^TX_0 M.
  \end{align}
\end{thm}

\begin{proof}
  Rearranging \Eq{eq:col21eq} for \(\NN^{s}_W\), we have  
  \begin{align*}
    \operatorname{Var}\left(\NN^{s}_W\right)
        		= \operatorname{Var}\left(\DeltaU M \sum_{t=1}^{s}\e_{\omegaVk_t} + z_s\right) \text{, }
  \end{align*}
  where \(z_s(\omegaVk ,u) \coloneqq M(u - \pi_2T^s(\omegaVk ,u))\). For all $\omegaVk \in \Sigma_N,u\in D$ 
  \(z_s(\omegaVk,u) < c \in \R \) with \(c\) independent of \(s\) since
  \(\pi_2T^s(\omegaVk ,u) \in D \subseteq [0,\VU_i)_{i\in [N]}\). We have
  \begin{align}\label{thm:VwTemp}
    \operatorname{Var}(\NN^{s}_W) &= \operatorname{Var}\left(\DeltaU M \sum_{t=1}^{s}\e_{\omegaVk_t} + z_s\right)\\
                                  &= \DeltaU^2 M^T\operatorname{Var}\left(\sum_{t=1}^{s}\e_{\omegaVk_t}\right)M + \operatorname{Var}(z_s) +
                                    2\DeltaU \operatorname{Cov}\left(M \sum_{t=1}^{s}\e_{\omegaVk_t},z_s\right)
  \end{align}
    The term \(\sum_{t=1}^s \e_{\omegaVk_t}\) is multinomially distributed
    with success probabilities \(p\) and \(s\) trials and thus  
    \[\operatorname{Var}\left(\sum_{t=1}^s \e_{\omegaVk_t}\right) = s\left(\diag(p)-pp^T\right) \text{ .}\]
  
  From the boundedness of \(z_s\), we have $\operatorname{Var}(z_s)< c_1 \in R$ independent of $s$, and with the Cauchy-Schwartz inequality we get
  \[\operatorname{Cov} \left(M \sum_{t=1}^{s}\e_{\omegaVk_t},z_s \right) < \sqrt{M^T\operatorname{Var}\left(\sum_{t=1}^{s}\e_{\omegaVk_t}\right)M c_1}
    = \sqrt{s c_1}\sqrt{M^T\left(\diag(p)-pp^T\right)M}\text{ .} \]
  Hence the two last terms in \Eq{thm:VwTemp} vanish in the limit \(s\rightarrow \infty\) and we get
    \begin{align*}
      \lim_{s\rightarrow \infty}\frac{\operatorname{Var}\left(\NN^{s}_W\right)}{s} = \DeltaU^2M^T\left ( \diag(p)-pp^T \right )M
    \end{align*}
\end{proof}

\subsection{Geometrical structure and self-similarity of the inhabited region}\label{sec:app-geomerty}

\subsubsection{Geometrical description and self-similarity of the noninhabited region \label{sec:self-similarity}}

In this section we show that the invariant space \(D\) (or more directly its complement) has a self-similar structure which will be used to simplify expressions for avalanche distributions considerably. We first introduce regions \(\Lambda_I\) and show that they have self-similar properties. We will call these regions 'non-inhabited' and justify this term by showing that \(\Lambda_{[N]} = C \setminus D\).
\begin{defn}\label{def:Lambda}
  Let \(\emptyset \neq H \subseteq [N]\) be an index set.
	Define the non-inhabited region $\Lambda_H$ along dimensions \(H\) by
  \begin{align}\label{eq:defvnoninh}
    \Gamma_I  &= \left[0, \sum_{j \in I}w_{i j} \right )_{i \in I} \\
    \Lambda_H &= \bigcup_{\emptyset \neq I \subseteq H} \Gamma_I \text{ .}
  \end{align}
\end{defn}

We proceed to show self-similar properties of \(\Lambda\) and relate it to the inhabited region \(D\) according to the following steps:

%\udo{The following enumeration is a good guide for the reader. Would it be possible to refer to the corresponding Lemmata etc. to make direct links?}

\begin{enumerate}
\item Perform a suitable decomposition of the state space \([0,U_i)_{i\in [N]} \) by disjoint hyperrectangles  in~\eqref{eq:decomp}.
\item Write \(\Lambda\) as a disjoint union of these lower-dimensional non-inhabited regions in Lemma~\ref{lem:rec-dec}.
\item Use this result to show that \(\Lambda\) is the complement of the inhabited region \(D\) in Theorem~\ref{thm:DGamma}.  
\end{enumerate}

Figure~\ref{fig:decomp_better} illustrates the geometrical structure of the noninhabited region and its self-similarity for dimensions 1 up to 3.

% This decomposition corresponds to the following expansion for the
% volume of the phase space:
% \begin{equation}
% \V([0, U_i]_{i\in H}) = \prod_{i\in H} U_i = \prod_{i\in H} (\sum_{j \in H} w_{i j} + (U_i - \sum_{j\in H} w_{i j})) \text{ .}
% \end{equation}

% Figure \ref{fig:decomp} graphically suggest that
% \label{sec:self-similarity} The overlap between a \(R(W,H,I)\) and
% \(\Lambda(W,H)\) is given by the overlap between \(R(W,H,I)\) and
% the non-inhabitated region along the dimensions \(I\). \\

The following Lemma will be used throughout this section and states that the noninhabited region along dimensions \(H_1\) is equal to the lower-dimensional  \(\Lambda_{H_1\setminus H_2}\) when intersected with a hyperrectangle which has lower boundaries along dimensions \(H_2\) which lie above the corresponding row sums in \(W\). 
\begin{lem}
  \label{lem:elimdim}
	For \(H_1,H_2 \subseteq [N]\) and \(a_i,b_i \geq \sum_{j\in H_1}w_{ij}\)
	for all \(i\in H_1 \cap H_2\) we have
  \[
		\Lambda_{H_1} \cap [a_i,b_i)_{H_2} = 
		\Lambda_{H_1\setminus H_2} \cap [a_i,b_i)_{H_2} \, .
	\]
\end{lem}

\begin{proof}
  Note that by definition \(\Lambda_H = \bigcup_{\emptyset \neq I \subseteq H}\Gamma_I\).
  The result follows if \([a_i,b_j)_{H_2} \cap \Gamma_J = \emptyset\) for all
	\(\emptyset \neq J \subseteq H_1\) such that \(J \cap H_2 \neq \emptyset\).
  We have
	\begin{align*}
		[a_i, b_j)_{H_2} \cap \Gamma_J
		= [a_i,b_j)_{H_2} \cap \left[0, \sum_{j \in J}w_{i j} \right )_{i \in J} 
		= \emptyset \text{, }
	\end{align*}
	since the intersection along the dimensions \(J \cap H_2\) is empty due to
	\(\sum_{j\in J}w_{ij} \leq \sum_{j\in H_1}w_{ij} < a_i\) for \(i\in J \cap H_2\).
\end{proof}

We introduce the following decomposition of the phase space along dimensions \(H\subseteq [N]\) into disjoint hyperrectangles:

\begin{equation}
	[0, \VU_i]_{i\in H}
	= \biguplus_{ I \subseteq H} \left ( \left [0, \sum_{j \in H}w_{i j}\right )_{i \in I} \cap 
		\left [\sum_{l\in H} w_{k l},\VU_k \right )_{k \in H\setminus I} \right )
              \label{eq:decomp}
\end{equation}

Figure~\ref{fig:decomp_better}, panel (d) shows this decomposition for the three-dimensional case $H=[N]=[3]$. Note that the intersection of the blue region with the noninhabited region is empty. Similarly the enclosed noninhabited region is just a single $\Gamma_\{k\}$ for $k\in [3]$ in each blue region and the union of $\Gamma_I$ generating two-dimensional noninhabited regions are enclosed in the red regions.
The next Lemma formalizes this self-similar structure of \(\Lambda_H\) for arbitrary subsets \(H\subseteq [N]\).

%\begin{figure*}
%  \centering
%  \begin{tabular}{cc}
%    \includegraphics[width=0.45\linewidth]{img/Gammas.pdf}&
%    \includegraphics[width=0.45\linewidth]{img/Rs.pdf}
%    \end{tabular}
%\caption{Left: Illustration of Non-inhabited region for the 3-dimensional EHE model formed by the union of non-inhabited regions generated by the union of \(\Gamma_I\) for \(\emptyset \neq I \subseteq \mathcal{N}\). Right: Decomposition of \([0,1)^3\) into the non-overlapping rectangles. \(R(W,\mathcal{N},I)\) contains the non-inhabited region generated by \(I\), which is formalized by Lemma~\ref{lem:rec-dec} \label{fig:decomp}.}
%\end{figure*}

\begin{lem}
	\label{lem:rec-dec}For \(\emptyset \neq H \subseteq [N]\) and
	\(\VV_{H} = (\VV_i)_{i\in H},\sum_{j\in H}w_{ij} < \VV_i \leq \VU_i\)
	for all \(i\in H\) we have
	\begin{align}
		\Lambda_H\cap [0,\VV_i)_{i\in H}
		= \biguplus_{\emptyset \neq I \subseteq H} \Lambda_I \cap \left [0,\sum_{j\in H}w_{ij}\right)_{i\in I}
			\cap \left [\sum_{l\in H} w_{k l}, \VV_k \right)_{k \in H\setminus I} \text{. }
			\label{eq:recursive-decomposition}
	\end{align}
\end{lem}

\begin{proof}
  With \((W\delta(H))_H < \VV_H \leq \VU_H\)
  and the decomposition from \Eq{eq:decomp} we have
  \begin{align*}
		\Lambda_H \cap [0,\VV_i)_{i\in H}
		&= \biguplus_{\emptyset \neq I \subseteq H} \left ( \left [0, \sum_{j \in H}w_{i j}\right )_{i \in I} 
		\cap \left [\sum_{l\in H} w_{k l},\VV_k \right )_{k \in H\setminus I} \cap \Lambda_H \right )
	\end{align*}
	Using Lemma~\ref{lem:elimdim} we have
	\(\left[\sum_{l\in H}w_{kl},\VV_k\right)_{k\in H\setminus I} \cap \Lambda_H
	= \left[\sum_{l\in H}w_{kl},\VV_k\right)_{k\in H\setminus I} \cap \Lambda_I \)
	and arrive at 
	\begin{align*}
		\Lambda_H \cap \left[0,\VV_i\right)_{i\in H}
		= \biguplus_{\emptyset \neq I \subseteq H} \Lambda_I\cap \left [0, \sum_{j \in H}w_{i j}\right )_{i \in I} 
		\cap \left [\sum_{l\in H} w_{k l}, \VV_k \right)_{k \in H\setminus I} \text{.}
	\end{align*}
	
\end{proof}

Lemma~\ref{lem:rec-dec} provides the direct generalization of the corresponding result for homogeneous systems \cite[Equation B5]{eurich2002finite} to non-negative weight matrices. This self-similar property of \(\Lambda\) will be used to show that it is the complement of the inhabited region \(D\). In the following Lemma, we give an alternative characterization of \(D\), which is more convenient to establish the relation to \(\Lambda\). The intuition behind this characterization is illustrated in~\Fig{fig:decomp_better}.

%\begin{figure*}
%  \includegraphics[width=0.9\textwidth]{img/D_Lambda_connection_illustration.png}
%  \caption{\label{fig:DL-illus}Shifts of \(M^{-1}n, 0\neq n\in {0,1}^N\) illustrated in two dimensions.
%    Left panel: For all points in \(D\), all three possible shifts lead to points outside of \(D\). This is formalized
%    in Lemma~\ref{lem:D}. Right panel: For all points in \(\Lambda_{[N]}\) exists a shift which leads to a point in \(D\). Exemplified with points \(\tilde{u}^{(1)} = u^{(1)}+M^{-1}(0,1)^T,\tilde{u}^{(2)} = u^{(2)}+M^{-1}(1,0)^T,\tilde{u}^{(3)} = u^{(1)}+M^{-1}(1,1)^T \). Together with the left panel, this implies that \(D = C \setminus \Lambda_{[N]}\). This is formalized in Theorem~\ref{thm:DGamma}.}
%\end{figure*}

\begin{lem}
	\label{lem:D}
	An equivalent characterization of the inhabited region is given by
	%Setting  $  \theta(z) \coloneqq \fixF (U+M^{-1}z)$ and $    D \coloneqq \theta(\T^n) $ we have
	\[D= \left\{u \in C  \mid  u + M^{-1}n \notin C,  n\in \{0,1\}^{[N]}\setminus \{0\}\right\}.\]
\end{lem}

\begin{proof}
  Denote the set on the right hand side by \(A\). Every \(u\in C\) can be written uniquely as \(u = W\1 +M^{-1}c\) since
  $M^{-1}=\diag(U)-W$ is bijective. Note that \(\VU = W\1 +M^{-1}\1 \), and if \(c_i \geq 1\) for some
  \(i\in [N]\) it follows that \(W\1 +M^{-1}c \notin C\).
	
	First we show that \(D\subseteq A\).
  Let \(u=\theta(z)\) be the image of \(z\in \TN\) in \(D\). Thus we have \(u = \fixF\left(W\1 +M^{-1}z\right) = W\1 +M^{-1}(z-n)\) for some \(n\in \{0,1\}^N\).
  Now suppose that there exists an \(n'\in \{0,1\}^N\) such that \(u+M^{-1}n' \in C\). We will show that this implies \(n'=0\).
  First, \(n' \leq n\) componentwise since \(u + M^{-1}n' = W\1 +M^{-1}(z-n+n')\) and $(z-n+n')_j \geq 1$ if \(n'_j > n_j\).
  However, using Lemma~\ref{lem:fixF-minimality} \(n' \leq n\) implies \(n'=0\).
  
	To show that \(A\subseteq D\) let \(x\in A\) be arbitrary. From the condition on \(A\) there is a unique way to write \(x\) as
  \(x = W\1 + M^{-1}(n(x)+z(x))\) for some \(n(x)\in \{0,1\}^N\) and \(z(x)\in [0,1)^N\) and for \(x_1,x_2\in A,x_1 \neq x_2\) we have \(z(x_1)\neq z(x_2)\).
  Since \(\fixF\) only subtracts integer combinations of \(M^{-1}\) columns this implies that \(x = \theta(z(x))\) and thus \(x\in D \).

  %% old approach with sim_M
  % To show that \(A\subseteq D\) note that from \(D = \phi(\TN) \subseteq A\) there is at least one \(x\in A\) auch that \(x = \fixF(W\1  + M^{-1}z) = W\1 + M^{-1}(z-n)\) for every \(z\in \TN\). The condition on \(A\) implies that there is at most one element \(x\in A\) with \(x\sim_M W\1  + M^{-1}z\) for every \(z\in \TN\). Taken together, this proves \(A=D\).
\end{proof}
% \udo{...schwer, der Argumentkette zu folgen. Verweise auf Formeln bzw. Definitionen an den entscheidenden Stellen? Hier und im folgenden Theorem m\"ussen wir Satz f\"ur Satz durchgehen.}

\begin{thm}\label{thm:DGamma} The inhabited region \(D\) is the complement of \(\Lambda_{[N]}\) in \(C = [0,\VU_i)_{i\in [N]}\)
  \begin{align}
    D = C \setminus \Lambda_{[N]}.
  \end{align}
\end{thm}

\begin{proof}
	Using Lemma~\ref{lem:D}, the complement of $D$ in $C$ is given by 
  \[
		B \coloneqq C \setminus D = \left\{u \in C  \mid  u + M^{-1}n \in  C
		\text{ for some } 0\neq n\in \{0,1\}^N\right\} \text{ .}
	\]
	We proceed to show that \(\Lambda_{[N]} = B\).
	
        We start by showing \(B\subseteq \Lambda_{[N]} \). Let \(x \in B\) and \(n \in \{0,1\}^N\) such that
        \(\tilde{x} = x + M^{-1}n \in C\). Now considering the coordinates \(i \in I = \{i\in [N]  \mid  n_i = 1\}\),
        we find
  \begin{align*}
    x_i < \VU_i - \left(M^{-1}n_I\right)_i \leq \sum_{k\in I}w_{ik}\text{, }
  \end{align*}
  thus we have \(x\in \Gamma_I \subseteq \Lambda_{[N]}\).\\

  To show \(\Lambda_{[N]} \subseteq B \) let \(x \in \Lambda_{[N]}\) be arbitrary.
  Using the decomposition into disjoint sets in Lemma~\ref{lem:rec-dec} there exists exactly one set \(\emptyset \neq I \subseteq [N]\) such that $x_i < \sum_{j\in [I]}w_{ij}$ for all \(i\in I\) and \(x_k\geq \sum_{l\in [N]}w_{kl}\) for $k\in [N]\setminus I$.\\ 
  Set \(n_I = \delta(I) \) and consider
  \(\tilde{x} = x + M^{-1}n_I = x + \diag(\VU)n_I - Wn_I\). From the choice of \(I\) we have  \(\tilde{x}_k = x_k - \sum_{i\in I}w_{ki} \in [0,\VU_k)\) for all \(k\in [N]\setminus I\).
  For the coordinates \(i\in I \), the choice of $I$ assures that 
  \(\tilde{x}_i = x_i + \VU_i - \sum_{j\in I}w_{ij} \in [0,\VU_i)\). Taken together, we have \(\tilde{x}\in C\), thus \(x\in B\) which completes the proof.
\end{proof}

\begin{cor}[Volume of inhabited region with upper boundaries $\VV$]\label{cor:volinhabited}
  Let \(0<\VV_i\leq \VU_i\) such that \(\sum_{j\in I}w_{ij} < \VV_i  \)
	for \(i=1,\ldots,N\) and let \(\pi_I\) be the projection from \(\mathbb{R}^{[N]}\) to
	\(\mathbb{R}^I\) and \(\lambda_I\) be the Lebesgue measure on \(\mathbb{R}^I\). Then, 
  \begin{align}\label{eq:vin}
    \V_{I}\left(\VV\right) \coloneqq \lambda_I\left(\pi_I\left(\left[0,\VV_i\right)_{i\in I} \setminus \Lambda_I\right)\right) = \left\rvert\diag(\VV_I)-W_I\right\lvert
  \end{align}
\end{cor}

\begin{proof}
  Consider the lower-dimensional subsystem of $T$ defined on the units
  in $I$, with coupling matrix \(W_I\) given by rows and columns in
  \(I\) from \(W\), and with firing thresholds given by 
	% \(\VV_I\).
  \(\VU_I\).
  From \(\sum_{j\in I}w_{ij} < \VV_i \), this system fulfills
  condition~\eqref{ass} for some \(\DeltaU\). With
  Theorem~\ref{thm:DGamma}, the inhabited region \(D_I\) of this
  subsystem is given by
  \(D_I = \pi_I\left(\left[0,\VV_i\right)_{i\in I} \setminus \Lambda_I\right)\). With
  Theorem~\ref{thm:D}, the inhabited region is the image
  \(\theta\left(\TN\right)\). Since \(\fixF\) only consists of translations it is
  volume-preserving and we have  
  \[\lambda_I(D_I) = \lambda_I\left(\theta\left(\mathbb{T}^I\right)\right) =
	\lambda_I\left(\left(\diag\left(\VV\right)_I-W_I\right)\mathbb{T}^{I}\right) = \left\lvert\diag(\VV)_I-W_I\right\rvert\]

  % \(W_I \) fulfills the corresponding assumption~\ref{ass} and the
  % corresponding phase space is \(\pi_I[0,\VV_i)_{i\in I}\).
  % Thus there exist the bijection \(\theta\) from \(\mathbb{T}^{|I|}\)
  % to
  % \(\pi_I\left ([0,\VV_i)_{i\in I} \setminus \Lambda_I
  % \right)\). \(\theta\) is volume preserving, which gives
  % \[\V(\pi_I\left ([0,\VV_i)_{i\in I} \setminus \Lambda_I\right )) = \V(\{(\diag(\VV)-W_I)z|z\in [0,1)^{|I|}\}) = |\diag(\VV)-W_I|\]
\end{proof}

\subsubsection{Phase-space regions leading to avalanches}

In the previous subsections, we have shown unique ergodicity relative to \(\mathbb{B}_p\) of the normalised Lebesgue measure on the inhabited region \(D = C\setminus \Lambda_{[N]}\) and established an understanding of the self-similar geometry of \(\Lambda_{[N]}\) as well as their corresponding Lebesgue volumina. These insights allow us now to derive probabilities for specific avalanches by identifying the pre-images of the avalanche function \(\av(k,u)\), and by calculating their phase space volumes with respect to the ergodic measure \(\P\):

\begin{defn}\label{def:A}
 We call a vector $a=(a_1,\ldots, a_d)$ of non-empty pairwise disjoint subsets $a_i\subset [N]$, $1\leq i\leq d$ with $a_1=\{k\}$ an avalanche with {\em duration} $d(\leq N)$ starting in $k$. 	The coordinate $a_j$ will be called \emph{generation} $j$ of the avalanche $a$. The set of all avalanches starting in $k\in N$ is denoted by \(\mathcal{A}_k\) and we define the set of all avalanches (including the empty avalanche $()$\,) by %\(\mathcal{A}\) and the set of non-empty avalanches  started by unit \(i\in [N]\) by
 % \begin{align}
%    \mathcal{A}_i&\coloneqq \big\{(a_1,\ldots,a_d ) \mid \varnothing \neq a_i\subset [N], a_i\cap a_j=\emptyset, 0<i,j\leq d\leq N,i\neq j, %\biguplus_{j=1}^d a_{j} \subseteq [N],
 %   a_1=\{i\}\big\}  \\
   \(    \mathcal{A} \coloneqq \biguplus_{k\in [N]} \mathcal{A}_k \uplus \{()\}\).
   %\text{ .}\end{align}

%  The empty avalanche is denoted by \(()\).

Since the first
  generation of an avalanche always contains exactly one element, we
  use for a particular avalanche \(a\in \mathcal{A}_k\)
  the notation \(a_1\) both to denote  the singleton set and its only member.

We define the phase space region leading to the
  avalanche \(a \in \mathcal{A}\) by
  \begin{align}
    R(a) \coloneqq \{u\in D \mid \av (u,a_1) = a \}
  \end{align}
\end{defn}

We introduce the shorthand 
\begin{align}
r^I \coloneqq W\delta(I) 
\end{align}
for the total recurrent activation distributed in an avalanche with assembly \(I\).

\begin{prop}\label{prop:rav}
  For \(a \in \mathcal{A}_k\), let $I_j = \U_j(\av)$ for $j=1,\ldots,\dur(a)$ and $I=\U(\av)$. We have
  \begin{align}
    R(a) &= [\VU_{a_1}-\DeltaU ,\VU_{a_1})_{a_1} \cap R_\U(a) \cap R_{\U^c}(I) 
  \end{align}
  with
    \begin{align}
    R_\U(a)
			&\coloneqq \bigcap_{j=2}^{\dur(a)}\left [\VU_i - r^{I_{j-1}}_i,
			\VU_i - r^{I_{j}}_i \right  )_{i\in a_j} \nonumber \\
		R_{\U^c}(I)
			&\coloneqq \left [0,\VU_i-r^I_i\right )_{i\in [N]\setminus I}
			\setminus \Lambda_{[N]\setminus I}  \nonumber \text{ .}
  \end{align}
\end{prop}

\begin{proof}
	Since \(a_1 = \{k\}\), it follows that
	\[
		A(u+ \DeltaU\e_k )_{k} = 
		1 \iff u_{k} + \DeltaU \geq \VU_k \iff u_{k} \in [\VU_k-\DeltaU ,\VU_k) \text{ .}
	\]
	Using Lemma~\ref{prop:at-most-once}, the condition that unit \(i\) spikes only in step \(j\) in the avalanche,
	% \(k\in \av _j\) 
	$i\in a_j$
	reduces to \(A(F^{j-1}(u + \DeltaU \e_k) ))_i = 1\), and we have
	% \[(F^{j-1}(u + \DeltaU \e_k ))_i \geq 1 > (F^{j-2}(u + \DeltaU \e_k ))_i \text{ .}\]
	\[
		(F^{j-1}(u + \DeltaU \e_k ))_i \geq U_i > (F^{j-2}(u + \DeltaU \e_k ))_i \text{ .}
	\]
	This is fulfilled if and only if
	\(\VU_i-r^{I_j}_i \leq u_i < \VU_i - r^{I_{j-1}}_i\).
	Similarly, unit \(u_l\) does not fire in the avalanche if and only if
	\(A(F^{i-1} (u + \DeltaU \e_k) )_l = 0\) for all \(i=1,\ldots,\dur(a) \),
	so \(u_l\in \left[0,\VU_l-r^I_l\right)\).
	Using Lemma~\ref{lem:elimdim}, we have
	\begin{align*}
		R(a) 
			&= \left[\VU_k-\DeltaU ,\VU_k\right)_{\{k\}} \cap R_\U(a) \cap
				\left[0,\VU_i-r^I_i\right)_{i\in [N]\setminus I} \cap \Lambda_{[N]} \\ 
			&= \left[\VU_k-\DeltaU ,\VU_k\right)_{\{k\}} \cap R_\U(a) \cap
				\left[0,\VU_i-r^I_i\right)_{i\in [N]\setminus I} 
				\cap \Lambda_{[N]\setminus I} \text{ ,}                                       
	\end{align*}
	since \(\VU_i-r^I_i \geq \sum_{l\in [N]\setminus I}w_{il}\) for \(i\in [N]\setminus I\).
	
\end{proof}

\subsection{Avalanche distributions}\label{sec:app-avalanche}

To arrive at probabilities \(\P(\av = a)\), the volumes of the preimage \(R(a)\) has to be normalized by the volume of the region where external input does not result in an avalanche, i.e. the preimages for which \(\av(u,a_1) = ()\). The following Lemma specifies these regions and the probability of an empty avalanche given external input to unit \(k\in [N]\).

\begin{lem}[Empty avalanches]\label{lem:empty}
  The phase space region on which external input to unit \(k\) leads to the empty avalanche is given by
  \begin{align}\label{eq:empty-region}
    % \{u \in D | \av(u,k) = ()\} = [0,U_j-\DeltaU \e_k^T\e_j )_{j\in [N]} \cap \Lambda_{[N]\setminus \{k\}} \text{ ,}
    \{u \in D  \mid  \av(u,k) = ()\} = [0,U_j-\DeltaU \delta_{k,j} )_{j\in [N]} \setminus \Lambda_{[N]\setminus \{k\}} \text{ ,}
  \end{align}
  and for the probability of the empty avalanche given external input to unit \(k \) we have
  \begin{align}\label{eq:prob-empty-avalanche}
    \P(\av(u,\omegaVk_1) = ()  \mid  \omegaVk_1= k)
		&= \frac{\V_{[N]}(\VU-\DeltaU \e_{k})}{\V_{[N]}(\VU)}
			= 1 - \frac{\DeltaU \V_{[N]\setminus \{k\}}(\VU)}{\V_{[N]}(\VU)} \text{ .} 
		\end{align}
\end{lem}
\begin{proof}
  The empty avalanche results upon external input to unit $k$ if and only if \(u_k < \VU_k-\DeltaU\). Since
  \(\VU_k-\DeltaU > \sum_{\ell\in [N]\setminus \{k\}}w_{k\ell}\) we can apply Lemma~\ref{lem:elimdim} to
  get~\eqref{eq:empty-region}. \Eq{eq:prob-empty-avalanche} follows from~\Eq{eq:empty-region} with~\Eq{eq:vin}.
\end{proof}

\begin{prop}[Relation between $\P(a)$ and $\P_k(a)$]\label{prop:p-pk}
  For \(a\in \mathcal{A}_k,k\in L\) we have
  \begin{align*}
    \P_k(\av=a) 
		&= \frac{\P(\av=a \mid \omegaVk_1=k)\V_{[N]}(\VU)}{\DeltaU \V_{[N] \setminus \{k\}}(\VU)}
		= \frac{\P(\av=a)\V_{[N]}(\VU)}{p_k\DeltaU \V_{[N] \setminus \{k\}}(\VU)}
  \end{align*}
\end{prop}

\begin{proof}
	The condition that \(\av(u,\omegaVk_1) = a\) is equivalent to \(u\in R(a)\) and \(\omegaVk_1 = k\). By definition
  \(\P_k(\av = a) = \P(\av = a \mid  a_1 = \{k\})\) and the condition \(a_1 = \{k\}\) is equivalent to \(a \neq (),\omegaVk_1 = k\).
  Thus, we have \(\P_k(\av = a) = \frac{\P(\av=a\mid\omegaVk_1 = k)}{\P(\av \neq () \mid \omegaVk_1 = k)} =
	\frac{\P(\av=a\mid\omegaVk_1 = k)}{1 - \P(\av = () \mid \omegaVk_1 = k)}\)
	and using Lemma~\ref{lem:empty} this results in the first equality
  \[
		\P_k(\av=a) = \frac{\P(\av=a\mid\omegaVk_1=k)\V_{[N]}(\VU)}
			{\DeltaU \V_{[N] \setminus \{k\}}(\VU)} \text{ .}
	\]
  The second equality follows from \(\P(\omegaVk_1 = k) = p_k\).
\end{proof}

Proposition~\ref{prop:p-pk} allows to transform between \(\P(\av = a)\) and \(\P_k(\av = a) \). In the following we will
calculate the latter probabilities. Note that they depend neither on \(\DeltaU\) nor \(p_k\) but only on the coupling matrix \(W\).

\begin{thm}[Avalanche distributions]\label{theo:avdist}
  The probability distribution for a nonempty avalanche \(a \in \mathcal{A}_k,k\in L,\U(\av) = I\) is given by
  \begin{align}\label{eq:pav}
    \P_k(\av = a) = \frac{\V_{[N]\setminus \U(a)}\left(\VU-r^I\right)}
			{\V_{[N]\setminus \{k\}}(\VU)}
			\prod_{j=2}^{\dur(a)}\prod_{i\in a_j}\sum_{l\in a_{j-1}}w_{il} 
  \end{align}
\end{thm}

\begin{proof}
  Using Proposition~\ref{prop:rav}, the \(N\)-dimensional Lebesgue volume of \(R(a)\) is given by
  \begin{align*}
    \lambda(R(a)) &= \lambda\left(\left[\VU_{a_1}-\DeltaU ,\VU_{a_1}\right)_{a_1}\right) \lambda(R_\U(a)) \lambda(R_{\U^c}(I)) \\
                  &= \DeltaU  \V_{[N]\setminus \U(a)}\left(\VU-r^I\right)  \prod_{j=2}^{\dur(a)} \prod_{i\in a_j}\sum_{l\in a_{j-1}}w_{il} \text{ ,}
  \end{align*}
  where Corollary~\ref{cor:volinhabited} was used to compute the volume of \(R_{\U^c}(I) \).\\
  % \[
  %       	\lambda(R(a)) = \DeltaU  \V_{[N]\setminus \U(a)}(\VU-W\delta(\U(a)))  \prod_{j=2}^{\mathcal{D}(a)} \prod_{i\in a_j}\sum_{\red{l}\in a_{j-1}}w_{i\red{l}}
  %       		 \text{ .}
  %       \]
  This leads to
	% \( \P(\av= a \mid  \omegaVk_1 = k) = \lambda(R(a))/\V_{[N]}(\U)\) 
	\(\P(\av= a \mid  \omegaVk_1 = k) = \lambda(R(a))/\V_{[N]}(\VU)\) 
	and with Proposition~\ref{prop:p-pk} to
  \begin{align*}
		\P_k(\av = a) = \frac{\V_{[N]\setminus \U(a)}\left(\VU-r^I\right)}{\V_{[N]\setminus \{k\}}(\VU)}
		\prod_{j=2}^{\dur(a)}\prod_{i\in a_j}\sum_{l\in a_{j-1}}w_{il} 
  \end{align*}

\end{proof}

Avalanches \(a\) with the same set of participating units \(\U(a)\) thus have the same volume
along the \([N]\setminus \U(a)\) dimensions. We will derive a closed form expression
for the phase space volume of the union of all such avalanches.
% \udo{Die Struktur der Gleichungen ist durch die vielen Klammern oft nicht auf den ersten
% Blick zu erfassen. Insbesondere $(W\delta(I))_k$ wiederholt sich oft. Moeglicherweise waere es
% doch sinnvoll, hier eine zusaetzliche Variable einfzufuehren und zu benennen, z.B. den rekurrenten
% Input zu Unit $k$ bei Feuern der Units in $I$, $r_k(I)$ oder $r_k^I$ (hiermit haetten wir auch
% dem RKI in diesen Corona-Zeiten ein Denkmal gesetzt!}

% \ms{TODO: Hier Beweis ausfuehrlicher aufschreiben und $W(\delta(I))_k$ durch $r_k^I$ ersetzen}
\begin{thm}\label{thm:avu}
  The $I\setminus \{k\}$ components of the images of all avalanche regions \(R(a)\) with avalanche units \(\U(a)=I\) and
  started by unit \(k\) fill up the inhabited region along dimensions \(I \setminus\{k\}\)  up to the upper boundaries \(r(I)+\DeltaU\delta(\{k\})\):
  \begin{eqnarray*}
     \biguplus_{a \in \mathcal{A}_k,\U(a) = I} T_k(R(a)) &=& A \quad\text{with:}\\
    A &\coloneqq& \left[r^I_k,r^I_k+\DeltaU \right)_{\{k\}} \, \cap \, \left ( \left[0,r^I_i\right)_{i\in I\setminus \{k\}} \setminus \Lambda_{I\setminus \{k\}}\right ) \\
                                                         &\cap& \{u \mid \left(u-r^I\right)_{[N]\setminus I} \in \pi_{[N]\setminus I}(R_{\U^c}(I))\}
 \text{, }
  \end{eqnarray*}
  and the distribution of avalanche assemblies started by unit \(k\) is given by
  \begin{align}\label{eq:pavu}
    \P_k(\U(\av) = I)
		= \frac{\V_{I\setminus \{k\}}(r^{I})\V_{[N]\setminus I}(\VU-r^{I}))}
			{\V_{[N]\setminus \{k\}}(\VU)} \text{ .}
  \end{align}
\end{thm}

\begin{proof}
  By injectivity of \(T_k\), \(T_k(R(a))\) are disjoint for all \(a\in \mathcal{A}_k\).
        % We denote the right hand side by \(A\).
  First, we show that
  \[
        	\biguplus_{a \in \mathcal{A}_k,\U(a) = I} T_k(R(a))  \subseteq A \text{ .}
        \]
  By \Eq{cor:total-action}, \(T_k \) induces a shift by \(\DeltaU  \e_k - M^{-1}\delta(I) = \DeltaU \e_k - \diag(U)\delta(\U(a)) + r^{I}\) 
  on all states $u$ in \(R(a)\) with \(a  \in \mathcal{A}_k \). %, \U(a) = I \).
  Since \(\pi_{\{k\}}R(a)= [U_k-\DeltaU,U_k)\) for all \(a\in \mathcal{A}_k\) we have
  \[
        	\pi_{\{k\}}T_k(R(a)) = [r^I_k \, , \, r^I_k+\DeltaU ).
        \]
        The states of all the remaining units which do not participate in the avalanche are just shifted by
        \(r^I_{[N]\setminus I}\), so that for all %\(a\in \A_k\)
        \(\pi_{[N]\setminus I}T(R(a)) = \pi_{[N]\setminus I}(R_{\U^c}(I)+r^I)\).
        Finally, the states of all \(i\in I\setminus \{k\}\) must be sufficiently close to
	the threshold such that the recurrent input makes the units fire,
  \(\VU_i-r^{I}_i \leq u_i < \VU_i\), and thus
  \[\pi_{I\setminus \{k\}}T_k(R(a)) \in \pi_{I\setminus \{k\}}([0,r^I_i)_{i\in I \setminus \{k\}} \setminus \Lambda_{I\setminus \{k\}})\text{ .}\]

  which completes the proof of
              \(\biguplus_{a \in \mathcal{A}_k,\U(a) = I} T_k(R(a))\subseteq A\).
	% \begin{align*}
	% 	[(W\delta(\U(a)))_k \, , \, (W\delta(\U(a)))_k+\DeltaU )_{\{k\}}
	% 	\cap [0,(W\delta(\U(a)))_i)_{i\in I \setminus \{k\}} \setminus \Lambda_{[N]} = \\
	% 	[(W\delta(\U(a)))_k \, ,\, (W\delta(\U(a)))_k+\DeltaU )_{\{k\}}
	% 	\cap \left ( [0,(W\delta(\U(a)))_i)_{i\in I \setminus \{k\}} \setminus \Lambda_{I\setminus \{k\}} \right )
	% \end{align*}Finall

        We continue to show that       
	\[
		A \subseteq \biguplus_{a \in \mathcal{A}_k,\U(a) = I} T_k(R(a)) \text{ .}
	\]
        Since \(A \subset D\) but \(u-\e_k\DeltaU \in \Gamma_I \subseteq \Lambda_I\) for \(u \in A\) we have
        \(k\in \tilde{I}\coloneqq \U(\av (\tilde{u},k))\), where \(\tilde{u} = T_k^{-1}(u)\).
        We will show \(\tilde{I} = I\) by contradiction.\\
        Suppose that \(\tilde{I}\neq I\), then either \(I \setminus \tilde{I}\) or \(\tilde{I} \setminus I\) has to be nonempty.
        We proceed by a case distinction:\\
        Let \(j\in I \setminus \tilde{I} \neq \emptyset \) be arbitrary. We have
	\[
		u_j = \tilde{u}_j + r^{\tilde{I}}_j < r^I_j
		\implies \tilde{u}_j < r^{I\setminus \tilde{I}}  \text{, }
	\]
	thus \(\tilde{u} \in \Gamma_{I\setminus \tilde{I}}\) and hence \(\tilde{u} \notin D\).
        This is a contradiction.\\
        Now let \(m\in \tilde{I}\setminus I \neq \emptyset\) be arbitrary.
        We have \(u_m = \tilde{u}_m - \VU_m + r^{\tilde{I}}< r^{\tilde{I}}\) and thus 
        % However, \(u\in A \) implies
        % \[(u-r^I)_{i\in [N]\setminus I} \in R_{\U^c}(I)_{i\in [N]\setminus I} \implies (u-r^I)_{i\in [N]\setminus I}<  \VU_i-r^I_i\]
        % and we have
	\[
		u_m - r^I_m < r^{\tilde{I}}_m - r^I_m < r^{\tilde{I}\setminus I}_m \text{, }
              \]
              
              which implies that \((u- r^I)_{\tilde{I}\setminus I} \in \pi_{\tilde{I}\setminus I} \Gamma_{\tilde{I}\setminus I} \) and
              contradicts $(u-r^I)_{[N]\setminus I} \in \pi_{[N]\setminus I}R_{\U^c}(I)$.
              This completes the proof of 
	\[
		A  = \biguplus_{a \in \mathcal{A}_k,\U(a) = I} T_k(R(a)) \text{ .}
              \]

              % To show \Eq{eq:pavu}, we have
              % \begin{align*}
              %   \P(\U(\av) = I,\omegaVk_1 = k) &= \frac{\sum_{a \in \A_k,\U(a)=I} \lambda(R(a))}{\V_{[N]}(U)} \\
              %   &= \frac{\sum_{a \in \A_k,\U(a)=I} \lambda(T(R(a)))}{\V_{[N]}(U)}\\
              %   &= \sum_{a \in \A_k,\U(a)=I} \lambda([0,r^I_i)_{i\in I \setminus \{k\}) \frac{\DeltaU \V_{[N]\setminus I}(\VU-r^I)}{\V_{[N]}(U)}
              % \end{align*}

              % \[
              %   A \setminus \Lambda_{[N]} = (([r^I_k,r^I_k+\DeltaU )_{\{k\}} \, \cap \, [0,r^I_i)_{i\in I\setminus \{k\}}) \setminus \Lambda_I) \cap (\{u \mid (u-r^I)_{[N]\setminus I} \in (R_{\U^c}(I))_{[N]\setminus I} \setminus \Lambda_{[N]\setminus I})
              % \]
              With Corollary~\ref{cor:volinhabited} we have
              \begin{align*}
                \P(\U(\av) = I,\omegaVk_1 = k) &= \lambda(A)/\V_{[N]}(\VU) = \frac{\DeltaU \V_{I\setminus \{k\}}(r^I)\V_{[N]\setminus I}(\VU-r^I)}{\V_{[N]}(\VU)}
              \end{align*}
              \Eq{eq:pavu} follows with Proposition~\ref{prop:p-pk}.

\end{proof}

\subsection{Relation to graph topology}\label{sec:app-topology}

\subsubsection{Graph properties determine phase space volumes and avalanche probabilities}

In addition to the geometrical proof of \Eq{eq:pavu}, we give a combinatorial proof invoking Kirchhoff's theorem which generalizes the corresponding proof for the homogeneous EHE model \cite{denker2014ergodicity}.

\begin{proof}[Combinatorial proof of \Eq{eq:pavu}:]
	\label{proof:comb}
	From Theorem~\eqref{theo:avdist} we have 
	\begin{align}
		\P_{k}(\U(\av) = I)
		&= \sum_{a\in \mathcal{A}_{k}, \U(a)=I}\P_k(\av = a) \\
		&= \frac{\V_{[N]\setminus I}(\VU-r^I)}{\V_{[N]\setminus \{k\}}(\VU)}
			\sum_{a \in \mathcal{A}_{k}, \U(a)=I}\prod_{j=2}^{\dur(a)}\prod_{i\in a_j}\sum_{l\in a_{j-1}}w_{il}
			\text{ .}
	\end{align}
	In order to show \Eq{eq:pavu}, we need to show
	\begin{align*}
		\sum_{a \in \mathcal{A}_{k}, \U(a)=I}\prod_{j=2}^{\dur(a)}
			\prod_{i\in a_j}\sum_{l\in a_{j-1}}w_{il}
		&= \V_{I\setminus \{k\}}\left(r^I\right) \\
		&= \left \lvert(\diag\left(r^I\right)_{I\setminus \{k\}})-W_{I\setminus \{k\}}\right \rvert \text{ .}
	\end{align*}
	The right hand side is the \((k,k)\) cofactor of the \(W_I\) graph Laplacian.
	By Kirchhoff's Theorem \cite{chaiken1978matrix}, this determinant equals the number of spanning trees
	rooted at \(k\), weighted by the product of weights along their arcs.
	There is a natural correspondence between an avalanche
	\(a = (a_i)_{i=1,\ldots,\dur(a)}, a_1=\{k\},
	\biguplus_{i=1}^{\dur(a)} a_i = I\)
	and spanning trees of the vertices \(I \) rooted at \(k\).
	For \(j\in I\), \(j\in a_s\) corresponds to vertex \(j\) being separated
	from the root \(k\) by \(s \) steps. In this way, \(a \in \mathcal{A}_{\omegaVk _1}, \U(a)=I\)
	partition the spanning trees by their level-structure, i.e. which sets of units are separated
	from the root by the same number of steps. Let \(\mathcal{T}(I,k)\) denote the set of weighted
	spanning trees in \(I\) rooted at \(k\). What remains to be shown is that 
	\[
		\prod_{j=2}^{\dur(a)}\prod_{i\in a_j}\sum_{l\in a_{j-1}}w_{il}
		= \sum_{\substack{t\in \mathcal{T}(I,k) \\ \operatorname{dist}(i,k) = j
		\text{ for } i\in a_j}}\prod_{(i,j)\in t}w_{ij} \text{.}
	\]
	By expanding the products on the left hand side iteratively, we enumerate all ways
	to connect elements in level/generation \(j-1\) with elements in level/generation \(j\)
	weighted by the corresponding edge weight. Thus, we have
	\[
		\sum_{a\in \mathcal{A}_{k}, \U(a)=I}\prod_{j=2}^{\dur(a)}
			\prod_{i\in a_j}\sum_{l\in a_{j-1}}w_{il}
		= \sum_{t\in \mathcal{T}(I,k)}\prod_{(i,j)\in t}w_{ij}
		= \left| \diag\left(\sum_{j\in I}w_{ij}\right)_{i\in I\setminus \{k\}} - W_{I\setminus \{k\}} \right|
	\]
\end{proof}

% In this section 
Next we expand on the implications of the equations \Eq{eq:pav}, \Eq{eq:pavu}, and \Eq{eq:vin}.
% To simplify the statements, we assume without loss of generality that \(\VU_i=1\) for all \(i\).
% \udo{Das ist in einigen Formeln im Folgenden nicht der Fall, da ist wieder $U$ drin...}

% \begin{cor}[Phase space volume in dependence of eigenvalues]
%   \label{cor:eig}
%   Let \(\VU_i=1\) for all \(i\) and let \(\lambda_1,\ldots,\lambda_N\) be the (possibly zero) eigenvalues of
%   \(W\). The volume of the inhabited region depends on the eigenvalues as follows:
%   \begin{align}
%     \lambda(D) = \V_{[N]}(\mathbf{1}) = |\mathds{1}-W| = 1 + \sum_{k=1}^N(-1)^k\sum_{I\subseteq [N],|I|=k}\prod_{i\in I}\lambda_i
%   \end{align}
%   In particular \(\lambda(D) = 1 - \lambda_1\) if \(W\) has rank one, and \(\lambda(D) = 1 - \lambda_1 - \lambda_2 + \lambda_1 \lambda_2\) if \(W\) has rank two.
% \end{cor}

% \begin{proof}
%   Let \(C(\lambda) = |\lambda \mathds{1} - W| = \prod_{i=1}^N(\lambda - \lambda_i)= \lambda^N + a_1 \lambda^{N-1} + \ldots + a_N  \) be the characteristic polynomial of \(W\). Using~\cite{BROOKS2006511}, the coefficient \(a_k\) is given by
%   \(a_k = (-1)^k\sum_{I\subseteq [N],|I|=k}\prod_{i\in I}\lambda_i\). The corollary follows from
%   \(|\mathds{1}-W| = C(1) = 1 + \sum_{i=1}^N a_i\).
% \end{proof}

\begin{cor}[Phase space volume in dependence of loops]
  \label{cor:loop}
  Let \(\VU_i=1\) for all \(i\).
  The volume of the inhabited region depends on the set of all linear directed subgraphs \(\mathscr{L}\) of \(W\) 
  weighted by the product of their arc weights. Every component of a linear directed subgraph is a directed cycle.
  Let \(\mathscr{L}(W)\) be the set of all linear directed subgraphs \(\mathcal{L}\) of \(W\) with \(i\) nodes and
  \(\#(\mathcal{L})\) be its number of components. The product of all arc weights in \(\mathcal{L}\) is denoted by
  \(\prod (\mathcal{L})\).
  \begin{align}\label{eq:ivw2}
    \lambda(D) = \V_{[N]}(\mathbf{1}) = |\mathds{1}-W| = 1 + \sum_{i=1}^N\sum_{\mathcal{L}\in\mathscr{L}_i(W)}(-1)^{\#(\mathcal{L})}\prod (\mathcal{L}) 
  \end{align}
\end{cor}

\begin{proof}
  \Eq{eq:ivw2} follows from the combinatorial interpretation of the characteristic polynomial of a weighted
  digraph \(W\) \cite{cvetkovic1980spectra}[Section 1.4]
  \begin{align}
    \label{eq:char-poly}
    |\lambda \mathds{1} - W| = \lambda^N + a_1\lambda^{N-1}+\ldots +a_N \text{, }
  \end{align}
  with \(a_i = \sum_{L\in \mathcal{L}_i} (-1)^{\#(L)}\prod (L)\). 
\end{proof}

\begin{prop}[Impact of self-loops on the inhabited phase-space volume]
\label{prop:self-weights}
  Let $W'$ be the coupling matrix equivalent to $W$ without self-loops, \(W' = W-\diag(W) \).
  \begin{align}
    \V_{W}(\VU,I) = \V_{W'}(\VU-\diag(W),I) \text{ ,}
  \end{align}
  and
  \begin{align}
    [0,\VU_i)_{i\in [N]} \setminus \Lambda_{[N]}^W
    = \{u+\diag(W)  \mid  u \in  [0,\VU_i-w_{ii})_{i\in [N]}\setminus \Lambda_{[N]}^{W'} \}
  \end{align}
\end{prop}

\begin{proof}
  The first equation follows immediately from corollary~\ref{cor:volinhabited} by
  \[
		\V_{W}(\VU,I) = |\diag(\VU)_I - W_I|
		= |\diag(\VU) - \diag(W)_I - W'_I| =  \V_{W'}(\VU-\diag(W),I) \text{ .}
	\]
  The second identity follows from the effect of adding diagonal entries to each \(\Gamma_I\),
  \[
		\Gamma^{W'+\diag(W)}_I = \left [0,w_{ii}+\sum_{j\in I,j \neq i}w_{ij} \right )_{i\in I} \text{ .}
	\]
\end{proof}

% \begin{rem}
%   Proposition~\ref{prop:self-weights} describes the effect of self-weights on the inhabited region by a series decomposition when expanding the product on the right hand side. If all diagonal entries added are the same \(w_d\) i.e. \(W_{i,i} = w_d\), and  we set \(\VU_i = V\) for all \(i\in I \)
%   \[\V_{W'}(U,I) = \V_W(U,I) + \prod_{n=1}^{|I|}(-1)^{n-1}V^{|I|-n}w^n_d \text{ .}\]
%   In general, one can approximate the effect of self-weights by expanding the product up to terms containing at most \(k\) multiplicants from \(W\) and neglect all higher order \(W\) terms.
% \end{rem}

\subsubsection{Stochastic dependencies between units}
% {states of different units}

In addition to firing-rate correlations, we can use the geometric
structure of the inhabited volume to analyze stochastic dependencies
between the states of units in relation to the network topology.
We denote the set of units forming the strongly connected components
of the graph with adjacency matrix \(W\) by \(\operatorname{scc}(W)\).

\begin{thm}
  \label{thm:unit-correlations}
  For every \( H\subseteq [N]\), the inhabited region decomposes into a direct product
	of inhabited regions along the strongly connected components of the subgraph $H$
	with adjacency matrix \(W_H\). 
	% \ms{Is $\bigtimes$ the correct notation here?, all sets on the right are elements of \(\R^{[N]}\).
	% I think $\bigcap$ is the correct operator.},
	% \udo{Was soll $\bigtimes$ denn bedeuten?}
  \begin{align*}
		& \!\!\!\!\!\!\!\ [0,\VU_h)_{h\in H} \setminus \Lambda_H
		= \bigcap_{J \in \operatorname{scc}(W_H)} [0,\VU_j)_{j\in J}\setminus \Lambda_J.
  \end{align*}
\end{thm}

\begin{proof}
  We denote the right hand side by A.
  \([0,\VU_h)_{h\in H} \setminus \Lambda_H \subseteq A\) is trivial since
  \[
		\bigcup_{I\in \operatorname{scc}(W_H)} \Lambda_I \subseteq \bigcup_{J\subseteq H}\Gamma_H
		= \Lambda_H \text{ .}
	\]
  To show \(A \subseteq [0,\VU_h)_{h\in H} \setminus \Lambda_H\), it now suffices to show
  \(\V(A) = \V\left(\left[0,\VU_h\right)_{h\in H} \setminus \Lambda_H\right)\), since the complement of \(A\) in $C$
	is a union of cylinder sets. We have
  \[
		\V(A) 
		= \prod_{J\in \operatorname{scc}\left(W_H\right)}\rvert\diag\left(\VU_J\right)-W_J\lvert
		= \rvert\diag\left(\VU_H\right)-W_H\lvert = \V\left(\left[0,\VU_h\right)_{h\in H} \setminus \Lambda_H\right) \text{, }
	\]
  where the second equality holds since \(\diag(\VU_H)-W_H\) can be reordered to form an upper
  triangular block matrix with respect to the strongly connected components 
\end{proof}

The direct product structure implies stochastic independence between units in different strongly connected
components:

\begin{cor}
	For two index sets \(I,J \subseteq [N]\) which do not share a common strongly connected component
	in the graph with adjacency matrix \(W\), the components \(u_I = \pi_I u \) and %\red{\coloneqq \delta(I)^T u}
	\(u_J = \pi_J u\) are stochastically independent with respect to the measure \(\PW\).
\end{cor}

\begin{proof}
  We have to show that the multivariate random variables $u_I$ and $u_J$ are independent.
  The cumulative distribution functions of $u_I$ and $u_J$ are obtained by marginalizations of \(\PW\).
  Since \(\PW\) is the normalised Lebesgue measure supported on \(D\),
  independence follows if \(D\) factorizes into a product of subspaces
  and no two units \(i\in I,j\in J\) share a common subspace. This is ensured by
  Theorem~\ref{thm:unit-correlations} if $I$ and $J$ do not share a common
  strongly connected component in the graph with adjacency matrix \(W\).
\end{proof}

\subsubsection{Avalanche branching process}

In this section, we study how the transition probabilities from step \(i \) to step \(i+1 \) during an avalanche are influenced by network
topology. 

% \begin{defn}
%   Each unit \(i\) can either be off, active or refractory. Consider an
%   avalanche \(\av \in \mathcal{A}\). At step
%   \(2 \leq j\leq |\av |\), these sets are given by
%   \begin{align*}
%     R &= \biguplus_{i=1}^{j-2}\av _i \\
%     A &= \av _{j-1}\\
%     O &= [N]\setminus (A \uplus R) 
%   \end{align*}
%   We split the set of units in the off state into two other sets
%   \(O = O^+\uplus O^- \), with \(O^+ \) being the units that
%   transition to the active state
%   \[O^+ = \av _j = O \setminus O^-\]
% \end{defn}

% Hier war ein off by one error drin! \U_{j-2} muss zu \U_{j-1} werden und so weiter!
% hab das jetzt geändert

\begin{thm}
  \label{thm:branching}
  Let \(a \in \mathcal{A}\) with \(\P(\av=a) > 0\) and let $I_j = \U_j(\av)$ for $j=1,\ldots,\dur(a)$ and $I=\U(\av)$. 
  The probability of the generation \(\av_j\) of an avalanche for 
  \(2 \leq j \leq \dur(a)\) given the previous steps of the avalanche is
	\begin{align}\label{eq:branching-theorem}
		\P(\av_j=a_{j} \mid \av_1=a_{1},\ldots,\av_{j-1}=a_{j-1}) = \prod_{k \in a_j}\left (\sum_{\ell\in a_{j-1}}w_{k\ell} \right )
			\times \frac{\V_{[N]\setminus I_j}\left(\VU - r^{I_{j-1}}\right)}{
			\V_{[N] \setminus I_{j-1}}\left(\VU  - r^{I_{j-2}}\right)}
	\end{align}
%  The probability that units $J\in [N]\setminus \U_{j-1}(a)$ fire in step $j$ of the avalanche is 
%	\begin{align}\label{eq:branching-theorem-J}
%		& \!\!\!\!\!\!\!\!\!\!\!\!\!\!\!\!  \P(J\subseteq \av_j|\av_1=a_{1},\ldots,\av_{j-1}=a_{j-1}) \nonumber\\
%		&= \prod_{k \in J}\left (\sum_{\ell\in a_{j-1}}w_{k\ell} \right )
%			\times \frac{\V_{[N]\setminus (\U_{j-1}(a)\uplus J)}(\VU - W\delta(\U_{j-1}(a)))}{
%			\V_{[N] \setminus \U_{j-1}(a)}(\VU  - W\delta(\U_{j-1}(a)))}
%	\end{align}
\end{thm}

\begin{proof}
  The region of states consistent with \(\av_1=a_1,\ldots,\av_{j-1}=a_{j-1}\) is like in Proposition~\ref{prop:rav} given by a hyperrectangle
  along dimensions \(I_{j-1}\), while the remaining coordinates \(O = [N]\setminus I_{j-1}\)
	are below their firing thresholds:
  \begin{align*}
  \left [\VU_{a_1}-\DeltaU,\VU_{a_1} \right )_{a_1} \cap \bigcap_{\ell=2}^{j-1} \left [\VU_i - r^{I_{\ell - 1}}_i,\VU_i - r^{I_{\ell}}_i\right )_{i\in a_\ell} 
  \cap \left [0,\VU_i-r^{I_{j-1}}_i\right)_{i\in O} \setminus \Lambda_O
  \end{align*}
Specifying which units fire at the next step of the avalanche leads to a smaller consistent region of states, which is the same along the dimensions $U_{j-1}(a)$ but splits up the region of states along the dimensions $O$ into 
\[\left [\VU_i-r^{I_j}_i,\VU_i-r^{I_{j-1}}_i \right)_{i\in a_j} \cap \left [0,\VU_i-r^{I_{j-1}}_i \right )_{i\in O\setminus a_j}\setminus \Lambda_{O\setminus a_j} \text{ ,}\]
since exactly the units in $a_j$ cross the firing threshold.

%With the same reasoning, the region along dimensions $O$ is split up by the event $J\in a_j$ into 
%\[\left [\VU_i-(W\delta(\U_{j-1}(a))\uplus J)_i,\VU_i-(W(\delta(\U_{j-1}(a))))_i \right)_{i\in J} \cap \left [0,\VU_i-(W\delta(\U_{j-1}(a)))_i \right )_{i\in O\setminus J} \text{ ,}\]
%where the original upper boundaries remain along dimensions $O\setminus a_j$.
The probability \Eq{eq:branching-theorem} is thus the quotient of the consistent volumes along dimensions $O$ which follow by using Corollary~\ref{cor:volinhabited}. 
%\[
%		\bigg[\VU_i-\sum_{k\in \U_{j-2}(a)}w_{ik},\VU_i\bigg)_{i\in O}\setminus \Lambda(W,O) 
%		\text{ .}
%	\] 
%	The condition for the units in \(O\setminus a_j\) to become active in the next step is that
%  the current activation from the units in \(a_j\) pushes them over the threshold.
%	This condition is met exactly on the following region
%  \begin{align*}
%		\bigg[\VU_i-\sum_{k\in \U_{j}(a)}w_{ik} \,
%			, \, \VU_i-\sum_{k\in \U_{j-1}(a)}w_{ik}\bigg)_{i\in a_j} 
%			, \, \VU_i\bigg)_{i\in O\setminus a_j} \setminus \Lambda(W,O\setminus a_j) \text{ .}
%  \end{align*}
	% \udo{Die langen Summenindex-Specs separieren die Ausdruecke optisch viel staerker als das trennende
	% Komma zwischen der unteren/oberen Intervallgrenze. Kann man das verbessern? Z.B. Summenindex in zwei
	% Zeilen umbrechen?}
  %The probability $\P(\av_j=a_j|\av_1=a_1,\ldots,\av_{j-1}=a_{j-1})$ is given by the quotient
%	of the Lebesgue volumes of these sets, which leads to \Eq{eq:branching-theorem}.
\end{proof}

This branching process needs memory of which units are refractory.

\subsection{Application to structurally simple networks}\label{sec:app-simple}
In this section we apply our framework to homogeneous and non-homogeneous networks with regular structures whose symmetries allow to simplify the measures and distributions derived in this paper. In consequence the avalanche size distributions can be given 
in closed form since assembly probabilities of a given size do not depend on the detailed assembly subgraph(s) but only on few global parameters. In this section we set \(\VU = \mathbf{1}, p = (1/N) \mathbf{1}\) unless otherwise specified.

\subsubsection{Homogeneous network}\label{sec:app-homogeneous}
The homogeneous network is the classical setting for the EHE-model, which was introduced and analyzed in \cite{eurich2002finite}.
In the following we will describe in detail how the known avalanche size distribution and its expected value \cite{levina2014abelian} arise naturally from our framework when the coupling matrix is homogeneous.  

Let \(W^{\text{hom}} = (w_{ij})_{i,j\in [N]}\), with \(w_{ij} = \frac{\alpha}{N} \) for all \(i,j\in [N]\) and \(\alpha +\DeltaU < 1\). We use the shorthand \(\P = \P^{W^{\text{hom}}}\) in this subsection.

For the special choice of \(W^{\text{hom}}\), the inverse \(M^{\text{hom}} = \left(1-W^{\text{hom}}\right)^{-1}\) is given in closed form by
\[M^{\text{hom}} = \mathds{1}  + \frac{1}{1-\alpha}W^{\text{hom}} \text{ .}\]
Thus, the probability that unit \(i\)  fires in an avalanche started by unit \(k\) is given by \Eq{eq:pavu_inline}
\[\P_k(i\in \U(\av)) = \frac{M^{\text{hom}}_{ik}}{M^{\text{hom}}_{kk}} = \frac{\alpha}{N-(N-1)\alpha} \text{ .}\]

The mean firing rate of the homogeneous network is
\[Y_W = \DeltaU M^{\text{hom}} p =  \left(1+\frac{\alpha}{1-\alpha}\right)\frac{\DeltaU}{N}\mathbf{1} = \frac{\DeltaU}{N(1-\alpha)}\mathbf{1} \text{ .}\]

The mean nonempty avalanche size is given by \Eq{eq:mavs}
\begin{align}\label{eq:mavs-hom-with-selfweights}
\mathbb{E}(\s(\av) \mid \av_1 = \{k\}) = 1 + \frac{\sum_{j\in[N]\setminus \{k\}} M_{jk}}{M_{kk}} = 1 + \frac{(N-1)\alpha}{N-(N-1)\alpha} = \frac{N}{N-(N-1)\alpha}
\end{align}
In order to calculate the avalanche size distributions, we start to simplify the expression for the volume of the inhabited region
\(\lambda(D) = \V_{[N]}(\mathbf{1}) = |\mathds{1}-W^{\text{hom}}|\).
Since \(W^{\text{hom}}\) has only one nonzero eigenvalue equal to \(\alpha\), Equation~\Eq{eq:psvev} gives
$\lambda(D) = 1-\alpha$.

The more general expression \(\V_{I}(\VV)\), with constant vector \(\VV_i = v \) simplifies similarly to
\begin{align}\label{eq:V-hom}
\V_{I}(\VV) = \left|\diag(\VV)_I-W^{\text{hom}}_I\right| = \left|\diag(\VV)_I\right|\left |\mathds{1}-W^{\text{hom}}_I/v\right| = v^{|I|}\left(1-\frac{|I|\alpha}{vN}\right)
\end{align}
With these simplifications, the probability of an empty avalanche (Lemma~\ref{lem:empty}) is given by
\[\P(\av(u,\omegaVk_1) = ()  \mid  \omegaVk_1= k)
  = 1 - \frac{\DeltaU \V_{[N]\setminus \{k\}}(\mathbf{1})}{\V_{[N]}(\mathbf{1})} = 1-\DeltaU \frac{N-(N-1)\alpha}{N-N\alpha} =
  \P(\av = ())\text{ .}\]

We will now consider \Eq{eq:pavu}, where $r^I$ is for this network given by \(r^I = W^{\text{hom}}\delta(I) = |I|\alpha/N\mathbf{1}\):

\[\P_k(\U(\av) = I) = \frac{\V_{I\setminus \{k\}}\left(r^I\right)\V_{[N]\setminus I}\left(\mathbf{1}-r^I\right)}{\V_{[N]\setminus \{k\}}(\mathbf{1})}.\]

The first term simplifies to
\[\V_{I\setminus \{k\}}\left(r^I\right) = \left(\frac{\alpha}{N}\right)^{|I|-1}|I|^{|I|-2} \text{ ,}\] which is the number of spanning trees in a complete graph of \(|I|\) units (Cayley's formula) weighted by the product of the \(|I|-1\) edge weights of each spanning tree.
The second term and the denominator are given by
\begin{align*}
  \V_{[N]\setminus I}\left(r^I\right) &= \left(1-\frac{|I|\alpha}{N}\right)^{N-|I|}\left(1-\frac{(N-|I|)\alpha}{N\left(1-|I|\frac{\alpha}{N}\right)}\right) = \left(1-\frac{|I|\alpha}{N}\right)^{(N-|I|-1)}(1-\alpha)\\
  \V_{[N]\setminus \{k\}}(\mathbf{1}) &= \left(1-\frac{(N-1)\alpha}{N}\right)
\end{align*}

Thus, \(\P_k(\U(\av) = I)\) depends in the homogeneous network only on \(|I|\) and is independent of the starting unit \(k\).
Putting these results together, the distribution of nonempty avalanches is given by
\begin{align*}
  \P(\s(\av)=n \mid \s(\av)>0)&= \sum_{\emptyset \neq I\subseteq N,|I|=n}\sum_{k\in I}\P_k(\U(\av)= I)\\
                         &= n\binom{N}{n}\left(\frac{\alpha}{N}\right)^{(n-1)}n^{(n-2)}
                        \left(1-\frac{n\alpha}{N}\right)^{N-n-1}
                        \frac{1-\alpha}{N}\left/\left(1-\frac{(N-1)\alpha}{N}\right)\right. \\
                         &= \binom{N}{n}\left(\frac{n\alpha}{N}\right)^{(n-1)}
                         \left(1-\frac{n\alpha}{N}\right)^{N-n-1} \frac{1-\alpha}{N-(N-1)\alpha}
\end{align*}                                   
which is equal to \cite[equation (8)]{eurich2002finite}, using \(\binom{N}{n}=\frac{N}{n}\binom{N-1}{n-1}\), and termed Abelian distribution \cite{levina2014abelian}. \\

%---------- analytical finite size critical coupling strength ---------------
For the homogeneous EHE-model it was shown~\cite{levina2008mathematical} that the avalanche size statistics converges in distribution to the statistics obtained from a Watson-Galton branching process. In this way, the homogeneous EHE-model behaves like a branching process and we may use the branching factor, approximated by the expected number of units in the second step of the avalanche $\mathds{E}(\s(\av_2) \mid  \av_1 \neq ())$, to find the critical coupling $\alpha_c$ at which large (but finite) networks display power-law avalanche size statistics. At this coupling, the branching factor should be one, i.e one unit causes on average one additional unit to fire in the next step of the avalanche. The expected number of units in the second step of the avalanche can be calculated using \Eq{eq:branching-theorem} for homogeneous networks to be
\[\mathds{E}\left(\rvert\av_2\lvert \mid  \av_1 = \{k\}\right) = \frac{(N-1)\alpha (N-(N-2)\alpha)}{N(N-(N-1)\alpha)} =  \mathds{E}\left(\rvert\av_2\lvert \mid  \av_1 \neq ()\right)\]

Setting $\mathds{E}(|\av_2| \mid  \av_1 \neq ())=1$ and solving for $\alpha<1$ we obtain 
\begin{align}
\alpha_c = \frac{N^2 - N\sqrt{N - 1} - N}{N^2 - 3N + 2}
\end{align}
For large $N$, this expression scales like $(1 - N^{-1/2})/N$, consistent with the numerical evidence for the homogeneous EHE-model \cite{eurich2002finite}.

The same calculation for the homogeneous coupling matrix without self-weights $W^{h}$, which was studied in section~\ref{sec:shift-invariant-networks} and for which $\V_{I}^{W^h}(\VV)= \V_{I}^{W^{\text{hom}}}(\VV+\alpha/N)$ (Proposition~\ref{prop:self-weights}), leads to  
\[\mathds{E}^{W^h}\left(\rvert\av_2\lvert \mid  \av_1 = \{k\}\right) = \frac{(N-1)\alpha((N-3)\alpha-N)}{(N-2)\alpha^2+N(N-3)\alpha-N^2} =  \mathds{E}^{W^h}\left(\rvert\av_2\lvert \mid  \av_1 \neq ()\right)\]
and to the critical coupling strength
\begin{align}\label{eq:alpha-crit-hom-ws}
\alpha_c^{W^h} = \frac{N^2 - N\sqrt{N - 1} - N}{N^2 - 5N + 5}
\end{align}

\subsubsection{Coupled homogeneous networks}\label{sec:hom-coupled}

In this subsection we will generalize the avalanche distribution of the homogeneous network to coupled homogeneous networks. 
Let \(W^{\text{block}} \in \mathbb{R}^{[N\times N]}\) be a \(c\times c\) block matrix, with each block being a homogeneous
matrix. Let \(0 \leq w_{ij} \) be the weight between units belonging to subnetworks (blocks) $i$ and $j$, \(0 < i,j \leq c\) and \(N_i>0\) be the total
number of units in subnetwork \(i\) with $\sum_{i=1}^c N_i=N$. We denote the \(c\times c\) matrix with entries \(w_{ij}\) by \(W^c\) and require $\sum_{j=1}^c w_{ij}N_j + \DeltaU < 1$ for all $i=1,\ldots,c$.
In this section will use the shorthand $\P = \P^{W^{\text{block}}}$ and set $\U = \mathbf{1}$.
Due to the block matrix structure, each assembly $I\subseteq [N]$ is characterized by the number of participating units in each subnetwork, which we denote by $\mathbf{n}(I) = (|I \cap \{1,\ldots,N_1\}|,\ldots,|I\cap \{N-N_c+1,\ldots,N\}|)$.
The index set of positive entries in the pattern is given by $\operatorname{pos}(\mathbf{n}) = \{i\in [c] \mid \mathbf{n}_i>0\}$.
For coupled homogeneous networks, avalanche assemblies are described by the vector containing the number of participating units in each subnetwork. 
%Our goal is to find a closed form expression for the
%assembly distribution
%\[P_k(\mathbf{n}_1,\ldots,\mathbf{n}_c) \coloneqq \P(|\U(\av)\cap {0,\ldots,N_1}|=\mathbf{n}_1,\ldots,|\U(\av)\cap {N-N_c,\ldots,N}|=\mathbf{n}_c) \text{ .}\]

Note that the rank of the matrix \(W^{\text{block}}\) is the same as the rank of \(W^c\diag{\mathbf{n}([N])}\) and both matrices have the same set of nonzero eigenvalues, thus
the volume of the inhabited region can be calculated by an \(c\times c\) determinant 
\[\lambda(D) = \V_{[N]}(\mathbf{1}) = \left|\mathds{1}_N-W^{\text{block}}\right| = \left|\mathds{1}_c-W^c\diag(\mathbf{n}([N]))\right| \text{. }\]

%The avalanche statistics for the assembly \(I\subseteq [N]\) only depends on the number of units in \(I\) in
%each subnetwork \(n(I)=(|I\cap\{0,\ldots,N_1\}|,\ldots,|I\cap \{N-N_c,\ldots,N\}|)^T\).

For simplification of the assembly probabilities, we need to compute phase space volumes \(\mathcal{V}_I(\VV)\) for block constant
vectors $\VV$, which we can characterize by a vector $v\in \R^c$ with $\VV_{\{1,\ldots,N_1\}}=v_1$,\ldots,$\VV_{\{N-N_c,\ldots,N\}}=v_c$.
As for the volume of the inhabited region, we have, assuming $v>0$ component wise, 
\begin{align*}
\V_I(\VV) = \left|\diag{\VV}_I-W^{\text{block}}_I\right| &= \left|\diag{\VV}_I\right|\left|\mathds{1}_I-\diag(\VV(I))^{-1}W^{\text{block}}_I\right| \\
&= \prod_{i\in \operatorname{pos}(\mathbf{n}(I))}\left(v_i^c\right)^{\mathbf{n}(I)} \left|\mathds{1}_{\operatorname{pos}(\mathbf{n}(I))}- 
W^c_{\operatorname{pos}(\mathbf{n}(I))}\diag(\mathbf{n}(I)/v_i^c)_{\operatorname{pos}(\mathbf{n}(I))}\right|\\
&= \prod_{i\in \operatorname{pos}(\mathbf{n}(I))}\left(v_i^c\right)^{\mathbf{n}(I)-1} \left|\diag(v)_{\operatorname{pos}(\mathbf{n}(I))}- 
W^c_{\operatorname{pos}(\mathbf{n}(I))}\diag(\mathbf{n}(I))_{\operatorname{pos}(\mathbf{n}(I))}\right|\\
&=: \V^c(v,\mathbf{n}(I))\text{ .}
\end{align*}

%\(V_{I}(\VV)\) with block-constant vector \(\VV\), which we represent by $v$ where $V_{\{1,\ldots,\mathbf{n}_1\}}=v_1$,\ldots,$V_{\{N-\mathbf{n}_c,\ldots,N\}}=v_c$.Taken together, we have

%\begin{align*}
%  V_I(\VV) &= |\diag{\VV_I}-W^{\text{block}}_I| = |\diag{\VV_I}||\mathds{1}_I-W^{\text{block}}_I| \\
%                  &= \prod_{i\in c(I)}^c(v_i)^{n(I)_i} |\mathds{1}_{c(I)} -W^c_{c(I)}\diag(n(I)_{c(I)}/v_{c(I)})| \\
%                  &= \prod_{i\in c(I)}^c(v_i)^{n(I)_i-1}|\diag{v}_{c(I)} -W^c_{c(I)}\diag(N(I)_{c(I)}) =:  \V^c(v,I) 
%\end{align*}

%\begin{align*}
%  V_I(\VV) &= |\diag{\VV_I}-W^{\text{block}}_I| = |\diag{\VV_I}||\mathds{1}_I-W^{\text{block}}_I| \\
%                  &= \prod_{i\in c(I)}^c(v_i)^{n(I)_i} |\mathds{1}_{c(I)} -W^c_{c(I)}\diag(n(I)_{c(I)}/v_{c(I)})| \\
%                  &= \prod_{i\in c(I)}^c(v_i)^{n(I)_i-1}|\diag{v}_{c(I)} -W^c_{c(I)}\diag(N(I)_{c(I)}) =:  \V^c(v,I) 
%\end{align*}
%where \(c(I)\) is the index set of the positive entries in \(n(I)\) and we assumed that $\VV_I>0$.

The assembly distribution is thus given by 
\[\P_{k}(\U(\av) = I) = \begin{cases}
    0 \mbox{ if } (W^c\mathbf{n}(I))_j = 0 \text{ for some } j\in c(I) \\
    \V^c(W^c\mathbf{n}(I),I\setminus \{k\})\V^c(\mathbf{1}_c-W^c\mathbf{n}(I))/\V^c(\mathbf{1}_c,[N]\setminus \{k\}) \mbox{ otherwise} 
\end{cases}\]
The condition in the first case is true if and only if the subgraph formed by nodes \(I\) is not connected and in this case there are no spanning trees of the assembly network. Calculating phase space volumes with \(\V^c\) only needs to evaluate determinants of matrices with at most \(c\times c\) dimensions.

If the graph \(W^c\) is fully connected, i.e. \(W^c>0\) component wise, the case distinction in the assembly distribution is not needed. In this case, the probability distribution of \(\P_k(\mathbf{n}(\U(\av)))\) reduces to the expression reported in \cite[equation 7]{leleu2015unambiguous}.

\subsubsection{Two homogeneously coupled subnetworks}\label{sec:two-hom-coupled}

As a prototypical example for coupled homogeneous networks we consider two coupled subnetworks with \(N_s\) units each
and block coupling matrix \(W^c\) given by
\[W^c = \frac{1}{N_s}\begin{pmatrix}\alpha & \beta \\ \beta & \alpha \end{pmatrix}\]

Note that each unit in this network receives internal activation of $\alpha+\beta$ in a global avalanche. Thus we require $\alpha+\beta+\DeltaU < 1$.
% 
% To shorten the notation, we use $\alpha' \coloneqq \alpha/N_s,\beta' \coloneqq \beta/N_s$ in the following equations.
%
Simplifying the avalanche statistics according to the steps above and explicitly calculating the determinants for $\beta'>0$ leads to
the following distribution for sizes of non-empty avalanches:
% \begin{align}\label{eq:avs-two-coupled-hom}
% \P^{W^c}(\s(\av) = n \mid \s(\av) > 0) &= \frac{1}{P_0}\sum_{k=0}^n\binom{N_s}{k}\binom{N_s}{n-k}\frac{n\beta' s_1^ks_2^{n-k}S_1^{l_1-1}S_2^{l_2-1}((S_1-\alpha' l_1)(S_2-\alpha' l_2)-l_1l_2\beta'^2)}{2N_s(k(n-k)(\alpha'^2+\beta'^2)+\alpha'\beta' (k^2+(n-k)^2))}\end{align}
% where
% \begin{align*}
% s_1 &= k\alpha' + (n-k)\alpha'_{c}\text{, }s_2 = k\beta' + (n-k)\alpha'\\  
% S_1 &= 1-s_1\text{, } S_2 = 1-s_2 \\
% l_1 &= N_s-k\text{, }l_2 =N_s-n+k\\ 
% P_{0} &= 1-\alpha'(2N_s-1)+(\alpha'^2-\beta'^2(N_s(N_s-1))) \text{ .}
% \end{align*}

% \begin{align}\label{eq:avs-two-coupled-hom}
%     \P^{W^c}(&\s(\av) = n \mid \s(\av) > 0) \nonumber\\
%     &= \frac{\beta}{2 P_0 N_s^{2N_s}}
%     \sum_{k=0}^n\binom{N_s}{k}\binom{N_s}{n-k}
%     \frac{n x_1^k x_2^{n-k} X_1^{l_1-1} X_2^{l_2-1}
%     ((X_1-\alpha l_1)(X_2-\alpha l_2)-l_1 l_2\beta^2)}
% %    {k(n-k)(\alpha^2+\beta^2)+\alpha\beta (k^2+(n-k)^2)}
%     {n^2 \alpha \beta + k(n-k)(\alpha-\beta)^2}
% \end{align}
% where
% \begin{align*}
%     x_1 &= k\alpha + (n-k)\beta\text{, \hspace*{0.3cm}} x_2 = k\beta + (n-k)\alpha\\
%     X_1 &= 1-x_1\text{, \hspace*{0.3cm}} X_2 = 1-x_2 \\
%     l_1 &= N_s-k\text{, \hspace*{0.3cm}}l_2 =N_s-n+k\\ 
%     P_{0} &= 1-\alpha\left(2-\frac{1}{N_s}\right)
%         +\left(\alpha^2-\beta^2 \left(1-\frac{1}{N_s}\right)\right)
%         \text{ .}
% \end{align*}

\begin{align}\label{eq:avs-two-coupled-hom}
    \P^{W^c}(&\s(\av) = n \mid \s(\av) > 0) \nonumber\\
    &= \frac{\beta}{P_0 N_s^{2N_s-1}}
    \sum_{k=0}^n\binom{N_s}{k}\binom{N_s}{n-k}
    \frac{n x_1^k x_2^{n-k} X_1^{l_1-1} X_2^{l_2-1}
    ((X_1-\alpha l_1)(X_2-\alpha l_2)-l_1 l_2\beta^2)}
    {\alpha \beta n^2 -(\alpha-\beta)^2 k(k-n)}
%    {\alpha \beta (k^2+(k-n)^2)-(\alpha^2+\beta^2)k(k-n)}
\end{align}
where
\begin{align*}
    x_1 &= k\alpha + (n-k)\beta\text{, \hspace*{0.3cm}} x_2 = k\beta + (n-k)\alpha\\
    X_1 &= N_s-x_1\text{, \hspace*{0.3cm}} X_2 = N_s-x_2 \\
    l_1 &= N_s-k\text{, \hspace*{0.3cm}}l_2 =N_s-n+k\\ 
    P_{0} &= 2N_s - 2\alpha(2N_s-1)+2(N_s-1)(\alpha^2-\beta^2)
        \text{ .}
\end{align*}
There is an intuitive explanation for the terms in the simplification:
\(k\) indicates the number of units from one subnetwork participating in the avalanche and \(n-k\) the corresponding number of units from the other subnetwork. $x_1/N_s$ and $x_2/N_s$ represent the input given to a unit in the subnetworks, while $X_1/N_s$ and $X_2/N_s$ denote the upper boundaries for the states of units in the subnetworks not participating in the avalanche. $l_1$ and $l_2$ are the numbers of silent units in the subnetworks, and \(P_0\) is a normalization constant.

\subsubsection{One-dimensional ring and line networks}

Efficiently calculating the avalanche size distribution is possible if all avalanche assemblies of a given size (or pattern as in the coupled subnetwork case) have the same distribution or if the number of possible assemblies is restricted by the network. The latter is the case in sparsely coupled network, like one-dimensional ring or line networks.

In the one-dimensional ring network with $N$ units, each unit is connected bidirectionally to its two nearest neighbors with coupling weight \(\alpha/2 \). Thus, the coupling matrix $W^{\text{ring}}$ is a circulant matrix with just two positive entries $\alpha/2$ in each column. Thus we require $\alpha+\DeltaU < 1$. This simple form of $W^{\text{ring}}$ allows to specify the volume $\lambda(D)$ of the inhabited region in closed form:
% \[\lambda(D) = 1-\prod_{j=0}^{N-1}\left(1-2\alpha'\cos\left({2\pi j}/{N}\right) \right)\text{ .}\]
\[
    \lambda(D) = 1-\prod_{j=0}^{N-1}\left(1-\alpha\cos\left({2\pi j}/{N}\right) \right)\text{ .}
\]

Due to the sparsity of the networks, the connected assemblies are always simple line segments. This can be used to find a formula for the avalanche size distribution in the ring and line networks:

% \begin{align}\label{eq:avs_ring}
%     P^{W_{\text{ring}}}(\s(\av) = n \mid \s(\av)>0) = \frac{n\alpha'^{n-1}}{P^r_0}v^r(N-n)\, ,
% \end{align}
% where $P^r_0$ and $v(n)$ are given by 
% \[P^r_0 = (1-l_2)/(l_1-l_2)l_1^{N-1}+(l_1-1)/(l_1-l_2)l_2^{N-1}\]
% %    = \frac{\sqrt{1-\alpha^2}}{4}( l_1^N - l_2^N )\]
% \[v^r(n) = \begin{cases} 
%             1 &\mbox{ if } n=0 \\
%             1-2\alpha' &\mbox{ if } n=1 \\
%             (1-\alpha')(Cl_1^{n-1}+(1-C)l_2^{n-1})-\alpha'^2(Cl_1^{n-2}+(1-C)l_2^{n-2})
%             &\mbox{ otherwise}
%         \end{cases}\]
% and $l_1 = 1/2 + (1/4 - \alpha'^2)^{1/2},l_2 = 1/2-(1/4-\alpha'^2)^{1/2},C=(1-\alpha'-l_2)/(l_1-l_2)$.

\begin{align}\label{eq:avs_ring}
\P^{W_{\text{ring}}}(\s(\av) = n \mid \s(\av)>0) = n\left(\frac{\alpha}{2}\right)^{n-1}\frac{v^r(N-n)}{P^r_0}\, ,
\end{align}
where $P^r_0$ and $v(n)$ are given by
\begin{align*}
P^r_0 &= \frac{(1+a)^N-(1-a)^N}{a2^N}\\
v^r(n) &= \begin{cases}
            1 &\mbox{ if } n=0 \\
            1-\alpha &\mbox{ if } 0<n<3 \\
            a\dfrac{(1+a)^n-(1-a)^n}{(\alpha+1)2^n} &\mbox{ otherwise}
\end{cases}
\end{align*}
and $a=\sqrt{1-\alpha^2}$.
Note that there are $N$ line segments with size $n<N$ on the ring. For $n=N$, there is only one line segment which is the full ring. However, there are now $N$ possible spanning trees instead of only one spanning tree for each line segment with $n<N$.

The coupling matrix $W^{\text{line}}$ of the line network, which arises from the ring network by deletion of a single (undirected) edge, is a tridiagonal matrix with zeros on the diagonal and $\alpha/2$ on the off diagonals. The avalanche size distribution for the line network has a similar form as for the ring network, but note that in contrast to the ring network, the factor $\V_{[N]\setminus I}(W\delta(I))$ in \Eq{eq:pavu_inline} depends in the line network on the number of units to the left and to the right of the line segment corresponding to an assembly: 
% \begin{align}\label{eq:avs_line}
%     P^{W_{\text{line}}}(\s(\av) = n \mid \s(\av)>0) = \frac{n\alpha'^{n-1}}{P^l_0}\sum_{j=0}^{j=N-n} v^l(j)v^l(N-n-j)\, ,
% \end{align}
% where $v^l(n)$ and $P^l_0$ are given by
% \[P^l_0 = \sum_{j=0}^{j=N-1} \left(\frac{1-l_2}{l_1-l_2} l_1^j + \frac{l_1-1}{l_1-l_2}l_2^j \right)\left( \frac{1-l_2}{l_1-l_2} l_1^{N-j-1} + \frac{l_1-1}{l_1-l_2}l_2^{N-j-1} \right) \\
%      = \sum_{j=0}^{j=N-1} (l_1^{j+1} - l_2^{j+1}) (l_1^{N-j} - l_2^{N-j})
% \]
% \[\mbox{with} \,\,\, v^l(n) = \begin{cases}
% 1 &\mbox{ if } n=0 \\
% 1-\alpha' &\mbox{ if } n = 1\\
% Cl_1^n+(1-C)l_2^n &\mbox{ otherwise}
% \end{cases}\]

\begin{align}\label{eq:avs_line}
\P^{W_{\text{line}}}(\s(\av) = n \mid \s(\av)>0) =
\frac{n}{P^l_0}\left(\frac{\alpha}{2}\right)^{n-1}\sum_{j=0}^{j=N-n} v^l(j)v^l(N-n-j)\, ,
\end{align}
where $v^l(n)$ and $P^l_0$ are given by
\begin{align*}
    P^l_0 &= \sum_{j=0}^{j=N-1}\frac{\left((1+a)^{j+1}-(1-a)^{j+1})((1+a)^{N-j}-(1-a)^{N-j}\right)}{a^22^{N+1}}\\
    v^l(n) &= \begin{cases}
        1 &\mbox{ if } n=0\\
        \dfrac{(1-\alpha+a)(1+a)^n+(\alpha+a-1)(1-a)^n}{a2^{n+1}}  &\mbox{ otherwise}
    \end{cases}
\end{align*}

\subsubsection{Erdős–Rényi network}
For random graphs in which edges are independently sampled from a distribution, here exemplified by an (undirected) Erdős–Rényi graph, the expected avalanche size distribution can be well approximated by the expected
probability of an assembly of size \(n\). In this graph, each undirected edge occurs with probability $p$ and weight $\alpha/N$ independently of all other edges.

In order to compute this expected assembly distribution, the expected values
of the assembly Laplacian and of $\V_I(\VV)$ have to be determined. For an Erdős–Rényi graph with \(n\)
nodes, connection probability \(p\) and weight \(\alpha/N\), the expected graph Laplacian is just \((\alpha/N)^{n-1}\)
times the expected number of spanning trees in the random graph, which is particularly simple since there are \(n^{n-2}\)
spanning trees in the complete graph and each of the spanning trees occurs in the random graph with probability
\(p^{n-1}\). Taken together, we have \(\E(\V_{I\setminus \{k\}}(W\delta(I))) = (p\alpha/N)^{|I|-1}|I|^{|I|-2}\).

The expected determinant in the more general expression $\V_I(\VV)$ is more difficult to determine, since the diagonal elements \(\VV_i-w_{ii}\) have different moments than the off diagonal entries \(-w_{ij}\). A consequence of these different statistics is that in the Leibniz formula of the determinant, expected values for cycles in permutations differ depending on the cycle length (since cycles of length one involve a diagonal element, length two cycles the same edge twice, and longer cycles independent edges). Thus, the expected determinant is given by a \emph{cycle index} of the permutation group \(S_n\) for which generating functions are known (see \cite[Eq.  (5.30)]{stanleyenumerativeVol2}).
With these combinatorial results, the  expected value of $\V_I(\VV)$ for independent entries \(\VV_i\) can be given in terms of Hermite polynomials \(H_n\). As an example, we supply the expression for the expected volume of the inhabited region. Implementation of the analytical avalanche distribution using this technique will be made available by the authors upon reasonable request. With $z \coloneqq (1+p\alpha/N)/\sqrt{2(\alpha/N)^2(p-p^2)}$ we have
\begin{align*}
  \E(\lambda(D)) = \E(\V_{[N]}(\mathbf{1})) %&=
%2^{-N/2}((\alpha/N)^2(p-p^2))^{N/2}H_N(z)
%                                               -2^{-(N-1)/2}p(\alpha/N) N ((\alpha/N)^2(p-p^2))^{(N-1)/2}H_{N-1}(z).\\
    &= \left(\frac{\alpha}{N}\sqrt{\frac{p(1-p)}{2}} \right)^N
        \left( H_N(z) + N \sqrt{\frac{2p}{1-p}} H_{N-1}(z) \right)
\end{align*}

Note that unlike the expected number of spanning trees, the expected volume of the inhabited region is different than the corresponding volume for the homogeneous matrix with entries \(p(\alpha/N)\).

% numerical code for reference
% n = 300
% alpha = 1/n
% a = 1; b = p*alpha**2;c = (-p*alpha)
% d = a-c;v = b-c**2
% xh = (d/mp.sqrt(v))/mp.sqrt(2)
% edet  = (mp.power(2,-(n-1)/2)*c*n*mp.power(v,(n-1)/2)*mp.hermite(n-1,xh)+
% mp.power(2,-n/2)*mp.power(v,n/2)*mp.hermite(n,xh))
% simplifications:
% v = p\alpha^2 - p^2\alpha^2 = \alpha^2(p-p^2)
% d = 1+p\alpha
% xh = (1+p\alpha)/\sqrt{(2\alpha^2(p-p^2))}

%\bibliography{ehereferences}% Produces the bibliography via BibTeX.
%\end{widetext}

\pagebreak
\newpage

\subsection{List of variables and notation}

{
{\singlespacing
\begin{longtable}{ll}
$N,[N]$ & number of  neurons/units, set   $\{1,\ldots,N\}$ \\
$I,J,H$ & non-empty subsets of [N]\\
$X^Y$ & set of functions from $Y$ to $X$, e.~g.\ $\R^{I}$, $\TN$\\
$i, j, k$				&   indices for units\\
$u,z$				& states in phase space $u\in C,z\in \TN$\\
$s, t$					& iteration indices\\
$W, w_{ij}$			& coupling matrix, interaction weight from $j$ to $i$\\
$U_I,W_{I}$ & restriction of the vector $U$ to the index set $I$ and matrix $W$ to the index set ${I\times I}$\\
$G(W),E(W)$ & directed graph and edge set induced by $W$ \\
$S,S_k$ & set of directed (outgoing) spanning trees (rooted at $k$) \\
$\w(S)$ & sum of product of edge weights for all trees in $S$ \\
$\cut(I)$ & weight of directed graph cut $\cut(I)= (W\delta(I))_{[N]\setminus I}$\\
$\mathcal{L}(W)$ & directed graph Laplacian $W\1-W$\\
$\mathcal{L}^{(k)}(W)$ & $(k,k)$ cofactor of graph Laplacian \\
$\Omega^k$ & matrix of generalized effective $k$ resistances \Eq{eq:resistance}\\
%$\mathcal{W}$		& set of coupling matrices\\
$|\cdot|$ & cardinality of sets, determinant of square matrices \\
$\mathcal{A},\mathcal{A}_k$ & set of all avalanches/nonempty avalanches started by unit $k$\\
$a, a_i$				& avalanche, generation $i$ of an avalanche\\
$\dur(a),\s(a)$		& avalanche duration, size\\
$\U(a)$, $\U_j(a)$		& set of units (assembly) in avalanche $a$ (until generation $j$) \\
$\e_i,\delta(I)$					& unit vector in $\R^{[N]}$ in direction $i$,$\sum_{i\in I}\e_i$\\
$\1,\mathds{1}$    & constant $1$-vector $\sum_{i\in [N]}\e_i$, identity matrix $\diag{(\1)}$\\
$U,\mathbb{U}$							& firing thresholds $U\in \R^{[N]}$,$\mathbb{U}\coloneqq\diag{U}$ \\
$M$							& $M=(\diag(\mathbb{U})-W)^{-1}$; maps from state space to Torus\\
$r^I$ & vector $W(\delta(I))$ of internal activation during avalanche with assembly $I$ \\
$C$							& phase space $C \coloneqq \bigtimes_{i\in [N]} [0, U_i)$\\
$D$							& inhabited region $D=C\setminus \Lambda = \Theta(\TN)$\\
$\Lambda$				& non-inhabited region \Eq{eq:Dmain}\\
$A$ & indicator vector of supra-threshold units $A\in \R^{[N]}$\\ 
$F,\fixF$ & (one generation of) avalanche dynamics\\
$T_k$ & $T_k(u) = \fixF(u+\e_k \DeltaU)$, maps to new state after external input to unit $k$\\
$\Sigma_N$			& space of right-infinite (input unit index) sequences $\Sigma_N = [N]^{\N}$\\
$\mathcal{B}$   & Borel $\sigma$-algebra \\
$\Vk$  & element of $\Sigma_N$, $\Vk=(k_1,k_2,\ldots)$\\
%$\omega$ & element of $\Sigma_N$, $\omega=(\omega_1,\omega_2,\ldots)$ (used in Appendix)\\
$\sigma$ & left shift operator on $\Sigma_N$\\
$T$				& model dynamics formalised as skew product $T(\Vk,u)=T(\sigma(\Vk),T_{k_1}(u))$\\
$\TN$						& $N$-torus\\
$\Theta$				& quotient map from $N$-torus to inhabited region $D$\\
$\hat{T}, \hat{T}_g$ & ($g$-extended) equivalent dynamics on $N$-torus\\
$p,\mathbb{B}_p$ & vector of input probabilities,   Bernoulli measure on $\Sigma_N$ with respect to $p$\\
$L$ & support of $p$ (set of units receiving external input)\\
$\lambda,\lambda_D$				& Lebesgue measure, normalized Lebesgue measure supported on $D$\\
$\P,\E,\operatorname{cov}$ & probability, expectation, covariance operator on $(\Sigma_N \times C,\mathcal{B},\mathbb{B}_p \times \lambda_D$)\\
$\av(k, u)$			& avalanche function returning the avalanche upon input to $k$ from state $u$\\
$\av(\Vk,u)$        & $\av(\Vk,u)\coloneqq \av(k_1,u)$ for $\Vk=(k_1,k_2,\ldots) $ (random variable on $(\Sigma_N \times C,\mathcal{B})$)\\
$\NN^{s}(\Vk,u)$				& spike count vector (random variable) after $s$ iterations from state $(\Vk,u)$\\
$\mathcal{V}_I(\VV)$ & volume of inhabited region $\mathcal{V}_I(\VV)=| \diag(\VV)_I - W_I |$ \Eq{eq:def_vi}\\
$\pi_1, \pi_2$	& projection to first/second component of input-state space $\Sigma_N \times C$\\
$\pi_I$ & natural projection from $C$ to $\R^I$
\end{longtable}
}}
%%% Local Variables:
%%% mode: latex
%%% TeX-master: "EHEpaper"
%%% End:

%\end{document}
%%% Local Variables:
%%% mode: latex
%%% TeX-master: "EHEpaper"
%%% End:

\pagebreak

\bibliography{ehereferences}% Produces the bibliography via BibTeX.

\pagebreak

%\begin{figure*}
%  \includegraphics[width=0.9\textwidth]{img/D_Lambda_connection_illustration.png}
%  \caption{\label{fig:DL-illus}Shifts of \(M^{-1}n, 0\neq n\in {0,1}^N\) illustrated in two dimensions.
%    Left panel: For all points in \(D\), all three possible shifts lead to points outside of \(D\). This is formalized
%    in Lemma~\ref{lem:D}. Right panel: For all points in \(\Lambda_{[N]}\) exists a shift which leads to a point in \(D\). Exemplified with points \(\tilde{u}^{(1)} = u^{(1)}+M^{-1}(0,1)^T,\tilde{u}^{(2)} = u^{(2)}+M^{-1}(1,0)^T,\tilde{u}^{(3)} = u^{(1)}+M^{-1}(1,1)^T \). Together with the left panel, this implies that \(D = C \setminus \Lambda_{[N]}\). This is formalized in Theorem~\ref{thm:DGamma}.}
%\end{figure*}

\end{document}